\documentclass[11pt]{article}
\usepackage{fullpage}
\usepackage{times}

\AtBeginDocument{
  }

\usepackage{xcolor}
\usepackage{colortbl}
\usepackage{amsmath}
\usepackage{amsfonts}
\usepackage{amssymb}
\usepackage{graphicx}
\usepackage{url}
\usepackage{hyperref}
\usepackage{babel}

\usepackage{wrapfig}

\usepackage{prettyref}
\newcommand{\rref}[2][]{\prettyref{#2}}
\newrefformat{sec}{Sect.\,\ref{#1}}
\newrefformat{def}{Definition\,\ref{#1}}
\newrefformat{thm}{Theorem\,\ref{#1}}
\newrefformat{app}{Appendix\,\ref{#1}}
\newrefformat{prop}{Proposition\,\ref{#1}}
\newrefformat{rem}{Remark\,\ref{#1}}
\newrefformat{lem}{Lemma\,\ref{#1}}
\newrefformat{cor}{Corollary\,\ref{#1}}
\newrefformat{ex}{Example\,\ref{#1}}
\newrefformat{tab}{Table\,\ref{#1}}
\newrefformat{fig}{Fig.\,\ref{#1}}
\newrefformat{case}{case\,\ref{#1}}
\newrefformat{foot}{Footnote\,\ref{#1}}
\newrefformat{para}{Paragraph\,\ref{#1}}

\usepackage{isabelle}
\usepackage{isabellesym}

\newcommand{\snip}[4]{\expandafter\newcommand\csname #1\endcsname{#4}}
\snip{releaseuntildual}{1}{2}{%
\isacommand{lemma}\isamarkupfalse%
\ release{\isacharunderscore}{\kern0pt}until{\isacharunderscore}{\kern0pt}dual{\isacharcolon}{\kern0pt}\ \isanewline
\ \ \isakeyword{fixes}\ a\ b{\isacharcolon}{\kern0pt}{\isacharcolon}{\kern0pt}{\isachardoublequoteopen}nat{\isachardoublequoteclose}\ \isakeyword{assumes}\ a{\isacharunderscore}{\kern0pt}leq{\isacharunderscore}{\kern0pt}b{\isacharcolon}{\kern0pt}\ {\isachardoublequoteopen}a\ {\isasymle}\ b{\isachardoublequoteclose}\isanewline
\ \ \isakeyword{shows}\ {\isachardoublequoteopen}{\isasymphi}\ R\isactrlsub m{\isacharbrackleft}{\kern0pt}a{\isacharcomma}{\kern0pt}b{\isacharbrackright}{\kern0pt}\ {\isasympsi}\ {\isasymequiv}\isactrlsub m\ Not\isactrlsub m\ {\isacharparenleft}{\kern0pt}{\isacharparenleft}{\kern0pt}Not\isactrlsub m\ {\isasymphi}{\isacharparenright}{\kern0pt}\ U\isactrlsub m{\isacharbrackleft}{\kern0pt}a{\isacharcomma}{\kern0pt}b{\isacharbrackright}{\kern0pt}\ {\isacharparenleft}{\kern0pt}Not\isactrlsub m\ {\isasympsi}{\isacharparenright}{\kern0pt}{\isacharparenright}{\kern0pt}{\isachardoublequoteclose}%
}
\snip{futureasuntil}{1}{2}{%
\isacommand{lemma}\isamarkupfalse%
\ future{\isacharunderscore}{\kern0pt}as{\isacharunderscore}{\kern0pt}until{\isacharcolon}{\kern0pt}\isanewline
\ \ \isakeyword{fixes}\ a\ b{\isacharcolon}{\kern0pt}{\isacharcolon}{\kern0pt}{\isachardoublequoteopen}nat{\isachardoublequoteclose}\ \isakeyword{assumes}\ {\isachardoublequoteopen}a\ {\isasymle}\ b{\isachardoublequoteclose}\isanewline
\ \ \isakeyword{shows}\ {\isachardoublequoteopen}F\isactrlsub m{\isacharbrackleft}{\kern0pt}a{\isacharcomma}{\kern0pt}b{\isacharbrackright}{\kern0pt}\ {\isasymphi}\ {\isasymequiv}\isactrlsub m\ True\isactrlsub m\ U\isactrlsub m{\isacharbrackleft}{\kern0pt}a{\isacharcomma}{\kern0pt}b{\isacharbrackright}{\kern0pt}\ {\isasymphi}{\isachardoublequoteclose}%
}
\snip{globallyasrelease}{1}{2}{%
\isacommand{lemma}\isamarkupfalse%
\ globally{\isacharunderscore}{\kern0pt}as{\isacharunderscore}{\kern0pt}release{\isacharcolon}{\kern0pt}\isanewline
\ \ \isakeyword{fixes}\ a\ b{\isacharcolon}{\kern0pt}{\isacharcolon}{\kern0pt}{\isachardoublequoteopen}nat{\isachardoublequoteclose}\ \isakeyword{assumes}\ {\isachardoublequoteopen}a\ {\isasymle}\ b{\isachardoublequoteclose}\isanewline
\ \ \isakeyword{shows}\ {\isachardoublequoteopen}G\isactrlsub m{\isacharbrackleft}{\kern0pt}a{\isacharcomma}{\kern0pt}b{\isacharbrackright}{\kern0pt}\ {\isasymphi}\ {\isasymequiv}\isactrlsub m\ False\isactrlsub m\ R\isactrlsub m{\isacharbrackleft}{\kern0pt}a{\isacharcomma}{\kern0pt}b{\isacharbrackright}{\kern0pt}\ {\isasymphi}{\isachardoublequoteclose}%
}
\snip{globallyfuturedual}{1}{2}{%
\isacommand{lemma}\isamarkupfalse%
\ globally{\isacharunderscore}{\kern0pt}future{\isacharunderscore}{\kern0pt}dual{\isacharcolon}{\kern0pt}\isanewline
\ \ \isakeyword{fixes}\ a\ b{\isacharcolon}{\kern0pt}{\isacharcolon}{\kern0pt}{\isachardoublequoteopen}nat{\isachardoublequoteclose}\ \isakeyword{assumes}\ {\isachardoublequoteopen}a\ {\isasymle}\ b{\isachardoublequoteclose}\isanewline
\ \ \isakeyword{shows}\ {\isachardoublequoteopen}G\isactrlsub m{\isacharbrackleft}{\kern0pt}a{\isacharcomma}{\kern0pt}b{\isacharbrackright}{\kern0pt}\ {\isasymphi}\ {\isasymequiv}\isactrlsub m\ Not\isactrlsub m\ {\isacharparenleft}{\kern0pt}F\isactrlsub m{\isacharbrackleft}{\kern0pt}a{\isacharcomma}{\kern0pt}b{\isacharbrackright}{\kern0pt}\ {\isacharparenleft}{\kern0pt}Not\isactrlsub m\ {\isasymphi}{\isacharparenright}{\kern0pt}{\isacharparenright}{\kern0pt}{\isachardoublequoteclose}%
}
\snip{theoremone}{1}{2}{%
\isacommand{theorem}\isamarkupfalse%
\ formula{\isacharunderscore}{\kern0pt}progression{\isacharunderscore}{\kern0pt}decomposition{\isacharcolon}{\kern0pt}\isanewline
\ \ \isakeyword{fixes}\ {\isasymphi}{\isacharcolon}{\kern0pt}{\isacharcolon}{\kern0pt}{\isachardoublequoteopen}{\isacharprime}{\kern0pt}a\ mltl{\isachardoublequoteclose}\ \isakeyword{assumes}\ {\isachardoublequoteopen}k\ {\isasymge}\ {\isadigit{1}}{\isachardoublequoteclose}\ \isakeyword{and}\ {\isachardoublequoteopen}k\ {\isasymle}\ length\ {\isasympi}{\isachardoublequoteclose}\isanewline
\ \ \isakeyword{shows}\ {\isachardoublequoteopen}formula{\isacharunderscore}{\kern0pt}progression\ {\isacharparenleft}{\kern0pt}formula{\isacharunderscore}{\kern0pt}progression\ {\isasymphi}\ {\isacharparenleft}{\kern0pt}take\ k\ {\isasympi}{\isacharparenright}{\kern0pt}{\isacharparenright}{\kern0pt}\ {\isacharparenleft}{\kern0pt}drop\ k\ {\isasympi}{\isacharparenright}{\kern0pt}\isanewline
\ \ \ \ \ \ \ \ \ {\isacharequal}{\kern0pt}\ formula{\isacharunderscore}{\kern0pt}progression\ {\isasymphi}\ {\isasympi}{\isachardoublequoteclose}%
}
\snip{theoremtwo}{1}{2}{%
\isacommand{theorem}\isamarkupfalse%
\ satisfiability{\isacharunderscore}{\kern0pt}preservation{\isacharcolon}{\kern0pt}\isanewline
\ \ \isakeyword{fixes}\ {\isasymphi}{\isacharcolon}{\kern0pt}{\isacharcolon}{\kern0pt}{\isachardoublequoteopen}{\isacharprime}{\kern0pt}a\ mltl{\isachardoublequoteclose}\isanewline
\ \ \isakeyword{assumes}\ {\isachardoublequoteopen}k\ {\isasymge}\ {\isadigit{1}}{\isachardoublequoteclose}\ \isakeyword{and}\ {\isachardoublequoteopen}k\ {\isacharless}{\kern0pt}\ length\ {\isasympi}{\isachardoublequoteclose}\ \isakeyword{and}\ {\isachardoublequoteopen}intervals{\isacharunderscore}{\kern0pt}welldef\ {\isasymphi}{\isachardoublequoteclose}\isanewline
\ \ \isakeyword{shows}\ {\isachardoublequoteopen}{\isacharparenleft}{\kern0pt}drop\ k\ {\isasympi}{\isacharparenright}{\kern0pt}\ {\isasymTurnstile}\isactrlsub m\ {\isacharparenleft}{\kern0pt}formula{\isacharunderscore}{\kern0pt}progression\ {\isasymphi}\ {\isacharparenleft}{\kern0pt}take\ k\ {\isasympi}{\isacharparenright}{\kern0pt}{\isacharparenright}{\kern0pt}\ {\isasymlongleftrightarrow}\ {\isasympi}\ {\isasymTurnstile}\isactrlsub m\ {\isasymphi}{\isachardoublequoteclose}%
}
\snip{theoremthree}{1}{2}{%
\isacommand{theorem}\isamarkupfalse%
\ formula{\isacharunderscore}{\kern0pt}progression{\isacharunderscore}{\kern0pt}correctness{\isacharcolon}{\kern0pt}\isanewline
\ \ \isakeyword{fixes}\ {\isasymphi}{\isacharcolon}{\kern0pt}{\isacharcolon}{\kern0pt}{\isachardoublequoteopen}{\isacharprime}{\kern0pt}a\ mltl{\isachardoublequoteclose}\isanewline
\ \ \isakeyword{assumes}\ {\isachardoublequoteopen}intervals{\isacharunderscore}{\kern0pt}welldef\ {\isasymphi}{\isachardoublequoteclose}\ \isakeyword{and}\ {\isachardoublequoteopen}length\ {\isasympi}\ {\isasymge}\ complen{\isacharunderscore}{\kern0pt}mltl\ {\isasymphi}{\isachardoublequoteclose}\isanewline
\ \ \isakeyword{shows}\ {\isachardoublequoteopen}{\isacharparenleft}{\kern0pt}{\isacharparenleft}{\kern0pt}formula{\isacharunderscore}{\kern0pt}progression\ {\isasymphi}\ {\isasympi}{\isacharparenright}{\kern0pt}\ {\isasymequiv}\isactrlsub m\ True{\isacharunderscore}{\kern0pt}mltl{\isacharparenright}{\kern0pt}\ \ {\isasymlongleftrightarrow}\ {\isasympi}\ {\isasymTurnstile}\isactrlsub m\ {\isasymphi}{\isachardoublequoteclose}%
}
\snip{complenboundedbyone}{1}{2}{%
\isacommand{lemma}\isamarkupfalse%
\ complen{\isacharunderscore}{\kern0pt}bounded{\isacharunderscore}{\kern0pt}by{\isacharunderscore}{\kern0pt}{\isadigit{1}}{\isacharcolon}{\kern0pt}\isanewline
\ \ \isakeyword{fixes}\ {\isasympi}{\isacharcolon}{\kern0pt}{\isacharcolon}{\kern0pt}{\isachardoublequoteopen}{\isacharprime}{\kern0pt}a\ set\ list{\isachardoublequoteclose}\isanewline
\ \ \isakeyword{assumes}\ {\isachardoublequoteopen}intervals{\isacharunderscore}{\kern0pt}welldef\ {\isasymphi}{\isachardoublequoteclose}\ \isakeyword{and}\ {\isachardoublequoteopen}{\isadigit{1}}\ {\isasymge}\ complen{\isacharunderscore}{\kern0pt}mltl\ {\isasymphi}{\isachardoublequoteclose}\isanewline
\ \ \isakeyword{shows}\ {\isachardoublequoteopen}{\isacharparenleft}{\kern0pt}{\isasymforall}{\isasymeta}{\isachardot}{\kern0pt}\ semantics{\isacharunderscore}{\kern0pt}mltl\ {\isasymeta}\ {\isacharparenleft}{\kern0pt}formula{\isacharunderscore}{\kern0pt}progression{\isacharunderscore}{\kern0pt}len{\isadigit{1}}\ {\isasymphi}\ {\isacharparenleft}{\kern0pt}{\isasympi}{\isacharbang}{\kern0pt}{\isadigit{0}}{\isacharparenright}{\kern0pt}{\isacharparenright}{\kern0pt}\ {\isacharequal}{\kern0pt}\ True{\isacharparenright}{\kern0pt}\ \isanewline
\ \ \ \ \ \ {\isasymor}\ {\isacharparenleft}{\kern0pt}{\isasymforall}{\isasymeta}{\isachardot}{\kern0pt}\ semantics{\isacharunderscore}{\kern0pt}mltl\ {\isasymeta}\ {\isacharparenleft}{\kern0pt}formula{\isacharunderscore}{\kern0pt}progression{\isacharunderscore}{\kern0pt}len{\isadigit{1}}\ {\isasymphi}\ {\isacharparenleft}{\kern0pt}{\isasympi}{\isacharbang}{\kern0pt}{\isadigit{0}}{\isacharparenright}{\kern0pt}{\isacharparenright}{\kern0pt}\ {\isacharequal}{\kern0pt}\ False{\isacharparenright}{\kern0pt}{\isachardoublequoteclose}%
}
\snip{convertnnfpreservessemantics}{1}{2}{%
\isacommand{lemma}\isamarkupfalse%
\ convert{\isacharunderscore}{\kern0pt}nnf{\isacharunderscore}{\kern0pt}preserves{\isacharunderscore}{\kern0pt}semantics{\isacharcolon}{\kern0pt}\isanewline
\ \ \isakeyword{assumes}\ {\isachardoublequoteopen}intervals{\isacharunderscore}{\kern0pt}welldef\ {\isasymphi}{\isachardoublequoteclose}\ \isakeyword{shows}\ {\isachardoublequoteopen}{\isasymphi}\ {\isasymequiv}\isactrlsub m\ {\isacharparenleft}{\kern0pt}convert{\isacharunderscore}{\kern0pt}nnf\ {\isasymphi}{\isacharparenright}{\kern0pt}{\isachardoublequoteclose}%
}
\snip{convertnnfconvertnnf}{1}{2}{%
\isacommand{lemma}\isamarkupfalse%
\ convert{\isacharunderscore}{\kern0pt}nnf{\isacharunderscore}{\kern0pt}convert{\isacharunderscore}{\kern0pt}nnf{\isacharcolon}{\kern0pt}\isanewline
\ \ \isakeyword{shows}\ {\isachardoublequoteopen}convert{\isacharunderscore}{\kern0pt}nnf\ {\isacharparenleft}{\kern0pt}convert{\isacharunderscore}{\kern0pt}nnf\ {\isasymphi}{\isacharparenright}{\kern0pt}\ {\isacharequal}{\kern0pt}\ convert{\isacharunderscore}{\kern0pt}nnf\ {\isasymphi}{\isachardoublequoteclose}%
}
\snip{nnfsubformulas}{1}{2}{%
\isacommand{lemma}\isamarkupfalse%
\ nnf{\isacharunderscore}{\kern0pt}subformulas{\isacharcolon}{\kern0pt}\isanewline
\ \ \isakeyword{assumes}\ {\isachardoublequoteopen}{\isasymphi}\ {\isacharequal}{\kern0pt}\ convert{\isacharunderscore}{\kern0pt}nnf\ init{\isacharunderscore}{\kern0pt}{\isasymphi}{\isachardoublequoteclose}\ \isakeyword{and}\ {\isachardoublequoteopen}{\isasympsi}\ {\isasymin}\ subformulas\ {\isasymphi}{\isachardoublequoteclose}\isanewline
\ \ \isakeyword{shows}\ {\isachardoublequoteopen}{\isasymexists}\ init{\isacharunderscore}{\kern0pt}{\isasympsi}{\isachardot}{\kern0pt}\ {\isasympsi}\ {\isacharequal}{\kern0pt}\ convert{\isacharunderscore}{\kern0pt}nnf\ init{\isacharunderscore}{\kern0pt}{\isasympsi}{\isachardoublequoteclose}%
}
\snip{nnfinduct}{1}{2}{%
\isacommand{lemma}\isamarkupfalse%
\ nnf{\isacharunderscore}{\kern0pt}induct{\isacharbrackleft}{\kern0pt}case{\isacharunderscore}{\kern0pt}names\ nnf\ True\ False\ Prop\ And\ Or\ Final\ \isanewline
\ \ \ \ \ \ \ \ \ \ \ \ \ \ \ \ \ Global\ Until\ Release\ NotProp{\isacharbrackright}{\kern0pt}{\isacharcolon}{\kern0pt}\isanewline
\ \ \isakeyword{assumes}\ nnf{\isacharcolon}{\kern0pt}\ {\isachardoublequoteopen}{\isasymexists}{\isasymphi}{\isachardot}{\kern0pt}\ {\isasymeta}\ {\isacharequal}{\kern0pt}\ convert{\isacharunderscore}{\kern0pt}nnf\ {\isasymphi}{\isachardoublequoteclose}\ \isakeyword{and}\ True{\isacharcolon}{\kern0pt}\ {\isachardoublequoteopen}{\isasymP}\ True\isactrlsub m{\isachardoublequoteclose}\ \isanewline
\ \ \isakeyword{and}\ False{\isacharcolon}{\kern0pt}\ {\isachardoublequoteopen}{\isasymP}\ False\isactrlsub m{\isachardoublequoteclose}\ \isakeyword{and}\ Prop{\isacharcolon}{\kern0pt}\ {\isachardoublequoteopen}{\isasymAnd}\ p{\isachardot}{\kern0pt}\ {\isasymP}\ {\isacharparenleft}{\kern0pt}Prop\isactrlsub m\ {\isacharparenleft}{\kern0pt}p{\isacharparenright}{\kern0pt}{\isacharparenright}{\kern0pt}{\isachardoublequoteclose}\isanewline
\ \ \isakeyword{and}\ And{\isacharcolon}{\kern0pt}\ {\isachardoublequoteopen}{\isasymAnd}{\isasymeta}\ {\isasymphi}\ {\isasympsi}{\isachardot}{\kern0pt}\ {\isasymlbrakk}{\isasymeta}\ {\isacharequal}{\kern0pt}\ {\isasymphi}\ And\isactrlsub m\ {\isasympsi}{\isacharsemicolon}{\kern0pt}\ {\isasymP}\ {\isasymphi}{\isacharsemicolon}{\kern0pt}\ {\isasymP}\ {\isasympsi}{\isasymrbrakk}\ {\isasymLongrightarrow}\ {\isasymP}\ {\isasymeta}{\isachardoublequoteclose}\isanewline
\ \ \isakeyword{and}\ Or{\isacharcolon}{\kern0pt}\ {\isachardoublequoteopen}{\isasymAnd}{\isasymeta}\ {\isasymphi}\ {\isasympsi}{\isachardot}{\kern0pt}\ {\isasymlbrakk}{\isasymeta}\ {\isacharequal}{\kern0pt}\ {\isasymphi}\ Or\isactrlsub m\ {\isasympsi}{\isacharsemicolon}{\kern0pt}{\isasymP}\ {\isasymphi}{\isacharsemicolon}{\kern0pt}\ {\isasymP}\ {\isasympsi}{\isasymrbrakk}\ {\isasymLongrightarrow}\ {\isasymP}\ {\isasymeta}{\isachardoublequoteclose}\isanewline
\ \ \isakeyword{and}\ Final{\isacharcolon}{\kern0pt}\ {\isachardoublequoteopen}{\isasymAnd}{\isasymeta}\ {\isasymphi}\ \ a\ b{\isachardot}{\kern0pt}\ {\isasymlbrakk}{\isasymeta}\ {\isacharequal}{\kern0pt}\ F\isactrlsub m{\isacharbrackleft}{\kern0pt}a{\isacharcomma}{\kern0pt}b{\isacharbrackright}{\kern0pt}\ {\isasymphi}{\isacharsemicolon}{\kern0pt}\ {\isasymP}\ {\isasymphi}{\isasymrbrakk}\ {\isasymLongrightarrow}\ {\isasymP}\ {\isasymeta}{\isachardoublequoteclose}\isanewline
\ \ \isakeyword{and}\ Global{\isacharcolon}{\kern0pt}\ {\isachardoublequoteopen}{\isasymAnd}{\isasymeta}\ {\isasymphi}\ \ a\ b{\isachardot}{\kern0pt}\ {\isasymlbrakk}{\isasymeta}\ {\isacharequal}{\kern0pt}\ G\isactrlsub m{\isacharbrackleft}{\kern0pt}a{\isacharcomma}{\kern0pt}b{\isacharbrackright}{\kern0pt}\ {\isasymphi}{\isacharsemicolon}{\kern0pt}\ {\isasymP}\ {\isasymphi}{\isasymrbrakk}\ {\isasymLongrightarrow}\ {\isasymP}\ {\isasymeta}{\isachardoublequoteclose}\isanewline
\ \ \isakeyword{and}\ Until{\isacharcolon}{\kern0pt}\ {\isachardoublequoteopen}{\isasymAnd}{\isasymeta}\ {\isasymphi}\ {\isasympsi}\ \ a\ b{\isachardot}{\kern0pt}\ {\isasymlbrakk}{\isasymeta}\ {\isacharequal}{\kern0pt}\ {\isasymphi}\ U\isactrlsub m{\isacharbrackleft}{\kern0pt}a{\isacharcomma}{\kern0pt}b{\isacharbrackright}{\kern0pt}\ {\isasympsi}{\isacharsemicolon}{\kern0pt}\ {\isasymP}\ {\isasymphi}{\isacharsemicolon}{\kern0pt}\ {\isasymP}\ {\isasympsi}{\isasymrbrakk}\ {\isasymLongrightarrow}\ {\isasymP}\ {\isasymeta}{\isachardoublequoteclose}\isanewline
\ \ \isakeyword{and}\ Release{\isacharcolon}{\kern0pt}\ {\isachardoublequoteopen}{\isasymAnd}{\isasymeta}\ {\isasymphi}\ {\isasympsi}\ a\ b{\isachardot}{\kern0pt}\ {\isasymlbrakk}{\isasymeta}\ {\isacharequal}{\kern0pt}\ {\isasymphi}\ R\isactrlsub m{\isacharbrackleft}{\kern0pt}a{\isacharcomma}{\kern0pt}b{\isacharbrackright}{\kern0pt}\ {\isasympsi}{\isacharsemicolon}{\kern0pt}\ {\isasymP}\ {\isasymphi}{\isacharsemicolon}{\kern0pt}\ {\isasymP}\ {\isasympsi}{\isasymrbrakk}\ {\isasymLongrightarrow}\ {\isasymP}\ {\isasymeta}{\isachardoublequoteclose}\isanewline
\ \ \isakeyword{and}\ Not{\isacharunderscore}{\kern0pt}Prop{\isacharcolon}{\kern0pt}\ {\isachardoublequoteopen}{\isasymAnd}\ {\isasymeta}\ p{\isachardot}{\kern0pt}\ {\isasymeta}\ {\isacharequal}{\kern0pt}\ Not\isactrlsub m\ {\isacharparenleft}{\kern0pt}Prop\isactrlsub m\ {\isacharparenleft}{\kern0pt}p{\isacharparenright}{\kern0pt}{\isacharparenright}{\kern0pt}\ {\isasymLongrightarrow}\ {\isasymP}\ {\isasymeta}{\isachardoublequoteclose}\isanewline
\ \ \isakeyword{shows}\ {\isachardoublequoteopen}{\isasymP}\ {\isasymeta}{\isachardoublequoteclose}%
}
\snip{bnfinduct}{1}{2}{%
\isacommand{lemma}\isamarkupfalse%
\ bnf{\isacharunderscore}{\kern0pt}induct{\isacharbrackleft}{\kern0pt}case{\isacharunderscore}{\kern0pt}names\ IntervalsWellDef\ PProp\ True\ False\ Prop\ Not\ And\ Until{\isacharbrackright}{\kern0pt}{\isacharcolon}{\kern0pt}\isanewline
\ \ \isakeyword{assumes}\ IntervalsWellDef{\isacharcolon}{\kern0pt}\ {\isachardoublequoteopen}intervals{\isacharunderscore}{\kern0pt}welldef\ {\isasymeta}{\isachardoublequoteclose}\isanewline
\ \ \isakeyword{and}\ PProp{\isacharcolon}{\kern0pt}\ {\isachardoublequoteopen}{\isacharparenleft}{\kern0pt}{\isasymAnd}\ {\isasymeta}\ {\isasymphi}{\isachardot}{\kern0pt}\ {\isacharparenleft}{\kern0pt}{\isasymeta}\ {\isasymequiv}\isactrlsub m\ {\isasymphi}\ {\isasymlongrightarrow}\ {\isasymP}\ {\isasymeta}\ {\isacharequal}{\kern0pt}\ {\isasymP}\ {\isasymphi}{\isacharparenright}{\kern0pt}{\isacharparenright}{\kern0pt}{\isachardoublequoteclose}\ \isakeyword{and}\ True{\isacharcolon}{\kern0pt}\ {\isachardoublequoteopen}{\isasymP}\ True\isactrlsub m{\isachardoublequoteclose}\isanewline
\ \ \isakeyword{and}\ False{\isacharcolon}{\kern0pt}\ {\isachardoublequoteopen}{\isasymP}\ False\isactrlsub m{\isachardoublequoteclose}\ \isakeyword{and}\ Prop{\isacharcolon}{\kern0pt}\ {\isachardoublequoteopen}{\isasymAnd}\ p{\isachardot}{\kern0pt}\ {\isasymP}\ {\isacharparenleft}{\kern0pt}Prop\isactrlsub m\ {\isacharparenleft}{\kern0pt}p{\isacharparenright}{\kern0pt}{\isacharparenright}{\kern0pt}{\isachardoublequoteclose}\isanewline
\ \ \isakeyword{and}\ Not{\isacharcolon}{\kern0pt}\ {\isachardoublequoteopen}{\isasymAnd}\ {\isasymeta}\ {\isasymphi}{\isachardot}{\kern0pt}\ {\isasymlbrakk}{\isasymeta}\ {\isacharequal}{\kern0pt}\ Not\isactrlsub m\ {\isasymphi}{\isacharsemicolon}{\kern0pt}\ {\isasymP}\ {\isasymphi}{\isasymrbrakk}\ {\isasymLongrightarrow}\ {\isasymP}\ {\isasymeta}{\isachardoublequoteclose}\isanewline
\ \ \isakeyword{and}\ And{\isacharcolon}{\kern0pt}\ {\isachardoublequoteopen}{\isasymAnd}{\isasymeta}\ {\isasymphi}\ {\isasympsi}{\isachardot}{\kern0pt}\ \ {\isasymlbrakk}{\isasymeta}\ {\isacharequal}{\kern0pt}\ {\isasymphi}\ And\isactrlsub m\ {\isasympsi}{\isacharsemicolon}{\kern0pt}\ {\isasymP}\ {\isasymphi}{\isacharsemicolon}{\kern0pt}\ {\isasymP}\ {\isasympsi}{\isasymrbrakk}\ {\isasymLongrightarrow}\ {\isasymP}\ {\isasymeta}{\isachardoublequoteclose}\isanewline
\ \ \isakeyword{and}\ Until{\isacharcolon}{\kern0pt}{\isachardoublequoteopen}{\isasymAnd}{\isasymeta}\ {\isasymphi}\ {\isasympsi}\ a\ b{\isachardot}{\kern0pt}\ {\isasymlbrakk}{\isasymeta}\ {\isacharequal}{\kern0pt}\ {\isasymphi}\ U\isactrlsub m{\isacharbrackleft}{\kern0pt}a{\isacharcomma}{\kern0pt}b{\isacharbrackright}{\kern0pt}\ {\isasympsi}{\isacharsemicolon}{\kern0pt}\ {\isasymP}\ {\isasymphi}{\isacharsemicolon}{\kern0pt}\ {\isasymP}\ {\isasympsi}{\isasymrbrakk}\ {\isasymLongrightarrow}\ {\isasymP}\ {\isasymeta}{\isachardoublequoteclose}\isanewline
\ \ \isakeyword{shows}\ {\isachardoublequoteopen}{\isasymP}\ {\isasymeta}{\isachardoublequoteclose}%
}
\snip{topcorollary}{1}{2}{%
\isacommand{corollary}\isamarkupfalse%
\ trace{\isacharunderscore}{\kern0pt}append{\isacharcolon}{\kern0pt}\isanewline
\ \ \isakeyword{fixes}\ {\isasymphi}{\isacharcolon}{\kern0pt}{\isacharcolon}{\kern0pt}{\isachardoublequoteopen}{\isacharprime}{\kern0pt}a\ mltl{\isachardoublequoteclose}\ \isanewline
\ \ \isakeyword{assumes}\ {\isachardoublequoteopen}intervals{\isacharunderscore}{\kern0pt}welldef\ {\isasymphi}{\isachardoublequoteclose}\ \isakeyword{and}\ {\isachardoublequoteopen}length\ {\isasympi}\ {\isasymge}\ complen{\isacharunderscore}{\kern0pt}mltl\ {\isasymphi}{\isachardoublequoteclose}\ \isakeyword{and}\ {\isachardoublequoteopen}{\isasympi}\ {\isasymTurnstile}\isactrlsub m\ {\isasymphi}{\isachardoublequoteclose}\isanewline
\ \ \isakeyword{shows}\ {\isachardoublequoteopen}{\isacharparenleft}{\kern0pt}{\isasympi}\ {\isacharat}{\kern0pt}\ {\isasymzeta}{\isacharparenright}{\kern0pt}\ {\isasymTurnstile}\isactrlsub m\ {\isasymphi}{\isachardoublequoteclose}%
}
\snip{complenproperty}{1}{2}{%
\isacommand{corollary}\isamarkupfalse%
\ trace{\isacharunderscore}{\kern0pt}append{\isacharunderscore}{\kern0pt}iff{\isacharcolon}{\kern0pt}\isanewline
\ \ \isakeyword{fixes}\ {\isasymphi}{\isacharcolon}{\kern0pt}{\isacharcolon}{\kern0pt}{\isachardoublequoteopen}{\isacharprime}{\kern0pt}a\ mltl{\isachardoublequoteclose}\isanewline
\ \ \isakeyword{assumes}\ {\isachardoublequoteopen}intervals{\isacharunderscore}{\kern0pt}welldef\ {\isasymphi}{\isachardoublequoteclose}\ \isakeyword{and}\ {\isachardoublequoteopen}length\ {\isasympi}\ {\isasymge}\ complen{\isacharunderscore}{\kern0pt}mltl\ {\isasymphi}{\isachardoublequoteclose}\isanewline
\ \ \isakeyword{shows}\ {\isachardoublequoteopen}{\isasympi}\ {\isasymTurnstile}\isactrlsub m\ {\isasymphi}\ {\isasymlongleftrightarrow}\ {\isacharparenleft}{\kern0pt}{\isasymforall}{\isasymzeta}{\isachardot}{\kern0pt}\ {\isacharparenleft}{\kern0pt}{\isasympi}\ {\isacharat}{\kern0pt}\ {\isasymzeta}{\isacharparenright}{\kern0pt}\ {\isasymTurnstile}\isactrlsub m\ {\isasymphi}{\isacharparenright}{\kern0pt}{\isachardoublequoteclose}%
}
\snip{formulaprogressiondecreasescomplen}{1}{2}{%
\isacommand{lemma}\isamarkupfalse%
\ formula{\isacharunderscore}{\kern0pt}progression{\isacharunderscore}{\kern0pt}decreases{\isacharunderscore}{\kern0pt}complen{\isacharcolon}{\kern0pt}\isanewline
\ \ \isakeyword{assumes}\ {\isachardoublequoteopen}intervals{\isacharunderscore}{\kern0pt}welldef\ {\isasymphi}{\isachardoublequoteclose}\isanewline
\ \ \isakeyword{shows}\ {\isachardoublequoteopen}complen{\isacharunderscore}{\kern0pt}mltl\ {\isasymphi}\ {\isacharequal}{\kern0pt}\ {\isadigit{1}}\ {\isasymor}\ complen{\isacharunderscore}{\kern0pt}mltl\ {\isacharparenleft}{\kern0pt}formula{\isacharunderscore}{\kern0pt}progression\ {\isasymphi}\ {\isasympi}{\isacharparenright}{\kern0pt}\ {\isacharequal}{\kern0pt}\ {\isadigit{1}}\ {\isasymor}\ \isanewline
\ \ \ \ \ \ \ \ \ complen{\isacharunderscore}{\kern0pt}mltl\ {\isacharparenleft}{\kern0pt}formula{\isacharunderscore}{\kern0pt}progression\ {\isasymphi}\ {\isasympi}{\isacharparenright}{\kern0pt}\ {\isasymle}\ complen{\isacharunderscore}{\kern0pt}mltl\ {\isasymphi}\ {\isacharminus}{\kern0pt}\ {\isacharparenleft}{\kern0pt}length\ {\isasympi}{\isacharparenright}{\kern0pt}{\isachardoublequoteclose}%
}
\snip{complenoneimpliesone}{1}{2}{%
\isacommand{lemma}\isamarkupfalse%
\ complen{\isacharunderscore}{\kern0pt}one{\isacharunderscore}{\kern0pt}implies{\isacharunderscore}{\kern0pt}one{\isacharcolon}{\kern0pt}\isanewline
\ \ \isakeyword{assumes}\ {\isachardoublequoteopen}intervals{\isacharunderscore}{\kern0pt}welldef\ {\isasymphi}{\isachardoublequoteclose}\ \isakeyword{and}\ {\isachardoublequoteopen}complen{\isacharunderscore}{\kern0pt}mltl\ {\isasymphi}\ {\isacharequal}{\kern0pt}\ {\isadigit{1}}{\isachardoublequoteclose}\isanewline
\ \ \isakeyword{shows}\ {\isachardoublequoteopen}complen{\isacharunderscore}{\kern0pt}mltl\ {\isacharparenleft}{\kern0pt}formula{\isacharunderscore}{\kern0pt}progression\ {\isasymphi}\ {\isasympi}{\isacharparenright}{\kern0pt}\ {\isacharequal}{\kern0pt}\ {\isadigit{1}}{\isachardoublequoteclose}%
}
\snip{futureordistribute}{1}{2}{%
\isacommand{lemma}\isamarkupfalse%
\ future{\isacharunderscore}{\kern0pt}or{\isacharunderscore}{\kern0pt}distribute{\isacharcolon}{\kern0pt}\isanewline
\ \ \isakeyword{shows}\ {\isachardoublequoteopen}F\isactrlsub m{\isacharbrackleft}{\kern0pt}a{\isacharcomma}{\kern0pt}b{\isacharbrackright}{\kern0pt}\ {\isacharparenleft}{\kern0pt}{\isasymphi}{\isadigit{1}}\ Or\isactrlsub m\ {\isasymphi}{\isadigit{2}}{\isacharparenright}{\kern0pt}\ {\isasymequiv}\isactrlsub m\ {\isacharparenleft}{\kern0pt}F\isactrlsub m{\isacharbrackleft}{\kern0pt}a{\isacharcomma}{\kern0pt}b{\isacharbrackright}{\kern0pt}\ {\isasymphi}{\isadigit{1}}{\isacharparenright}{\kern0pt}\ Or\isactrlsub m\ {\isacharparenleft}{\kern0pt}F\isactrlsub m{\isacharbrackleft}{\kern0pt}a{\isacharcomma}{\kern0pt}b{\isacharbrackright}{\kern0pt}\ {\isasymphi}{\isadigit{2}}{\isacharparenright}{\kern0pt}{\isachardoublequoteclose}%
}
\snip{globalanddistribute}{1}{2}{%
\isacommand{lemma}\isamarkupfalse%
\ global{\isacharunderscore}{\kern0pt}and{\isacharunderscore}{\kern0pt}distribute{\isacharcolon}{\kern0pt}\isanewline
\ \ \isakeyword{shows}\ {\isachardoublequoteopen}G\isactrlsub m{\isacharbrackleft}{\kern0pt}a{\isacharcomma}{\kern0pt}b{\isacharbrackright}{\kern0pt}\ {\isacharparenleft}{\kern0pt}{\isasymphi}{\isadigit{1}}\ And\isactrlsub m\ {\isasymphi}{\isadigit{2}}{\isacharparenright}{\kern0pt}\ {\isasymequiv}\isactrlsub m\ {\isacharparenleft}{\kern0pt}G\isactrlsub m{\isacharbrackleft}{\kern0pt}a{\isacharcomma}{\kern0pt}b{\isacharbrackright}{\kern0pt}\ {\isasymphi}{\isadigit{1}}{\isacharparenright}{\kern0pt}\ And\isactrlsub m\ {\isacharparenleft}{\kern0pt}G\isactrlsub m{\isacharbrackleft}{\kern0pt}a{\isacharcomma}{\kern0pt}b{\isacharbrackright}{\kern0pt}\ {\isasymphi}{\isadigit{2}}{\isacharparenright}{\kern0pt}{\isachardoublequoteclose}%
}
\snip{untilordistribute}{1}{2}{%
\isacommand{lemma}\isamarkupfalse%
\ until{\isacharunderscore}{\kern0pt}or{\isacharunderscore}{\kern0pt}distribute{\isacharcolon}{\kern0pt}\ \isanewline
\isakeyword{fixes}\ a\ b\ {\isacharcolon}{\kern0pt}{\isacharcolon}{\kern0pt}{\isachardoublequoteopen}nat{\isachardoublequoteclose}\ \isakeyword{assumes}\ {\isachardoublequoteopen}a\ {\isasymle}\ b{\isachardoublequoteclose}\isanewline
\isakeyword{shows}\ {\isachardoublequoteopen}{\isasymphi}\ U\isactrlsub m{\isacharbrackleft}{\kern0pt}a{\isacharcomma}{\kern0pt}b{\isacharbrackright}{\kern0pt}\ {\isacharparenleft}{\kern0pt}{\isasymalpha}\ Or\isactrlsub m\ {\isasymbeta}{\isacharparenright}{\kern0pt}\ {\isasymequiv}\isactrlsub m\ {\isacharparenleft}{\kern0pt}{\isasymphi}\ U\isactrlsub m{\isacharbrackleft}{\kern0pt}a{\isacharcomma}{\kern0pt}b{\isacharbrackright}{\kern0pt}\ {\isasymalpha}{\isacharparenright}{\kern0pt}\ Or\isactrlsub m\ {\isacharparenleft}{\kern0pt}{\isasymphi}\ U\isactrlsub m{\isacharbrackleft}{\kern0pt}a{\isacharcomma}{\kern0pt}b{\isacharbrackright}{\kern0pt}\ {\isasymbeta}{\isacharparenright}{\kern0pt}{\isachardoublequoteclose}%
}
\snip{untilanddistribute}{1}{2}{%
\isacommand{lemma}\isamarkupfalse%
\ until{\isacharunderscore}{\kern0pt}and{\isacharunderscore}{\kern0pt}distribute{\isacharcolon}{\kern0pt}\ \isanewline
\isakeyword{fixes}\ a\ b\ {\isacharcolon}{\kern0pt}{\isacharcolon}{\kern0pt}{\isachardoublequoteopen}nat{\isachardoublequoteclose}\ \isakeyword{assumes}\ {\isachardoublequoteopen}a\ {\isasymle}\ b{\isachardoublequoteclose}\isanewline
\isakeyword{shows}\ {\isachardoublequoteopen}{\isacharparenleft}{\kern0pt}{\isasymalpha}\ And\isactrlsub m\ {\isasymbeta}{\isacharparenright}{\kern0pt}\ U\isactrlsub m\ {\isacharbrackleft}{\kern0pt}a{\isacharcomma}{\kern0pt}b{\isacharbrackright}{\kern0pt}\ {\isasymphi}\ {\isasymequiv}\isactrlsub m\ {\isacharparenleft}{\kern0pt}{\isasymalpha}\ U\isactrlsub m\ {\isacharbrackleft}{\kern0pt}a{\isacharcomma}{\kern0pt}b{\isacharbrackright}{\kern0pt}\ {\isasymphi}{\isacharparenright}{\kern0pt}\ And\isactrlsub m\ {\isacharparenleft}{\kern0pt}{\isasymbeta}\ U\isactrlsub m\ {\isacharbrackleft}{\kern0pt}a{\isacharcomma}{\kern0pt}b{\isacharbrackright}{\kern0pt}\ {\isasymphi}{\isacharparenright}{\kern0pt}{\isachardoublequoteclose}%
}
\snip{formulaprogressiontrueorfalse}{1}{2}{%
\isacommand{lemma}\isamarkupfalse%
\ formula{\isacharunderscore}{\kern0pt}progression{\isacharunderscore}{\kern0pt}true{\isacharunderscore}{\kern0pt}or{\isacharunderscore}{\kern0pt}false{\isacharcolon}{\kern0pt}\isanewline
\ \ \isakeyword{fixes}\ {\isasymphi}{\isacharcolon}{\kern0pt}{\isacharcolon}{\kern0pt}{\isachardoublequoteopen}{\isacharprime}{\kern0pt}a\ mltl{\isachardoublequoteclose}\isanewline
\ \ \isakeyword{assumes}\ {\isachardoublequoteopen}intervals{\isacharunderscore}{\kern0pt}welldef\ {\isasymphi}{\isachardoublequoteclose}\ \isakeyword{and}\ {\isachardoublequoteopen}length\ {\isasympi}\ {\isasymge}\ complen{\isacharunderscore}{\kern0pt}mltl\ {\isasymphi}{\isachardoublequoteclose}\isanewline
\ \ \isakeyword{and}\ {\isachardoublequoteopen}{\isasympsi}\ {\isacharequal}{\kern0pt}\ {\isacharparenleft}{\kern0pt}formula{\isacharunderscore}{\kern0pt}progression\ {\isasymphi}\ {\isasympi}{\isacharparenright}{\kern0pt}{\isachardoublequoteclose}\isanewline
\ \ \isakeyword{shows}\ {\isachardoublequoteopen}{\isacharparenleft}{\kern0pt}{\isasympsi}\ {\isasymequiv}\isactrlsub m\ False{\isacharunderscore}{\kern0pt}mltl{\isacharparenright}{\kern0pt}\ {\isasymor}\ {\isacharparenleft}{\kern0pt}{\isasympsi}\ {\isasymequiv}\isactrlsub m\ True{\isacharunderscore}{\kern0pt}mltl{\isacharparenright}{\kern0pt}{\isachardoublequoteclose}%
}
\snip{proglenone}{1}{2}{%
\isacommand{function}\isamarkupfalse%
\ formula{\isacharunderscore}{\kern0pt}progression{\isacharunderscore}{\kern0pt}len{\isadigit{1}}{\isacharcolon}{\kern0pt}{\isacharcolon}{\kern0pt}\ {\isachardoublequoteopen}{\isacharprime}{\kern0pt}a\ mltl\ {\isasymRightarrow}\ {\isacharprime}{\kern0pt}a\ set\ {\isasymRightarrow}\ {\isacharprime}{\kern0pt}a\ mltl{\isachardoublequoteclose}\ \ \isanewline
\isakeyword{where}\ {\isachardoublequoteopen}formula{\isacharunderscore}{\kern0pt}progression{\isacharunderscore}{\kern0pt}len{\isadigit{1}}\ True\isactrlsub m\ tr{\isacharunderscore}{\kern0pt}entry\ {\isacharequal}{\kern0pt}\ True\isactrlsub m{\isachardoublequoteclose}\ \ \ \ \ \ \ \ \ \ \ \ \ \ \ \ \ \ \ \ \ \ \ \ \ \ \ \ \ \ \ \isanewline
{\isacharbar}{\kern0pt}\ {\isachardoublequoteopen}formula{\isacharunderscore}{\kern0pt}progression{\isacharunderscore}{\kern0pt}len{\isadigit{1}}\ False\isactrlsub m\ tr{\isacharunderscore}{\kern0pt}entry\ {\isacharequal}{\kern0pt}\ False\isactrlsub m{\isachardoublequoteclose}\ \ \ \ \ \ \ \ \ \ \ \ \ \ \ \ \ \ \ \ \ \ \ \ \ \ \ \isanewline
{\isacharbar}{\kern0pt}\ {\isachardoublequoteopen}formula{\isacharunderscore}{\kern0pt}progression{\isacharunderscore}{\kern0pt}len{\isadigit{1}}\ {\isacharparenleft}{\kern0pt}Prop\isactrlsub m\ {\isacharparenleft}{\kern0pt}p{\isacharparenright}{\kern0pt}{\isacharparenright}{\kern0pt}\ tr{\isacharunderscore}{\kern0pt}entry\ {\isacharequal}{\kern0pt}\ \isanewline
\ \ \ {\isacharparenleft}{\kern0pt}if\ p\ {\isasymin}\ tr{\isacharunderscore}{\kern0pt}entry\ then\ True\isactrlsub m\ else\ False\isactrlsub m{\isacharparenright}{\kern0pt}{\isachardoublequoteclose}\ \ \ \ \ \ \ \ \ \ \ \ \ \ \ \ \ \ \ \ \ \ \ \ \isanewline
{\isacharbar}{\kern0pt}\ {\isachardoublequoteopen}formula{\isacharunderscore}{\kern0pt}progression{\isacharunderscore}{\kern0pt}len{\isadigit{1}}\ {\isacharparenleft}{\kern0pt}Not\isactrlsub m\ F{\isacharparenright}{\kern0pt}\ tr{\isacharunderscore}{\kern0pt}entry\ {\isacharequal}{\kern0pt}\ \isanewline
\ \ \ Not\isactrlsub m\ {\isacharparenleft}{\kern0pt}formula{\isacharunderscore}{\kern0pt}progression{\isacharunderscore}{\kern0pt}len{\isadigit{1}}\ F\ tr{\isacharunderscore}{\kern0pt}entry{\isacharparenright}{\kern0pt}{\isachardoublequoteclose}\isanewline
{\isacharbar}{\kern0pt}\ {\isachardoublequoteopen}formula{\isacharunderscore}{\kern0pt}progression{\isacharunderscore}{\kern0pt}len{\isadigit{1}}\ {\isacharparenleft}{\kern0pt}F{\isadigit{1}}\ And\isactrlsub m\ F{\isadigit{2}}{\isacharparenright}{\kern0pt}\ tr{\isacharunderscore}{\kern0pt}entry\ {\isacharequal}{\kern0pt}\ \isanewline
{\isacharparenleft}{\kern0pt}formula{\isacharunderscore}{\kern0pt}progression{\isacharunderscore}{\kern0pt}len{\isadigit{1}}\ F{\isadigit{1}}\ tr{\isacharunderscore}{\kern0pt}entry{\isacharparenright}{\kern0pt}\ And\isactrlsub m\ {\isacharparenleft}{\kern0pt}formula{\isacharunderscore}{\kern0pt}progression{\isacharunderscore}{\kern0pt}len{\isadigit{1}}\ F{\isadigit{2}}\ tr{\isacharunderscore}{\kern0pt}entry{\isacharparenright}{\kern0pt}{\isachardoublequoteclose}\isanewline
{\isacharbar}{\kern0pt}\ {\isachardoublequoteopen}formula{\isacharunderscore}{\kern0pt}progression{\isacharunderscore}{\kern0pt}len{\isadigit{1}}\ {\isacharparenleft}{\kern0pt}F{\isadigit{1}}\ Or\isactrlsub m\ F{\isadigit{2}}{\isacharparenright}{\kern0pt}\ tr{\isacharunderscore}{\kern0pt}entry\ {\isacharequal}{\kern0pt}\ \isanewline
{\isacharparenleft}{\kern0pt}formula{\isacharunderscore}{\kern0pt}progression{\isacharunderscore}{\kern0pt}len{\isadigit{1}}\ F{\isadigit{1}}\ tr{\isacharunderscore}{\kern0pt}entry{\isacharparenright}{\kern0pt}\ Or\isactrlsub m\ {\isacharparenleft}{\kern0pt}formula{\isacharunderscore}{\kern0pt}progression{\isacharunderscore}{\kern0pt}len{\isadigit{1}}\ F{\isadigit{2}}\ tr{\isacharunderscore}{\kern0pt}entry{\isacharparenright}{\kern0pt}{\isachardoublequoteclose}\isanewline
{\isacharbar}{\kern0pt}\ {\isachardoublequoteopen}formula{\isacharunderscore}{\kern0pt}progression{\isacharunderscore}{\kern0pt}len{\isadigit{1}}\ {\isacharparenleft}{\kern0pt}F{\isadigit{1}}\ U\isactrlsub m\ {\isacharbrackleft}{\kern0pt}a{\isacharcomma}{\kern0pt}b{\isacharbrackright}{\kern0pt}\ F{\isadigit{2}}{\isacharparenright}{\kern0pt}\ tr{\isacharunderscore}{\kern0pt}entry\ {\isacharequal}{\kern0pt}\ \ \isanewline
{\isacharparenleft}{\kern0pt}if\ {\isacharparenleft}{\kern0pt}{\isadigit{0}}{\isacharless}{\kern0pt}a\ {\isasymand}\ a{\isasymle}b{\isacharparenright}{\kern0pt}\ then\ {\isacharparenleft}{\kern0pt}F{\isadigit{1}}\ U\isactrlsub m\ {\isacharbrackleft}{\kern0pt}{\isacharparenleft}{\kern0pt}a{\isacharminus}{\kern0pt}{\isadigit{1}}{\isacharparenright}{\kern0pt}{\isacharcomma}{\kern0pt}\ {\isacharparenleft}{\kern0pt}b{\isacharminus}{\kern0pt}{\isadigit{1}}{\isacharparenright}{\kern0pt}{\isacharbrackright}{\kern0pt}\ F{\isadigit{2}}{\isacharparenright}{\kern0pt}\ else\ {\isacharparenleft}{\kern0pt}if\ {\isacharparenleft}{\kern0pt}{\isadigit{0}}{\isacharequal}{\kern0pt}a\ {\isasymand}\ a{\isacharless}{\kern0pt}b{\isacharparenright}{\kern0pt}\ \isanewline
\ \ then\ {\isacharparenleft}{\kern0pt}{\isacharparenleft}{\kern0pt}formula{\isacharunderscore}{\kern0pt}progression{\isacharunderscore}{\kern0pt}len{\isadigit{1}}\ F{\isadigit{2}}\ tr{\isacharunderscore}{\kern0pt}entry{\isacharparenright}{\kern0pt}\ Or\isactrlsub m\ \isanewline
\ \ \ \ {\isacharparenleft}{\kern0pt}{\isacharparenleft}{\kern0pt}formula{\isacharunderscore}{\kern0pt}progression{\isacharunderscore}{\kern0pt}len{\isadigit{1}}\ F{\isadigit{1}}\ tr{\isacharunderscore}{\kern0pt}entry{\isacharparenright}{\kern0pt}\ And\isactrlsub m\ {\isacharparenleft}{\kern0pt}F{\isadigit{1}}\ U\isactrlsub m\ {\isacharbrackleft}{\kern0pt}{\isadigit{0}}{\isacharcomma}{\kern0pt}\ {\isacharparenleft}{\kern0pt}b{\isacharminus}{\kern0pt}{\isadigit{1}}{\isacharparenright}{\kern0pt}{\isacharbrackright}{\kern0pt}\ F{\isadigit{2}}{\isacharparenright}{\kern0pt}{\isacharparenright}{\kern0pt}{\isacharparenright}{\kern0pt}\isanewline
\ \ else\ {\isacharparenleft}{\kern0pt}formula{\isacharunderscore}{\kern0pt}progression{\isacharunderscore}{\kern0pt}len{\isadigit{1}}\ F{\isadigit{2}}\ tr{\isacharunderscore}{\kern0pt}entry{\isacharparenright}{\kern0pt}{\isacharparenright}{\kern0pt}{\isacharparenright}{\kern0pt}{\isachardoublequoteclose}\isanewline
{\isacharbar}{\kern0pt}\ {\isachardoublequoteopen}formula{\isacharunderscore}{\kern0pt}progression{\isacharunderscore}{\kern0pt}len{\isadigit{1}}\ {\isacharparenleft}{\kern0pt}F{\isadigit{1}}\ R\isactrlsub m\ {\isacharbrackleft}{\kern0pt}a{\isacharcomma}{\kern0pt}b{\isacharbrackright}{\kern0pt}\ F{\isadigit{2}}{\isacharparenright}{\kern0pt}\ tr{\isacharunderscore}{\kern0pt}entry\ {\isacharequal}{\kern0pt}\ \isanewline
Not\isactrlsub m\ {\isacharparenleft}{\kern0pt}formula{\isacharunderscore}{\kern0pt}progression{\isacharunderscore}{\kern0pt}len{\isadigit{1}}\ {\isacharparenleft}{\kern0pt}{\isacharparenleft}{\kern0pt}Not\isactrlsub m\ F{\isadigit{1}}{\isacharparenright}{\kern0pt}\ U\isactrlsub m\ {\isacharbrackleft}{\kern0pt}a{\isacharcomma}{\kern0pt}b{\isacharbrackright}{\kern0pt}\ {\isacharparenleft}{\kern0pt}Not\isactrlsub m\ F{\isadigit{2}}{\isacharparenright}{\kern0pt}{\isacharparenright}{\kern0pt}\ tr{\isacharunderscore}{\kern0pt}entry{\isacharparenright}{\kern0pt}{\isachardoublequoteclose}\isanewline
{\isacharbar}{\kern0pt}\ {\isachardoublequoteopen}formula{\isacharunderscore}{\kern0pt}progression{\isacharunderscore}{\kern0pt}len{\isadigit{1}}\ {\isacharparenleft}{\kern0pt}G\isactrlsub m\ {\isacharbrackleft}{\kern0pt}a{\isacharcomma}{\kern0pt}b{\isacharbrackright}{\kern0pt}\ F{\isacharparenright}{\kern0pt}\ tr{\isacharunderscore}{\kern0pt}entry\ {\isacharequal}{\kern0pt}\ \isanewline
Not\isactrlsub m\ {\isacharparenleft}{\kern0pt}formula{\isacharunderscore}{\kern0pt}progression{\isacharunderscore}{\kern0pt}len{\isadigit{1}}\ {\isacharparenleft}{\kern0pt}F\isactrlsub m\ {\isacharbrackleft}{\kern0pt}a{\isacharcomma}{\kern0pt}b{\isacharbrackright}{\kern0pt}\ {\isacharparenleft}{\kern0pt}Not\isactrlsub m\ F{\isacharparenright}{\kern0pt}{\isacharparenright}{\kern0pt}\ tr{\isacharunderscore}{\kern0pt}entry{\isacharparenright}{\kern0pt}{\isachardoublequoteclose}\isanewline
{\isacharbar}{\kern0pt}\ {\isachardoublequoteopen}formula{\isacharunderscore}{\kern0pt}progression{\isacharunderscore}{\kern0pt}len{\isadigit{1}}\ {\isacharparenleft}{\kern0pt}F\isactrlsub m\ {\isacharbrackleft}{\kern0pt}a{\isacharcomma}{\kern0pt}b{\isacharbrackright}{\kern0pt}\ F{\isacharparenright}{\kern0pt}\ tr{\isacharunderscore}{\kern0pt}entry\ {\isacharequal}{\kern0pt}\ \ \isanewline
{\isacharparenleft}{\kern0pt}if\ {\isadigit{0}}{\isacharless}{\kern0pt}\ a\ {\isasymand}\ a{\isasymle}\ b\ then\ \ {\isacharparenleft}{\kern0pt}F\isactrlsub m\ {\isacharbrackleft}{\kern0pt}{\isacharparenleft}{\kern0pt}a{\isacharminus}{\kern0pt}{\isadigit{1}}{\isacharparenright}{\kern0pt}{\isacharcomma}{\kern0pt}\ {\isacharparenleft}{\kern0pt}b{\isacharminus}{\kern0pt}{\isadigit{1}}{\isacharparenright}{\kern0pt}{\isacharbrackright}{\kern0pt}\ F{\isacharparenright}{\kern0pt}\ else\ if\ {\isacharparenleft}{\kern0pt}{\isadigit{0}}\ {\isacharequal}{\kern0pt}\ a\ {\isasymand}\ a\ {\isacharless}{\kern0pt}\ b{\isacharparenright}{\kern0pt}\ \isanewline
then\ {\isacharparenleft}{\kern0pt}{\isacharparenleft}{\kern0pt}formula{\isacharunderscore}{\kern0pt}progression{\isacharunderscore}{\kern0pt}len{\isadigit{1}}\ F\ tr{\isacharunderscore}{\kern0pt}entry{\isacharparenright}{\kern0pt}\ Or\isactrlsub m\ {\isacharparenleft}{\kern0pt}F\isactrlsub m\ {\isacharbrackleft}{\kern0pt}{\isadigit{0}}{\isacharcomma}{\kern0pt}\ {\isacharparenleft}{\kern0pt}b{\isacharminus}{\kern0pt}{\isadigit{1}}{\isacharparenright}{\kern0pt}{\isacharbrackright}{\kern0pt}\ F{\isacharparenright}{\kern0pt}{\isacharparenright}{\kern0pt}\isanewline
else\ {\isacharparenleft}{\kern0pt}formula{\isacharunderscore}{\kern0pt}progression{\isacharunderscore}{\kern0pt}len{\isadigit{1}}\ F\ tr{\isacharunderscore}{\kern0pt}entry{\isacharparenright}{\kern0pt}{\isacharparenright}{\kern0pt}{\isachardoublequoteclose}%
}
\snip{prog}{1}{2}{%
\isacommand{fun}\isamarkupfalse%
\ formula{\isacharunderscore}{\kern0pt}progression{\isacharcolon}{\kern0pt}{\isacharcolon}{\kern0pt}\ {\isachardoublequoteopen}{\isacharprime}{\kern0pt}a\ mltl\ {\isasymRightarrow}\ {\isacharprime}{\kern0pt}a\ set\ list\ {\isasymRightarrow}\ {\isacharprime}{\kern0pt}a\ mltl{\isachardoublequoteclose}\ \isanewline
\isakeyword{where}\ {\isachardoublequoteopen}formula{\isacharunderscore}{\kern0pt}progression\ {\isasymphi}\ {\isasympi}\ {\isacharequal}{\kern0pt}\ {\isacharparenleft}{\kern0pt}if\ length\ {\isasympi}\ {\isacharequal}{\kern0pt}\ {\isadigit{0}}\ then\ {\isasymphi}\ \isanewline
\ \ else\ {\isacharparenleft}{\kern0pt}if\ length\ {\isasympi}\ {\isacharequal}{\kern0pt}\ {\isadigit{1}}\ then\ {\isacharparenleft}{\kern0pt}formula{\isacharunderscore}{\kern0pt}progression{\isacharunderscore}{\kern0pt}len{\isadigit{1}}\ {\isasymphi}\ {\isacharparenleft}{\kern0pt}{\isasympi}{\isacharbang}{\kern0pt}{\isadigit{0}}{\isacharparenright}{\kern0pt}{\isacharparenright}{\kern0pt}\isanewline
\ \ else\ {\isacharparenleft}{\kern0pt}formula{\isacharunderscore}{\kern0pt}progression\ {\isacharparenleft}{\kern0pt}formula{\isacharunderscore}{\kern0pt}progression{\isacharunderscore}{\kern0pt}len{\isadigit{1}}\ {\isasymphi}\ {\isacharparenleft}{\kern0pt}{\isasympi}{\isacharbang}{\kern0pt}{\isadigit{0}}{\isacharparenright}{\kern0pt}{\isacharparenright}{\kern0pt}\ {\isacharparenleft}{\kern0pt}drop\ {\isadigit{1}}\ {\isasympi}{\isacharparenright}{\kern0pt}{\isacharparenright}{\kern0pt}{\isacharparenright}{\kern0pt}{\isacharparenright}{\kern0pt}{\isachardoublequoteclose}%
}
\snip{mltlsemantics}{1}{2}{%
\isacommand{primrec}\isamarkupfalse%
\ semantics{\isacharunderscore}{\kern0pt}mltl\ {\isacharcolon}{\kern0pt}{\isacharcolon}{\kern0pt}\ {\isachardoublequoteopen}{\isacharbrackleft}{\kern0pt}{\isacharprime}{\kern0pt}a\ set\ list{\isacharcomma}{\kern0pt}\ {\isacharprime}{\kern0pt}a\ mltl{\isacharbrackright}{\kern0pt}\ {\isasymRightarrow}\ bool{\isachardoublequoteclose}\ {\isacharparenleft}{\kern0pt}{\isachardoublequoteopen}{\isacharunderscore}{\kern0pt}\ {\isasymTurnstile}\isactrlsub m\ {\isacharunderscore}{\kern0pt}{\isachardoublequoteclose}\ {\isacharbrackleft}{\kern0pt}{\isadigit{8}}{\isadigit{0}}{\isacharcomma}{\kern0pt}{\isadigit{8}}{\isadigit{0}}{\isacharbrackright}{\kern0pt}\ {\isadigit{8}}{\isadigit{0}}{\isacharparenright}{\kern0pt}\isanewline
\isakeyword{where}\ {\isachardoublequoteopen}{\isasympi}\ {\isasymTurnstile}\isactrlsub m\ True\isactrlsub m\ {\isacharequal}{\kern0pt}\ True{\isachardoublequoteclose}\isanewline
{\isacharbar}{\kern0pt}\ {\isachardoublequoteopen}{\isasympi}\ {\isasymTurnstile}\isactrlsub m\ False\isactrlsub m\ {\isacharequal}{\kern0pt}\ False{\isachardoublequoteclose}\isanewline
{\isacharbar}{\kern0pt}\ {\isachardoublequoteopen}{\isasympi}\ {\isasymTurnstile}\isactrlsub m\ Prop\isactrlsub m\ {\isacharparenleft}{\kern0pt}q{\isacharparenright}{\kern0pt}\ {\isacharequal}{\kern0pt}\ {\isacharparenleft}{\kern0pt}{\isasympi}\ {\isasymnoteq}\ {\isacharbrackleft}{\kern0pt}{\isacharbrackright}{\kern0pt}\ {\isasymand}\ q\ {\isasymin}\ {\isacharparenleft}{\kern0pt}{\isasympi}\ {\isacharbang}{\kern0pt}\ {\isadigit{0}}{\isacharparenright}{\kern0pt}{\isacharparenright}{\kern0pt}{\isachardoublequoteclose}\isanewline
{\isacharbar}{\kern0pt}\ {\isachardoublequoteopen}{\isasympi}\ {\isasymTurnstile}\isactrlsub m\ Not\isactrlsub m\ {\isasymphi}\ {\isacharequal}{\kern0pt}\ {\isacharparenleft}{\kern0pt}{\isasymnot}\ {\isasympi}\ {\isasymTurnstile}\isactrlsub m\ {\isasymphi}{\isacharparenright}{\kern0pt}{\isachardoublequoteclose}\isanewline
{\isacharbar}{\kern0pt}\ {\isachardoublequoteopen}{\isasympi}\ {\isasymTurnstile}\isactrlsub m\ {\isasymphi}\ And\isactrlsub m\ {\isasympsi}\ {\isacharequal}{\kern0pt}\ {\isacharparenleft}{\kern0pt}{\isasympi}\ {\isasymTurnstile}\isactrlsub m\ {\isasymphi}\ {\isasymand}\ {\isasympi}\ {\isasymTurnstile}\isactrlsub m\ {\isasympsi}{\isacharparenright}{\kern0pt}{\isachardoublequoteclose}\isanewline
{\isacharbar}{\kern0pt}\ {\isachardoublequoteopen}{\isasympi}\ {\isasymTurnstile}\isactrlsub m\ {\isasymphi}\ Or\isactrlsub m\ {\isasympsi}\ {\isacharequal}{\kern0pt}\ {\isacharparenleft}{\kern0pt}{\isasympi}\ {\isasymTurnstile}\isactrlsub m\ {\isasymphi}\ {\isasymor}\ {\isasympi}\ {\isasymTurnstile}\isactrlsub m\ {\isasympsi}{\isacharparenright}{\kern0pt}{\isachardoublequoteclose}\isanewline
{\isacharbar}{\kern0pt}\ {\isachardoublequoteopen}{\isasympi}\ {\isasymTurnstile}\isactrlsub m\ {\isacharparenleft}{\kern0pt}F\isactrlsub m\ {\isacharbrackleft}{\kern0pt}a{\isacharcomma}{\kern0pt}\ b{\isacharbrackright}{\kern0pt}\ {\isasymphi}{\isacharparenright}{\kern0pt}\ {\isacharequal}{\kern0pt}\ {\isacharparenleft}{\kern0pt}a\ {\isasymle}\ b\ {\isasymand}\ length\ {\isasympi}\ {\isachargreater}{\kern0pt}\ a\ {\isasymand}\ \isanewline
\ \ \ \ \ \ {\isacharparenleft}{\kern0pt}{\isasymexists}i{\isacharcolon}{\kern0pt}{\isacharcolon}{\kern0pt}nat{\isachardot}{\kern0pt}\ {\isacharparenleft}{\kern0pt}i\ {\isasymge}\ a\ {\isasymand}\ i\ {\isasymle}\ b{\isacharparenright}{\kern0pt}\ {\isasymand}\ {\isacharparenleft}{\kern0pt}drop\ i\ {\isasympi}{\isacharparenright}{\kern0pt}\ {\isasymTurnstile}\isactrlsub m\ {\isasymphi}{\isacharparenright}{\kern0pt}{\isacharparenright}{\kern0pt}{\isachardoublequoteclose}\isanewline
{\isacharbar}{\kern0pt}\ {\isachardoublequoteopen}{\isasympi}\ {\isasymTurnstile}\isactrlsub m\ {\isacharparenleft}{\kern0pt}G\isactrlsub m\ {\isacharbrackleft}{\kern0pt}a{\isacharcomma}{\kern0pt}\ b{\isacharbrackright}{\kern0pt}\ {\isasymphi}{\isacharparenright}{\kern0pt}\ {\isacharequal}{\kern0pt}\ {\isacharparenleft}{\kern0pt}a\ {\isasymle}\ b\ {\isasymand}\ {\isacharparenleft}{\kern0pt}length\ {\isasympi}\ {\isasymle}\ a\ {\isasymor}\ \isanewline
\ \ \ \ \ \ {\isacharparenleft}{\kern0pt}{\isasymforall}i{\isacharcolon}{\kern0pt}{\isacharcolon}{\kern0pt}nat{\isachardot}{\kern0pt}\ {\isacharparenleft}{\kern0pt}i\ {\isasymge}\ a\ {\isasymand}\ i\ {\isasymle}\ b{\isacharparenright}{\kern0pt}\ {\isasymlongrightarrow}\ {\isacharparenleft}{\kern0pt}drop\ i\ {\isasympi}{\isacharparenright}{\kern0pt}\ {\isasymTurnstile}\isactrlsub m\ {\isasymphi}{\isacharparenright}{\kern0pt}{\isacharparenright}{\kern0pt}{\isacharparenright}{\kern0pt}{\isachardoublequoteclose}\isanewline
{\isacharbar}{\kern0pt}\ {\isachardoublequoteopen}{\isasympi}\ {\isasymTurnstile}\isactrlsub m\ {\isacharparenleft}{\kern0pt}{\isasymphi}\ U\isactrlsub m\ {\isacharbrackleft}{\kern0pt}a{\isacharcomma}{\kern0pt}\ b{\isacharbrackright}{\kern0pt}\ {\isasympsi}{\isacharparenright}{\kern0pt}\ {\isacharequal}{\kern0pt}\ \ {\isacharparenleft}{\kern0pt}a\ {\isasymle}\ b\ {\isasymand}\ length\ {\isasympi}\ {\isachargreater}{\kern0pt}\ a\ {\isasymand}\ \isanewline
\ \ \ \ \ \ {\isacharparenleft}{\kern0pt}{\isasymexists}i{\isacharcolon}{\kern0pt}{\isacharcolon}{\kern0pt}nat{\isachardot}{\kern0pt}\ {\isacharparenleft}{\kern0pt}i\ {\isasymge}\ a\ {\isasymand}\ i\ {\isasymle}\ b{\isacharparenright}{\kern0pt}\ {\isasymand}\ {\isacharparenleft}{\kern0pt}{\isacharparenleft}{\kern0pt}drop\ i\ {\isasympi}{\isacharparenright}{\kern0pt}\ {\isasymTurnstile}\isactrlsub m\ {\isasympsi}\ \isanewline
\ \ \ \ \ \ \ \ {\isasymand}\ {\isacharparenleft}{\kern0pt}{\isasymforall}j{\isachardot}{\kern0pt}\ j\ {\isasymge}\ a\ {\isasymand}\ j{\isacharless}{\kern0pt}i\ {\isasymlongrightarrow}\ {\isacharparenleft}{\kern0pt}drop\ j\ {\isasympi}{\isacharparenright}{\kern0pt}\ {\isasymTurnstile}\isactrlsub m\ {\isasymphi}{\isacharparenright}{\kern0pt}{\isacharparenright}{\kern0pt}{\isacharparenright}{\kern0pt}{\isacharparenright}{\kern0pt}{\isachardoublequoteclose}\isanewline
{\isacharbar}{\kern0pt}\ {\isachardoublequoteopen}{\isasympi}\ {\isasymTurnstile}\isactrlsub m\ {\isacharparenleft}{\kern0pt}{\isasymphi}\ R\isactrlsub m\ {\isacharbrackleft}{\kern0pt}a{\isacharcomma}{\kern0pt}\ b{\isacharbrackright}{\kern0pt}\ {\isasympsi}{\isacharparenright}{\kern0pt}\ {\isacharequal}{\kern0pt}\ {\isacharparenleft}{\kern0pt}a\ {\isasymle}\ b\ {\isasymand}\ {\isacharparenleft}{\kern0pt}length\ {\isasympi}\ {\isasymle}\ a\ {\isasymor}\ \isanewline
\ \ \ \ \ \ {\isacharparenleft}{\kern0pt}{\isasymforall}i{\isacharcolon}{\kern0pt}{\isacharcolon}{\kern0pt}nat{\isachardot}{\kern0pt}\ {\isacharparenleft}{\kern0pt}i\ {\isasymge}\ a\ {\isasymand}\ i\ {\isasymle}\ b{\isacharparenright}{\kern0pt}\ {\isasymlongrightarrow}\ {\isacharparenleft}{\kern0pt}{\isacharparenleft}{\kern0pt}{\isacharparenleft}{\kern0pt}drop\ i\ {\isasympi}{\isacharparenright}{\kern0pt}\ {\isasymTurnstile}\isactrlsub m\ {\isasympsi}{\isacharparenright}{\kern0pt}{\isacharparenright}{\kern0pt}{\isacharparenright}{\kern0pt}\ {\isasymor}\ \isanewline
\ \ \ \ \ \ {\isacharparenleft}{\kern0pt}{\isasymexists}j{\isachardot}{\kern0pt}\ j\ {\isasymge}\ a\ {\isasymand}\ j\ {\isasymle}\ b{\isacharminus}{\kern0pt}{\isadigit{1}}\ {\isasymand}\ {\isacharparenleft}{\kern0pt}drop\ j\ {\isasympi}{\isacharparenright}{\kern0pt}\ {\isasymTurnstile}\isactrlsub m\ {\isasymphi}\ {\isasymand}\ \isanewline
\ \ \ \ \ \ \ \ \ {\isacharparenleft}{\kern0pt}{\isasymforall}k{\isachardot}{\kern0pt}\ a\ {\isasymle}\ k\ {\isasymand}\ k\ {\isasymle}\ j\ {\isasymlongrightarrow}\ {\isacharparenleft}{\kern0pt}drop\ k\ {\isasympi}{\isacharparenright}{\kern0pt}\ {\isasymTurnstile}\isactrlsub m\ {\isasympsi}{\isacharparenright}{\kern0pt}{\isacharparenright}{\kern0pt}{\isacharparenright}{\kern0pt}{\isacharparenright}{\kern0pt}{\isachardoublequoteclose}%
}
\snip{intervalswelldef}{1}{2}{%
\isacommand{fun}\isamarkupfalse%
\ intervals{\isacharunderscore}{\kern0pt}welldef{\isacharcolon}{\kern0pt}{\isacharcolon}{\kern0pt}\ {\isachardoublequoteopen}{\isacharprime}{\kern0pt}a\ mltl\ {\isasymRightarrow}\ bool{\isachardoublequoteclose}\isanewline
\ \ \isakeyword{where}\ {\isachardoublequoteopen}intervals{\isacharunderscore}{\kern0pt}welldef\ True\isactrlsub m\ {\isacharequal}{\kern0pt}\ True{\isachardoublequoteclose}\isanewline
\ \ {\isacharbar}{\kern0pt}\ {\isachardoublequoteopen}intervals{\isacharunderscore}{\kern0pt}welldef\ False\isactrlsub m\ {\isacharequal}{\kern0pt}\ True{\isachardoublequoteclose}\isanewline
\ \ {\isacharbar}{\kern0pt}\ {\isachardoublequoteopen}intervals{\isacharunderscore}{\kern0pt}welldef\ {\isacharparenleft}{\kern0pt}Prop\isactrlsub m\ {\isacharparenleft}{\kern0pt}p{\isacharparenright}{\kern0pt}{\isacharparenright}{\kern0pt}\ {\isacharequal}{\kern0pt}\ True{\isachardoublequoteclose}\ \isanewline
\ \ {\isacharbar}{\kern0pt}\ {\isachardoublequoteopen}intervals{\isacharunderscore}{\kern0pt}welldef\ {\isacharparenleft}{\kern0pt}Not\isactrlsub m\ {\isasymphi}{\isacharparenright}{\kern0pt}\ {\isacharequal}{\kern0pt}\ intervals{\isacharunderscore}{\kern0pt}welldef\ {\isasymphi}{\isachardoublequoteclose}\isanewline
\ \ {\isacharbar}{\kern0pt}\ {\isachardoublequoteopen}intervals{\isacharunderscore}{\kern0pt}welldef\ {\isacharparenleft}{\kern0pt}{\isasymphi}\ And\isactrlsub m\ {\isasympsi}{\isacharparenright}{\kern0pt}\ {\isacharequal}{\kern0pt}\ {\isacharparenleft}{\kern0pt}intervals{\isacharunderscore}{\kern0pt}welldef\ {\isasymphi}\ {\isasymand}\ intervals{\isacharunderscore}{\kern0pt}welldef\ {\isasympsi}{\isacharparenright}{\kern0pt}{\isachardoublequoteclose}\ \ \ \ \ \isanewline
\ \ {\isacharbar}{\kern0pt}\ {\isachardoublequoteopen}intervals{\isacharunderscore}{\kern0pt}welldef\ {\isacharparenleft}{\kern0pt}{\isasymphi}\ Or\isactrlsub m\ {\isasympsi}{\isacharparenright}{\kern0pt}\ {\isacharequal}{\kern0pt}\ {\isacharparenleft}{\kern0pt}intervals{\isacharunderscore}{\kern0pt}welldef\ {\isasymphi}\ {\isasymand}\ intervals{\isacharunderscore}{\kern0pt}welldef\ {\isasympsi}{\isacharparenright}{\kern0pt}{\isachardoublequoteclose}\ \ \ \ \ \isanewline
\ \ {\isacharbar}{\kern0pt}\ {\isachardoublequoteopen}intervals{\isacharunderscore}{\kern0pt}welldef\ {\isacharparenleft}{\kern0pt}F\isactrlsub m\ {\isacharbrackleft}{\kern0pt}a{\isacharcomma}{\kern0pt}b{\isacharbrackright}{\kern0pt}\ {\isasymphi}{\isacharparenright}{\kern0pt}\ \ {\isacharequal}{\kern0pt}\ {\isacharparenleft}{\kern0pt}a\ {\isasymle}\ b\ {\isasymand}\ intervals{\isacharunderscore}{\kern0pt}welldef\ {\isasymphi}{\isacharparenright}{\kern0pt}{\isachardoublequoteclose}\isanewline
\ \ {\isacharbar}{\kern0pt}\ {\isachardoublequoteopen}intervals{\isacharunderscore}{\kern0pt}welldef\ {\isacharparenleft}{\kern0pt}G\isactrlsub m\ {\isacharbrackleft}{\kern0pt}a{\isacharcomma}{\kern0pt}b{\isacharbrackright}{\kern0pt}\ {\isasymphi}{\isacharparenright}{\kern0pt}\ \ {\isacharequal}{\kern0pt}\ {\isacharparenleft}{\kern0pt}a\ {\isasymle}\ b\ {\isasymand}\ intervals{\isacharunderscore}{\kern0pt}welldef\ {\isasymphi}{\isacharparenright}{\kern0pt}{\isachardoublequoteclose}\isanewline
\ \ {\isacharbar}{\kern0pt}\ {\isachardoublequoteopen}intervals{\isacharunderscore}{\kern0pt}welldef\ {\isacharparenleft}{\kern0pt}{\isasymphi}\ U\isactrlsub m\ {\isacharbrackleft}{\kern0pt}a{\isacharcomma}{\kern0pt}b{\isacharbrackright}{\kern0pt}\ {\isasympsi}{\isacharparenright}{\kern0pt}\ \ {\isacharequal}{\kern0pt}\ \isanewline
\ \ \ \ \ {\isacharparenleft}{\kern0pt}a\ {\isasymle}\ b\ {\isasymand}\ intervals{\isacharunderscore}{\kern0pt}welldef\ {\isasymphi}\ {\isasymand}\ intervals{\isacharunderscore}{\kern0pt}welldef\ {\isasympsi}{\isacharparenright}{\kern0pt}{\isachardoublequoteclose}\isanewline
\ \ {\isacharbar}{\kern0pt}\ {\isachardoublequoteopen}intervals{\isacharunderscore}{\kern0pt}welldef\ {\isacharparenleft}{\kern0pt}{\isasymphi}\ R\isactrlsub m\ {\isacharbrackleft}{\kern0pt}a{\isacharcomma}{\kern0pt}b{\isacharbrackright}{\kern0pt}\ {\isasympsi}{\isacharparenright}{\kern0pt}\ \ {\isacharequal}{\kern0pt}\ \isanewline
\ \ \ \ \ {\isacharparenleft}{\kern0pt}a\ {\isasymle}\ b\ {\isasymand}\ intervals{\isacharunderscore}{\kern0pt}welldef\ {\isasymphi}\ {\isasymand}\ intervals{\isacharunderscore}{\kern0pt}welldef\ {\isasympsi}{\isacharparenright}{\kern0pt}{\isachardoublequoteclose}%
}
\snip{complenmltl}{1}{2}{%
\isacommand{fun}\isamarkupfalse%
\ complen{\isacharunderscore}{\kern0pt}mltl{\isacharcolon}{\kern0pt}{\isacharcolon}{\kern0pt}\ {\isachardoublequoteopen}{\isacharprime}{\kern0pt}a\ mltl\ {\isasymRightarrow}\ nat{\isachardoublequoteclose}\isanewline
\ \ \isakeyword{where}\ {\isachardoublequoteopen}complen{\isacharunderscore}{\kern0pt}mltl\ False\isactrlsub m\ {\isacharequal}{\kern0pt}\ {\isadigit{1}}{\isachardoublequoteclose}\isanewline
\ \ {\isacharbar}{\kern0pt}\ {\isachardoublequoteopen}complen{\isacharunderscore}{\kern0pt}mltl\ True\isactrlsub m\ {\isacharequal}{\kern0pt}\ {\isadigit{1}}{\isachardoublequoteclose}\isanewline
\ \ {\isacharbar}{\kern0pt}\ {\isachardoublequoteopen}complen{\isacharunderscore}{\kern0pt}mltl\ Prop\isactrlsub m\ {\isacharparenleft}{\kern0pt}p{\isacharparenright}{\kern0pt}\ {\isacharequal}{\kern0pt}\ {\isadigit{1}}{\isachardoublequoteclose}\isanewline
\ \ {\isacharbar}{\kern0pt}\ {\isachardoublequoteopen}complen{\isacharunderscore}{\kern0pt}mltl\ {\isacharparenleft}{\kern0pt}Not\isactrlsub m\ {\isasymphi}{\isacharparenright}{\kern0pt}\ {\isacharequal}{\kern0pt}\ complen{\isacharunderscore}{\kern0pt}mltl\ {\isasymphi}{\isachardoublequoteclose}\isanewline
\ \ {\isacharbar}{\kern0pt}\ {\isachardoublequoteopen}complen{\isacharunderscore}{\kern0pt}mltl\ {\isacharparenleft}{\kern0pt}{\isasymphi}\ And\isactrlsub m\ {\isasympsi}{\isacharparenright}{\kern0pt}\ {\isacharequal}{\kern0pt}\ max\ {\isacharparenleft}{\kern0pt}complen{\isacharunderscore}{\kern0pt}mltl\ {\isasymphi}{\isacharparenright}{\kern0pt}\ {\isacharparenleft}{\kern0pt}complen{\isacharunderscore}{\kern0pt}mltl\ {\isasympsi}{\isacharparenright}{\kern0pt}{\isachardoublequoteclose}\isanewline
\ \ {\isacharbar}{\kern0pt}\ {\isachardoublequoteopen}complen{\isacharunderscore}{\kern0pt}mltl\ {\isacharparenleft}{\kern0pt}{\isasymphi}\ Or\isactrlsub m\ {\isasympsi}{\isacharparenright}{\kern0pt}\ {\isacharequal}{\kern0pt}\ max\ {\isacharparenleft}{\kern0pt}complen{\isacharunderscore}{\kern0pt}mltl\ {\isasymphi}{\isacharparenright}{\kern0pt}\ {\isacharparenleft}{\kern0pt}complen{\isacharunderscore}{\kern0pt}mltl\ {\isasympsi}{\isacharparenright}{\kern0pt}{\isachardoublequoteclose}\isanewline
\ \ {\isacharbar}{\kern0pt}\ {\isachardoublequoteopen}complen{\isacharunderscore}{\kern0pt}mltl\ {\isacharparenleft}{\kern0pt}G\isactrlsub m\ {\isacharbrackleft}{\kern0pt}a{\isacharcomma}{\kern0pt}b{\isacharbrackright}{\kern0pt}\ {\isasymphi}{\isacharparenright}{\kern0pt}\ {\isacharequal}{\kern0pt}\ b\ {\isacharplus}{\kern0pt}\ {\isacharparenleft}{\kern0pt}complen{\isacharunderscore}{\kern0pt}mltl\ {\isasymphi}{\isacharparenright}{\kern0pt}{\isachardoublequoteclose}\isanewline
\ \ {\isacharbar}{\kern0pt}\ {\isachardoublequoteopen}complen{\isacharunderscore}{\kern0pt}mltl\ {\isacharparenleft}{\kern0pt}F\isactrlsub m\ {\isacharbrackleft}{\kern0pt}a{\isacharcomma}{\kern0pt}b{\isacharbrackright}{\kern0pt}\ {\isasymphi}{\isacharparenright}{\kern0pt}\ {\isacharequal}{\kern0pt}\ b\ {\isacharplus}{\kern0pt}\ {\isacharparenleft}{\kern0pt}complen{\isacharunderscore}{\kern0pt}mltl\ {\isasymphi}{\isacharparenright}{\kern0pt}{\isachardoublequoteclose}\isanewline
\ \ {\isacharbar}{\kern0pt}\ {\isachardoublequoteopen}complen{\isacharunderscore}{\kern0pt}mltl\ {\isacharparenleft}{\kern0pt}{\isasymphi}\ R\isactrlsub m\ {\isacharbrackleft}{\kern0pt}a{\isacharcomma}{\kern0pt}b{\isacharbrackright}{\kern0pt}\ {\isasympsi}{\isacharparenright}{\kern0pt}\ {\isacharequal}{\kern0pt}\ b\ {\isacharplus}{\kern0pt}\ {\isacharparenleft}{\kern0pt}max\ {\isacharparenleft}{\kern0pt}{\isacharparenleft}{\kern0pt}complen{\isacharunderscore}{\kern0pt}mltl\ {\isasymphi}{\isacharparenright}{\kern0pt}{\isacharminus}{\kern0pt}{\isadigit{1}}{\isacharparenright}{\kern0pt}\ {\isacharparenleft}{\kern0pt}complen{\isacharunderscore}{\kern0pt}mltl\ {\isasympsi}{\isacharparenright}{\kern0pt}{\isacharparenright}{\kern0pt}{\isachardoublequoteclose}\isanewline
\ \ {\isacharbar}{\kern0pt}\ {\isachardoublequoteopen}complen{\isacharunderscore}{\kern0pt}mltl\ {\isacharparenleft}{\kern0pt}{\isasymphi}\ U\isactrlsub m\ {\isacharbrackleft}{\kern0pt}a{\isacharcomma}{\kern0pt}b{\isacharbrackright}{\kern0pt}\ {\isasympsi}{\isacharparenright}{\kern0pt}\ {\isacharequal}{\kern0pt}\ b\ {\isacharplus}{\kern0pt}\ {\isacharparenleft}{\kern0pt}max\ {\isacharparenleft}{\kern0pt}{\isacharparenleft}{\kern0pt}complen{\isacharunderscore}{\kern0pt}mltl\ {\isasymphi}{\isacharparenright}{\kern0pt}{\isacharminus}{\kern0pt}{\isadigit{1}}{\isacharparenright}{\kern0pt}\ {\isacharparenleft}{\kern0pt}complen{\isacharunderscore}{\kern0pt}mltl\ {\isasympsi}{\isacharparenright}{\kern0pt}{\isacharparenright}{\kern0pt}{\isachardoublequoteclose}%
}
\snip{semanticequiv}{1}{2}{%
\isacommand{definition}\isamarkupfalse%
\ semantic{\isacharunderscore}{\kern0pt}equiv{\isacharcolon}{\kern0pt}{\isacharcolon}{\kern0pt}\ {\isachardoublequoteopen}{\isacharprime}{\kern0pt}a\ mltl\ {\isasymRightarrow}\ {\isacharprime}{\kern0pt}a\ mltl\ {\isasymRightarrow}\ bool{\isachardoublequoteclose}\ {\isacharparenleft}{\kern0pt}{\isachardoublequoteopen}{\isacharunderscore}{\kern0pt}\ {\isasymequiv}\isactrlsub m\ {\isacharunderscore}{\kern0pt}{\isachardoublequoteclose}\ {\isacharbrackleft}{\kern0pt}{\isadigit{8}}{\isadigit{0}}{\isacharcomma}{\kern0pt}\ {\isadigit{8}}{\isadigit{0}}{\isacharbrackright}{\kern0pt}\ {\isadigit{8}}{\isadigit{0}}{\isacharparenright}{\kern0pt}\isanewline
\ \ \isakeyword{where}\ {\isachardoublequoteopen}semantic{\isacharunderscore}{\kern0pt}equiv\ {\isasymphi}\ {\isasympsi}\ {\isacharequal}{\kern0pt}\ {\isacharparenleft}{\kern0pt}{\isasymforall}{\isasympi}{\isachardot}{\kern0pt}\ {\isacharparenleft}{\kern0pt}{\isasympi}\ {\isasymTurnstile}\isactrlsub m\ {\isasymphi}{\isacharparenright}{\kern0pt}\ {\isacharequal}{\kern0pt}\ {\isacharparenleft}{\kern0pt}{\isasympi}\ {\isasymTurnstile}\isactrlsub m\ {\isasympsi}{\isacharparenright}{\kern0pt}{\isacharparenright}{\kern0pt}{\isachardoublequoteclose}%
}
\snip{mltlsyntax}{1}{2}{%
\isacommand{datatype}\isamarkupfalse%
\ {\isacharparenleft}{\kern0pt}atoms{\isacharunderscore}{\kern0pt}mltl{\isacharcolon}{\kern0pt}\ {\isacharprime}{\kern0pt}a{\isacharparenright}{\kern0pt}\ mltl\ {\isacharequal}{\kern0pt}\isanewline
\ \ \ \ True{\isacharunderscore}{\kern0pt}mltl\ \ \ \ \ \ \ \ \ \ \ \ \ \ \ \ \ \ \ \ \ \ \ \ \ \ \ \ \ \ \ \ \ \ \ \ \ {\isacharparenleft}{\kern0pt}{\isachardoublequoteopen}True\isactrlsub m{\isachardoublequoteclose}{\isacharparenright}{\kern0pt}\ \ \ \ \ \ \ \ \ \ \ \isanewline
\ \ {\isacharbar}{\kern0pt}\ False{\isacharunderscore}{\kern0pt}mltl\ \ \ \ \ \ \ \ \ \ \ \ \ \ \ \ \ \ \ \ \ \ \ \ \ \ \ \ \ \ \ \ \ \ \ \ {\isacharparenleft}{\kern0pt}{\isachardoublequoteopen}False\isactrlsub m{\isachardoublequoteclose}{\isacharparenright}{\kern0pt}\isanewline
\ \ {\isacharbar}{\kern0pt}\ Prop{\isacharunderscore}{\kern0pt}mltl\ {\isacharprime}{\kern0pt}a\ \ \ \ \ \ \ \ \ \ \ \ \ \ \ \ \ \ \ \ \ \ \ \ \ \ \ \ \ \ \ \ \ \ {\isacharparenleft}{\kern0pt}{\isachardoublequoteopen}Prop\isactrlsub m\ {\isacharprime}{\kern0pt}{\isacharparenleft}{\kern0pt}{\isacharunderscore}{\kern0pt}{\isacharprime}{\kern0pt}{\isacharparenright}{\kern0pt}{\isachardoublequoteclose}{\isacharparenright}{\kern0pt}\isanewline
\ \ {\isacharbar}{\kern0pt}\ Not{\isacharunderscore}{\kern0pt}mltl\ {\isachardoublequoteopen}{\isacharprime}{\kern0pt}a\ mltl{\isachardoublequoteclose}\ \ \ \ \ \ \ \ \ \ \ \ \ \ \ \ \ \ \ \ \ \ \ \ \ \ \ \ {\isacharparenleft}{\kern0pt}{\isachardoublequoteopen}Not\isactrlsub m\ {\isacharunderscore}{\kern0pt}{\isachardoublequoteclose}\ {\isacharbrackleft}{\kern0pt}{\isadigit{8}}{\isadigit{5}}{\isacharbrackright}{\kern0pt}\ {\isadigit{8}}{\isadigit{5}}{\isacharparenright}{\kern0pt}\isanewline
\ \ {\isacharbar}{\kern0pt}\ And{\isacharunderscore}{\kern0pt}mltl\ {\isachardoublequoteopen}{\isacharprime}{\kern0pt}a\ mltl{\isachardoublequoteclose}\ {\isachardoublequoteopen}{\isacharprime}{\kern0pt}a\ mltl{\isachardoublequoteclose}\ \ \ \ \ \ \ \ \ \ \ \ \ \ \ \ \ \ {\isacharparenleft}{\kern0pt}{\isachardoublequoteopen}{\isacharunderscore}{\kern0pt}\ And\isactrlsub m\ {\isacharunderscore}{\kern0pt}{\isachardoublequoteclose}\ {\isacharbrackleft}{\kern0pt}{\isadigit{8}}{\isadigit{2}}{\isacharcomma}{\kern0pt}\ {\isadigit{8}}{\isadigit{2}}{\isacharbrackright}{\kern0pt}\ {\isadigit{8}}{\isadigit{1}}{\isacharparenright}{\kern0pt}\isanewline
\ \ {\isacharbar}{\kern0pt}\ Or{\isacharunderscore}{\kern0pt}mltl\ {\isachardoublequoteopen}{\isacharprime}{\kern0pt}a\ mltl{\isachardoublequoteclose}\ {\isachardoublequoteopen}{\isacharprime}{\kern0pt}a\ mltl{\isachardoublequoteclose}\ \ \ \ \ \ \ \ \ \ \ \ \ \ \ \ \ \ \ {\isacharparenleft}{\kern0pt}{\isachardoublequoteopen}{\isacharunderscore}{\kern0pt}\ Or\isactrlsub m\ {\isacharunderscore}{\kern0pt}{\isachardoublequoteclose}\ {\isacharbrackleft}{\kern0pt}{\isadigit{8}}{\isadigit{1}}{\isacharcomma}{\kern0pt}\ {\isadigit{8}}{\isadigit{1}}{\isacharbrackright}{\kern0pt}\ {\isadigit{8}}{\isadigit{0}}{\isacharparenright}{\kern0pt}\isanewline
\ \ {\isacharbar}{\kern0pt}\ Future{\isacharunderscore}{\kern0pt}mltl\ {\isachardoublequoteopen}nat{\isachardoublequoteclose}\ {\isachardoublequoteopen}nat{\isachardoublequoteclose}\ {\isachardoublequoteopen}{\isacharprime}{\kern0pt}a\ mltl{\isachardoublequoteclose}\ \ \ \ \ \ \ \ \ \ \ \ \ {\isacharparenleft}{\kern0pt}{\isachardoublequoteopen}F\isactrlsub m\ {\isacharprime}{\kern0pt}{\isacharbrackleft}{\kern0pt}{\isacharunderscore}{\kern0pt}{\isacharprime}{\kern0pt}{\isacharcomma}{\kern0pt}{\isacharunderscore}{\kern0pt}{\isacharprime}{\kern0pt}{\isacharbrackright}{\kern0pt}\ {\isacharunderscore}{\kern0pt}{\isachardoublequoteclose}\ {\isacharbrackleft}{\kern0pt}{\isadigit{8}}{\isadigit{8}}{\isacharcomma}{\kern0pt}\ {\isadigit{8}}{\isadigit{8}}{\isacharcomma}{\kern0pt}\ {\isadigit{8}}{\isadigit{8}}{\isacharbrackright}{\kern0pt}\ {\isadigit{8}}{\isadigit{7}}{\isacharparenright}{\kern0pt}\isanewline
\ \ {\isacharbar}{\kern0pt}\ Global{\isacharunderscore}{\kern0pt}mltl\ {\isachardoublequoteopen}nat{\isachardoublequoteclose}\ {\isachardoublequoteopen}nat{\isachardoublequoteclose}\ {\isachardoublequoteopen}{\isacharprime}{\kern0pt}a\ mltl{\isachardoublequoteclose}\ \ \ \ \ \ \ \ \ \ \ \ \ {\isacharparenleft}{\kern0pt}{\isachardoublequoteopen}G\isactrlsub m\ {\isacharprime}{\kern0pt}{\isacharbrackleft}{\kern0pt}{\isacharunderscore}{\kern0pt}{\isacharprime}{\kern0pt}{\isacharcomma}{\kern0pt}{\isacharunderscore}{\kern0pt}{\isacharprime}{\kern0pt}{\isacharbrackright}{\kern0pt}\ {\isacharunderscore}{\kern0pt}{\isachardoublequoteclose}\ {\isacharbrackleft}{\kern0pt}{\isadigit{8}}{\isadigit{8}}{\isacharcomma}{\kern0pt}\ {\isadigit{8}}{\isadigit{8}}{\isacharcomma}{\kern0pt}\ {\isadigit{8}}{\isadigit{8}}{\isacharbrackright}{\kern0pt}\ {\isadigit{8}}{\isadigit{7}}{\isacharparenright}{\kern0pt}\isanewline
\ \ {\isacharbar}{\kern0pt}\ Until{\isacharunderscore}{\kern0pt}mltl\ {\isachardoublequoteopen}{\isacharprime}{\kern0pt}a\ mltl{\isachardoublequoteclose}\ {\isachardoublequoteopen}nat{\isachardoublequoteclose}\ {\isachardoublequoteopen}nat{\isachardoublequoteclose}\ {\isachardoublequoteopen}{\isacharprime}{\kern0pt}a\ mltl{\isachardoublequoteclose}\ \ \ \ {\isacharparenleft}{\kern0pt}{\isachardoublequoteopen}{\isacharunderscore}{\kern0pt}\ U\isactrlsub m\ {\isacharprime}{\kern0pt}{\isacharbrackleft}{\kern0pt}{\isacharunderscore}{\kern0pt}{\isacharprime}{\kern0pt}{\isacharcomma}{\kern0pt}{\isacharunderscore}{\kern0pt}{\isacharprime}{\kern0pt}{\isacharbrackright}{\kern0pt}\ {\isacharunderscore}{\kern0pt}{\isachardoublequoteclose}\ {\isacharbrackleft}{\kern0pt}{\isadigit{8}}{\isadigit{4}}{\isacharcomma}{\kern0pt}\ {\isadigit{8}}{\isadigit{4}}{\isacharcomma}{\kern0pt}\ {\isadigit{8}}{\isadigit{4}}{\isacharcomma}{\kern0pt}\ {\isadigit{8}}{\isadigit{4}}{\isacharbrackright}{\kern0pt}\ {\isadigit{8}}{\isadigit{3}}{\isacharparenright}{\kern0pt}\ \ \isanewline
\ \ {\isacharbar}{\kern0pt}\ Release{\isacharunderscore}{\kern0pt}mltl\ {\isachardoublequoteopen}{\isacharprime}{\kern0pt}a\ mltl{\isachardoublequoteclose}\ {\isachardoublequoteopen}nat{\isachardoublequoteclose}\ {\isachardoublequoteopen}nat{\isachardoublequoteclose}\ {\isachardoublequoteopen}{\isacharprime}{\kern0pt}a\ mltl{\isachardoublequoteclose}\ \ {\isacharparenleft}{\kern0pt}{\isachardoublequoteopen}{\isacharunderscore}{\kern0pt}\ R\isactrlsub m\ {\isacharprime}{\kern0pt}{\isacharbrackleft}{\kern0pt}{\isacharunderscore}{\kern0pt}{\isacharprime}{\kern0pt}{\isacharcomma}{\kern0pt}{\isacharunderscore}{\kern0pt}{\isacharprime}{\kern0pt}{\isacharbrackright}{\kern0pt}\ {\isacharunderscore}{\kern0pt}{\isachardoublequoteclose}\ {\isacharbrackleft}{\kern0pt}{\isadigit{8}}{\isadigit{4}}{\isacharcomma}{\kern0pt}\ {\isadigit{8}}{\isadigit{4}}{\isacharcomma}{\kern0pt}\ {\isadigit{8}}{\isadigit{4}}{\isacharcomma}{\kern0pt}\ {\isadigit{8}}{\isadigit{4}}{\isacharbrackright}{\kern0pt}\ {\isadigit{8}}{\isadigit{3}}{\isacharparenright}{\kern0pt}%
}

\snip{mltlsemanticspartial}{1}{2}{%
\isacommand{fun}\isamarkupfalse%
\ semantics{\isacharunderscore}{\kern0pt}mltl{\isacharcolon}{\kern0pt}{\isacharcolon}{\kern0pt}{\isachardoublequoteopen}{\isacharbrackleft}{\kern0pt}{\isacharprime}{\kern0pt}a\ set\ list{\isacharcomma}{\kern0pt}\ {\isacharprime}{\kern0pt}a\ mltl{\isacharbrackright}{\kern0pt}\ {\isasymRightarrow}\ bool{\isachardoublequoteclose} \isakeyword{where}\isanewline
{\isachardoublequoteopen}{\isasympi}\ {\isasymTurnstile}\isactrlsub m\ Prop\isactrlsub m\ {\isacharparenleft}{\kern0pt}q{\isacharparenright}{\kern0pt}\ {\isacharequal}{\kern0pt}\ {\isacharparenleft}{\kern0pt}{\isasympi}\ {\isasymnoteq}\ {\isacharbrackleft}{\kern0pt}{\isacharbrackright}{\kern0pt}\ {\isasymand}\ q\ {\isasymin}\ {\isacharparenleft}{\kern0pt}{\isasympi}\ {\isacharbang}{\kern0pt}\ {\isadigit{0}}{\isacharparenright}{\kern0pt}{\isacharparenright}{\kern0pt}{\isachardoublequoteclose}\isanewline
{\isacharbar}{\kern0pt}\ {\isachardoublequoteopen}{\isasympi}\ {\isasymTurnstile}\isactrlsub m\ {\isasymphi}\ And\isactrlsub m\ {\isasympsi}\ {\isacharequal}{\kern0pt}\ {\isacharparenleft}{\kern0pt}{\isasympi}\ {\isasymTurnstile}\isactrlsub m\ {\isasymphi}\ {\isasymand}\ {\isasympi}\ {\isasymTurnstile}\isactrlsub m\ {\isasympsi}{\isacharparenright}{\kern0pt}{\isachardoublequoteclose}\isanewline
{\isacharbar}{\kern0pt}\ {\isachardoublequoteopen}{\isasympi}\ {\isasymTurnstile}\isactrlsub m\ {\isacharparenleft}{\kern0pt}G\isactrlsub m\ {\isacharbrackleft}{\kern0pt}a{\isacharcomma}{\kern0pt}\ b{\isacharbrackright}{\kern0pt}\ {\isasymphi}{\isacharparenright}{\kern0pt}\ {\isacharequal}{\kern0pt}\ {\isacharparenleft}{\kern0pt}a\ {\isasymle}\ b\ {\isasymand}\ {\isacharparenleft}{\kern0pt}length\ {\isasympi}\ {\isasymle}\ a\ {\isasymor}\ \isanewline
\ \ \ \ \ \ {\isacharparenleft}{\kern0pt}{\isasymforall}i{\isacharcolon}{\kern0pt}{\isacharcolon}{\kern0pt}nat{\isachardot}{\kern0pt}\ {\isacharparenleft}{\kern0pt}i\ {\isasymge}\ a\ {\isasymand}\ i\ {\isasymle}\ b{\isacharparenright}{\kern0pt}\ {\isasymlongrightarrow}\ {\isacharparenleft}{\kern0pt}drop\ i\ {\isasympi}{\isacharparenright}{\kern0pt}\ {\isasymTurnstile}\isactrlsub m\ {\isasymphi}{\isacharparenright}{\kern0pt}{\isacharparenright}{\kern0pt}{\isacharparenright}{\kern0pt}{\isachardoublequoteclose}\isanewline
{\isacharbar}{\kern0pt}\ {\isachardoublequoteopen}{\isasympi}\ {\isasymTurnstile}\isactrlsub m\ {\isacharparenleft}{\kern0pt}{\isasymphi}\ U\isactrlsub m\ {\isacharbrackleft}{\kern0pt}a{\isacharcomma}{\kern0pt}\ b{\isacharbrackright}{\kern0pt}\ {\isasympsi}{\isacharparenright}{\kern0pt}\ {\isacharequal}{\kern0pt}\ \ {\isacharparenleft}{\kern0pt}a\ {\isasymle}\ b\ {\isasymand}\ length\ {\isasympi}\ {\isachargreater}{\kern0pt}\ a\ {\isasymand}\ \isanewline
\ \ \ \ \ \ {\isacharparenleft}{\kern0pt}{\isasymexists}i{\isacharcolon}{\kern0pt}{\isacharcolon}{\kern0pt}nat{\isachardot}{\kern0pt}\ {\isacharparenleft}{\kern0pt}i\ {\isasymge}\ a\ {\isasymand}\ i\ {\isasymle}\ b{\isacharparenright}{\kern0pt}\ {\isasymand}\ {\isacharparenleft}{\kern0pt}{\isacharparenleft}{\kern0pt}drop\ i\ {\isasympi}{\isacharparenright}{\kern0pt}\ {\isasymTurnstile}\isactrlsub m\ {\isasympsi}\ \isanewline
\ \ \ \ \ \ \ \ {\isasymand}\ {\isacharparenleft}{\kern0pt}{\isasymforall}j{\isachardot}{\kern0pt}\ j\ {\isasymge}\ a\ {\isasymand}\ j{\isacharless}{\kern0pt}i\ {\isasymlongrightarrow}\ {\isacharparenleft}{\kern0pt}drop\ j\ {\isasympi}{\isacharparenright}{\kern0pt}\ {\isasymTurnstile}\isactrlsub m\ {\isasymphi}{\isacharparenright}{\kern0pt}{\isacharparenright}{\kern0pt}{\isacharparenright}{\kern0pt}{\isacharparenright}{\kern0pt}{\isachardoublequoteclose}
}

\snip{complenmltlpartial}{1}{2}{%
\isacommand{fun}\isamarkupfalse%
\ complen{\isacharunderscore}{\kern0pt}mltl{\isacharcolon}{\kern0pt}{\isacharcolon}{\kern0pt}\ {\isachardoublequoteopen}{\isacharprime}{\kern0pt}a\ mltl\ {\isasymRightarrow}\ nat{\isachardoublequoteclose}
\isakeyword{where}\ {\isachardoublequoteopen}complen{\isacharunderscore}{\kern0pt}mltl\ Prop\isactrlsub m\ {\isacharparenleft}{\kern0pt}p{\isacharparenright}{\kern0pt}\ {\isacharequal}{\kern0pt}\ {\isadigit{1}}{\isachardoublequoteclose}\isanewline
{\isacharbar}{\kern0pt}\ {\isachardoublequoteopen}complen{\isacharunderscore}{\kern0pt}mltl\ {\isacharparenleft}{\kern0pt}{\isasymphi}\ And\isactrlsub m\ {\isasympsi}{\isacharparenright}{\kern0pt}\ {\isacharequal}{\kern0pt}\ max\ {\isacharparenleft}{\kern0pt}complen{\isacharunderscore}{\kern0pt}mltl\ {\isasymphi}{\isacharparenright}{\kern0pt}\ {\isacharparenleft}{\kern0pt}complen{\isacharunderscore}{\kern0pt}mltl\ {\isasympsi}{\isacharparenright}{\kern0pt}{\isachardoublequoteclose}\isanewline
{\isacharbar}{\kern0pt}\ {\isachardoublequoteopen}complen{\isacharunderscore}{\kern0pt}mltl\ {\isacharparenleft}{\kern0pt}G\isactrlsub m\ {\isacharbrackleft}{\kern0pt}a{\isacharcomma}{\kern0pt}b{\isacharbrackright}{\kern0pt}\ {\isasymphi}{\isacharparenright}{\kern0pt}\ {\isacharequal}{\kern0pt}\ b\ {\isacharplus}{\kern0pt}\ {\isacharparenleft}{\kern0pt}complen{\isacharunderscore}{\kern0pt}mltl\ {\isasymphi}{\isacharparenright}{\kern0pt}{\isachardoublequoteclose}\isanewline
{\isacharbar}{\kern0pt}\ {\isachardoublequoteopen}complen{\isacharunderscore}{\kern0pt}mltl\ {\isacharparenleft}{\kern0pt}{\isasymphi}\ U\isactrlsub m\ {\isacharbrackleft}{\kern0pt}a{\isacharcomma}{\kern0pt}b{\isacharbrackright}{\kern0pt}\ {\isasympsi}{\isacharparenright}{\kern0pt}\ {\isacharequal}{\kern0pt}\ b\ {\isacharplus}{\kern0pt}\ {\isacharparenleft}{\kern0pt}max\ {\isacharparenleft}{\kern0pt}{\isacharparenleft}{\kern0pt}complen{\isacharunderscore}{\kern0pt}mltl\ {\isasymphi}{\isacharparenright}{\kern0pt}{\isacharminus}{\kern0pt}{\isadigit{1}}{\isacharparenright}{\kern0pt}\ {\isacharparenleft}{\kern0pt}complen{\isacharunderscore}{\kern0pt}mltl\ {\isasympsi}{\isacharparenright}{\kern0pt}{\isacharparenright}{\kern0pt}{\isachardoublequoteclose}%
}

\snip{proglenoneuntil}{1}{2}{
formula{\isacharunderscore}{\kern0pt}progression{\isacharunderscore}{\kern0pt}len{\isadigit{1}}\ {\isacharparenleft}{\kern0pt}\isasymphi\ U\isactrlsub m\ {\isacharbrackleft}{\kern0pt}a{\isacharcomma}{\kern0pt}b{\isacharbrackright}{\kern0pt}\ \isasympsi{\isacharparenright}{\kern0pt}\ tr{\isacharunderscore}{\kern0pt}entry\ {\isacharequal}{\kern0pt}\ \ \isanewline
{\isacharparenleft}{\kern0pt}if\ {\isacharparenleft}{\kern0pt}{\isadigit{0}}{\isacharless}{\kern0pt}a\ {\isasymand}\ a{\isasymle}b{\isacharparenright}{\kern0pt}\ then\ {\isacharparenleft}{\kern0pt}\isasymphi\ U\isactrlsub m\ {\isacharbrackleft}{\kern0pt}{\isacharparenleft}{\kern0pt}a{\isacharminus}{\kern0pt}{\isadigit{1}}{\isacharparenright}{\kern0pt}{\isacharcomma}{\kern0pt}\ {\isacharparenleft}{\kern0pt}b{\isacharminus}{\kern0pt}{\isadigit{1}}{\isacharparenright}{\kern0pt}{\isacharbrackright}{\kern0pt}\ \isasympsi{\isacharparenright}{\kern0pt}\ else\ {\isacharparenleft}{\kern0pt}if\ {\isacharparenleft}{\kern0pt}{\isadigit{0}}{\isacharequal}{\kern0pt}a\ {\isasymand}\ a{\isacharless}{\kern0pt}b{\isacharparenright}{\kern0pt}\ \isanewline
\ \ then\ {\isacharparenleft}{\kern0pt}{\isacharparenleft}{\kern0pt}formula{\isacharunderscore}{\kern0pt}progression{\isacharunderscore}{\kern0pt}len{\isadigit{1}}\ \isasympsi\ tr{\isacharunderscore}{\kern0pt}entry{\isacharparenright}{\kern0pt}\ Or\isactrlsub m\ \isanewline
\ \ \ \ {\isacharparenleft}{\kern0pt}{\isacharparenleft}{\kern0pt}formula{\isacharunderscore}{\kern0pt}progression{\isacharunderscore}{\kern0pt}len{\isadigit{1}}\ \isasymphi\ tr{\isacharunderscore}{\kern0pt}entry{\isacharparenright}{\kern0pt}\ And\isactrlsub m\ {\isacharparenleft}{\kern0pt}\isasymphi\ U\isactrlsub m\ {\isacharbrackleft}{\kern0pt}{\isadigit{0}}{\isacharcomma}{\kern0pt}\ {\isacharparenleft}{\kern0pt}b{\isacharminus}{\kern0pt}{\isadigit{1}}{\isacharparenright}{\kern0pt}{\isacharbrackright}{\kern0pt}\ \isasympsi{\isacharparenright}{\kern0pt}{\isacharparenright}{\kern0pt}{\isacharparenright}{\kern0pt}\isanewline
\ \ else\ {\isacharparenleft}{\kern0pt}formula{\isacharunderscore}{\kern0pt}progression{\isacharunderscore}{\kern0pt}len{\isadigit{1}}\ \isasympsi\ tr{\isacharunderscore}{\kern0pt}entry{\isacharparenright}{\kern0pt}{\isacharparenright}{\kern0pt}{\isacharparenright}{\kern0pt}{\isachardoublequoteclose}
}

\title{Formalizing MLTL Formula Progression in Isabelle/HOL}

\author{Katherine Kosaian\and Zili Wang\and Elizabeth Sloan\and Kristin Rozier \thanks{University of Iowa, Iowa City, USA (Kosaian); Iowa State University, Ames, USA (Wang, Sloan, Rozier) \newline $\{$katherine-kosaian$\}$@uiowa.edu; $\{$ziliw1,elisloan,kyrozier$\}$@iastate.edu}}
\date{}

\usepackage{ifpdf}
\ifpdf
\fi

\begin{document}

\maketitle

\begin{abstract}
\textit{Mission-time Linear Temporal Logic (MLTL)} is rapidly increasing in popularity as a specification logic, e.g., for runtime verification and model checking, driving a need for a trustworthy tool base for analyzing MLTL. 
In this work, we formalize the syntax and semantics of MLTL and a library of key properties, including useful custom induction rules.
We envision this library as being useful for future formalizations involving MLTL and as serving as a reference point for theoretical work using or developing MLTL.
We then formalize the algorithm and correctness theorems for MLTL \textit{formula progression}; along the way, we identify and fix several errors and gaps in the source material.
A main motivation for our work is \textit{tool validation}; we ensure the executability of our algorithms by using Isabelle's built-in code generator. 
\end{abstract}

\section{Introduction}
Mission-time Linear Temporal Logic (MLTL) \cite{RRS14,LVR19} adds discrete, closed-interval, natural number bounds to the temporal operators of LTL, providing a finite-trace specification logic that captures many common requirements of, e.g., embedded systems. MLTL is a widely-used subset of Signal Temporal Logic (STL) \cite{MN04} and Metric Temporal Logic (MTL) \cite{AH90} (see \cite{OW08}).

Labeled timelines are common representations of requirements for operational concepts of aerospace systems.
Due to its intuitive expressiveness of finite timeline properties, MLTL has seen wide adoption as a specification logic in this context.
After an extensive survey of verification tools and their associated specification languages, NASA's Lunar Gateway Vehicle System Manager (VSM) team selected MLTL to formalize their English Assume-Guarantee Contract requirements \cite{DBR21,Dab21,DRB22}.
The VSM team is also running R2U2 \cite{JJKRZ23}, a runtime verification engine that natively reasons over MLTL for on-board operational verification and on-ground timeline verification.
NASA previously used MLTL to specify fault disambiguation protocols embedded in the knee of Robonaut2 on the International Space Station \cite{KZJZR20}, for runtime requirements of NASA's S1000 octocopter \cite{LBQGRR17} in the Autonomy Operating System (AOS) for UAS \cite{LB15}, and to specify system health management properties of the Swift UAS \cite{SRRMMI15,RRS14,GRS14,RSI15} and the DragonEye UAS \cite{SRRMMI15,SMR16}. MLTL was the specification logic of choice for a UAS Traffic Management (UTM) system \cite{CHHJR20}, an open-source UAS \cite{JABDEGJKLMRVR21}, and a high-altitude balloon \cite{LJBHCLR22}. 
JAXA used MLTL for specification of a resource-limited autonomous satellite mission \cite{JAXA}. 
Other space systems with MLTL specifications include a CubeSat communications system \cite{LLR21}, a sounding rocket \cite{HLR21}, the CySat-I satellite's autonomous fault recovery system \cite{AJR22}, and a case study on small satellites and landers \cite{Roz16b}.

Accordingly, many formal methods tools natively analyze MLTL specifications. MLTL was first named as the input logic to the Realizable Responsive Unobtrusive Unit (R2U2) \cite{RRS14,RS17,JJKRZ23}. The Formal Requirements Elicitation Tool (FRET) provides a GUI with color-coded segments of structured natural language to elicit more accurate MLTL specifications from system designers \cite{GMRPSS20,NASA-FRET,Mav22}.
WEST \cite{DBLP:conf/ifm/ElwingGSTWR23} provides a GUI to interactively validate MLTL specifications via regular expressions and sets of satisfying and falsifying traces. The model checker {\textsc nuXmv} accepts a subset of MLTL for use in symbolic model checking, where the $G$ and $F$ operators of an \texttt{LTLSPEC} can have integer bounds \cite{nuXmv-v1.1.0}, though bounds cannot be placed on $U$ or $V$ (the Release operator of \textsc{nuxmv}).
Ogma is a command-line tool to produce monitoring applications from MLTL formulas \cite{PMPGG22,Per23,ogma}. 
As MLTL represents a common core of logics used for runtime monitoring, it was featured in the 2018 Runtime Verification Benchmark Competition \cite{Roz17,RVBC2018}. We closely examine MLTL \emph{formula progression} \cite{LR18}, an algorithm submitted to generate MLTL benchmarks in that competition; later, the Formula PROGression Generator (FPROGG) expanded upon formula progression to create an (unverified) automated MLTL benchmark generation tool \cite{RR25}. 

Despite the rapid emergence of tools that analyze MLTL specifications, there is still a shortage of provably correct formal foundations from which to validate and verify those tools and their algorithms. We have SAT solvers for MLTL that work through translations to other logics, proved correct via pencil-and-paper proofs and experimental validation \cite{LVR19}. There is a native MLTL MAX-SAT algorithm and implementation that also relies on hand-proofs and experimental demonstrations of consistency \cite{HJRW23}. In contrast, LTL has benefitted from multiple mechanized formalizations, including an Isabelle/HOL library \cite{LTL-AFP}, and a sizeable body of work \cite{LTL_Master_Theorem-AFP,LTL_Normal_Form-AFP,LTL_to_DRA-AFP,LTL_to_GBA-AFP,CAVA_LTL_Modelchecker-AFP} that has built on this entry to establish a formalized collection of results in Isabelle/HOL. There is even a formalization of formula progression for the 3-valued variant LTL3 \cite{LTL3_Semantics-AFP}.
LTL is also formalized in Coq \cite{DBLP:journals/logcom/Coupet-Grimal03} and in PVS, where a library has been developed to facilitate modeling and verification for systems using LTL \cite{DBLP:conf/birthday/PnueliA03}. Similarly, MTL is formalized in PVS; this library was used to verify a translation from a structured natural language to MTL \cite{DBLP:conf/cpp/ConradTGPD22,fret-proof-framework}.\footnote{While this library contains some notions relevant for MLTL, it is not specialized for MLTL.} A formalization of MTL in Coq was used to generate OCaml code implementing a past-time MTL monitoring engine \cite{MTL_monitor_coq}. 
VeriMon \cite{DBLP:conf/ictac/BasinDHHMKKMRST22}, a monitoring tool for metric first-order temporal logic (MFOTL), was also developed and formally verified in Isabelle/HOL; Isabelle's code export is used to generate OCaml code for the tool. 
Similarly, Vydra \cite{DBLP:conf/atva/RaszykBT20}, a monitoring tool for metric dynamic logic (MDL), is formalized in Isabelle/HOL.
Notably, these logics and their associated algorithms can be quite intricate.
For example, in the case of Metric Interval Temporal Logic (MITL), recent work \cite{DBLP:conf/vmcai/Roohi018} found that the original semantics was incorrectly specified; the authors correct some of the original algorithms and avail themselves of formalization in PVS for an extra layer of trustworthiness.

The time has come to formalize MLTL. The remainder of this paper lays out our contributions as follows. 
We formalize the syntax and semantics of MLTL in the theorem prover Isabelle/HOL \cite{DBLP:books/sp/NipkowPW02,DBLP:journals/jar/Paulson89} and develop a formal library of key properties and useful lemmas for this encoding, including duality properties, an NNF transformation, and custom induction rules; see \rref{sec:encoding}.
We then verify formula progression \cite{LR18} in \rref{sec:fp}, fixing errors in the original correctness proofs; our verified algorithm is \textit{executable} and can be exported to code in SML, OCaml, Haskell, or Scala, using Isabelle/HOL's code generator.
\rref{sec:challenges} details the challenges and insights that emerged from our formalization. Our code is approximately 3300 lines in Isabelle/HOL and is available on \href{https://drive.google.com/drive/folders/15u-NjKc7GHcrd8g1xnO4Be9AbHC3wnwV?usp=sharing}{Google Drive} at https://tinyurl.com/fscd25artifact.\footnote{We plan to make our formula progression formalization available on Isabelle/HOL's Archive of Formal Proofs alongside our core MLTL library, which is already posted
\cite{Mission_Time_LTL-AFP}.} 
\rref{sec:conclusion} concludes with a high-level discussion of impact and future directions.

\section{Encoding MLTL in Isabelle/HOL}\label{sec:encoding}
We overview the syntax and semantics of MLTL before discussing our formalization thereof.
Let $\texttt{AP}$ be a finite set of atomic propositions. The grammar for MLTL formulas \cite{LVR19, RRS14} is:
\begin{align*}
\centering
\varphi, \psi ::=
\ \text{True} \ | 
\ \text{False} \ | 
\ p \ | 
\ \neg \varphi \ | 
\ \varphi \land \psi \ | 
\ \varphi \lor \psi \ |
\ \texttt{F}_{[a,b]} \ \varphi \ | 
\ \texttt{G}_{[a,b]} \ \varphi \ | 
\ \varphi \ \texttt{U}_{[a,b]} \ \psi \ | 
\ \varphi \ \texttt{R}_{[a,b]} \ \psi, 
\end{align*}
where $p \in \texttt{AP}$, and $a, b \in \mathbb{Z}$ with $0 \leq a \leq b$. 

\begin{wrapfigure}{r}{0.5\textwidth}  
\centering
\includegraphics[width=0.47\textwidth]{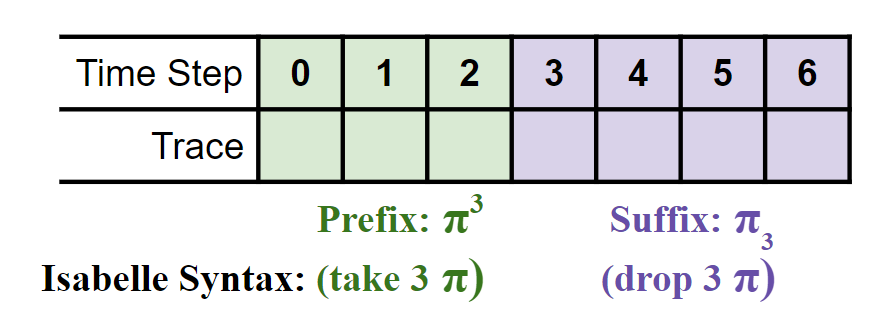}
\caption{Prefix and suffix notation for traces.}
\label{fig:PrefixSuffix}
\end{wrapfigure}

A \textit{trace} $\pi$ is a finite sequence of sets of atomic propositions, i.e., $\pi = \pi[0], \pi[1], \ldots, \pi[n]$ where $\pi[i] \subseteq \texttt{AP}$ for all $i \in \{0, 1, \ldots, n\}$.
Each $\pi[i]$ is called a \textit{state} of the trace $\pi$ and represents the set of atomic propositions that are true at time $i$.
The \text{length} of trace $\pi$ is denoted by $|\pi|$.
 We also use the following notation \cite{LVR19,LR18}: $\pi^k$ denotes the prefix of trace $\pi$ from 0 to $k-1$, i.e., $\pi^k = \pi[0], \pi[1], \ldots, \pi[k-1]$, and $\pi_k$ is the suffix of $\pi$ from $k$ onwards, i.e., $\pi_k = \pi[k], \pi[i+1], \ldots, \pi[n]$.
For easy reference, we visually summarize this notation in \rref{fig:PrefixSuffix}.

Trace $\pi$ satisfies MLTL formula $\varphi$ iff $\pi \models \varphi$, where $\models$ is defined inductively \cite{DBLP:conf/ifm/ElwingGSTWR23}:
\begin{table}[h]
\centering
\begin{tabular}{ll}
$\pi \models p$ iff $p \in \pi[0]$ &
$\pi \models \neg \varphi$ iff $\pi \not\models \varphi$ \\[4pt] 
$\pi \models \varphi \wedge \psi$ iff $\pi \models \varphi$ and $\pi \models \psi$   &
$\pi \models \varphi \vee \psi$ iff $\pi \models \varphi$ or $\pi \models \psi$     \\[4pt]
$\pi \models \texttt{F}_{[a,b]} \ \varphi \text{ iff } |\pi| > a \text{ and } \exists i \in [a,b]. \pi_i \models \varphi$  \ \ \ \ \
& $\pi \models \texttt{G}_{[a,b]} \ \varphi \text{ iff } |\pi| \leq a \text{ or } \forall i \in [a,b] \ \pi_i \models \varphi$
 \end{tabular}
\begin{tabular}{l}
$\pi \models \varphi \ \texttt{U}_{[a,b]} \ \psi \text{ iff } |\pi| > a \text{ and } \exists i \in [a,b]. \pi_i \models \psi$
$\text{ and }\forall j \in [a, i-1]\ \pi_j \models \varphi$ \\[4pt]
$\pi \models \varphi \ \texttt{R}_{[a,b]} \ \psi \text{ iff } |\pi| \leq a \text{ or } (\forall i \in [a,b] \ \pi_i \models \psi) \text{ or }$ $(\exists j \in [a,b]. \pi_j \models \varphi \text{ and } \forall k \in [a, j] \ \pi_k \models \psi$)
\end{tabular}
\end{table}

The temporal operators $\texttt{F}, \texttt{G}, \texttt{U}, \texttt{R}$ 
are commonly referred to as ``Future,'' (see, e.g., \cite{Pnu77,Var08,Roz11}) 
``Globally,'' ``Until,'' and ``Release,'' respectively.\footnote{The literature sometimes uses $\Diamond$ and $\Box$ to denote Future and Globally, but we use $\texttt{F}$ and $\texttt{G}$ to match the other temporal operators. The Future operator is also called ``Finally'' or ``Eventually'' \cite{LVR19,DBLP:conf/ifm/ElwingGSTWR23}.}
While more temporal operators, such as a Next operator and variations on Until and Release (which can be easily defined with the above grammar), could be included in the syntax of MLTL, we choose to use the above grammar as some of the algorithms we are most interested in formalizing (particularly formula progression \cite{LVR19} and MLTL-to-regular expressions \cite{DBLP:conf/ifm/ElwingGSTWR23}) use it.
However, in a development that necessitates frequent use of additional temporal operators, it would be straightforward to expand our core syntax either by directly creating an alternative expanded version of the syntax or by defining abbreviations for additional temporal operators.

We present our syntactical encoding of MLTL in \rref{sec:syntax}.
Our formalization follows the mathematical presentation of MLTL; we define each of the temporal operators $\texttt{F}, \texttt{G}, \texttt{U}, \texttt{R}$ and then formalize important properties of these operators, following existing pencil-and-paper proofs \cite{DBLP:conf/ifm/ElwingGSTWR23} (but deviating when necessary to make the proofs work in Isabelle/HOL).
\rref{sec:semantics} overviews our formalization of MLTL semantics, and \rref{sec:props}, \rref{sec:functions} present our formalization of basic properties and useful functions for MLTL, respectively. 
Although our development is standalone in that it only imports the standard HOL library, we drew some inspiration from the formalization of LTL in Isabelle/HOL \cite{LTL-AFP}.

\subsection{Syntax}\label{sec:syntax} 
In Isabelle/HOL, we formalize the grammar for MLTL formulas in the \isa{mltl} datatype, with syntactic sugar for pretty printing.
The atomic formulas are \isa{True\isactrlsub m}, \isa{False\isactrlsub m}, and atomic propositions \isa{Prop\isactrlsub m} which take a variable with arbitrary type \isa{'a} as an argument (the arbitrary type here provides flexibility).
The logical connectives are \isa{Not\isactrlsub m} ($\lnot$), \isa{And\isactrlsub m} ($\land$), and \isa{Or\isactrlsub m} ($\lor$).
Lastly, the temporal operators are \isa{F\isactrlsub m} (future), \isa{G\isactrlsub m} (globally), \isa{U\isactrlsub m} (until), and \isa{R\isactrlsub m} (release), each taking an interval $[a,b]$ of two naturals to specify the associated time bounds.
For the syntactic sugar, we mirror the operator precedence of the existing LTL formalization \cite{LTL-AFP}.

Notably, our syntax does not include the standard well-definedness assumption on intervals for temporal operators. 
For instance, our \textit{syntax} allows the formula ``\isa{F\isactrlsub m[5,3] True\isactrlsub m}'', even though this is not a well-defined formula in MLTL.
Instead, our \textit{semantics} requires that these intervals have well-defined bounds, and we include an assumption to this effect in our correctness theorems when necessary.

\subsection{Semantics}\label
{sec:semantics}
\begin{wrapfigure}{l}{0.53\textwidth}
\centering
\includegraphics[width=0.4\textwidth]{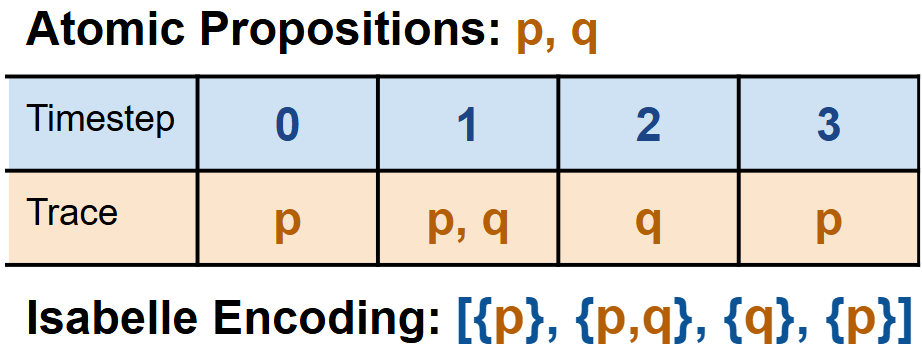}
\caption{Here, we have two Boolean variables, $p$ and $q$, and four timesteps. At time $0$, $p$ is true (and $q$ is false), at time $1$, $p$ $q$ are both true, etc.
In Isabelle, we represent this in the list $[\{p\},\{p, q\}, \{q\},\{p\}]$.
}
\label{fig:trace}
\end{wrapfigure}
MLTL traces are finite, and so we encode them as lists of sets of variables of arbitrary type (in Isabelle/HOL, this is type \isa{'a\ set\ list}).
\rref{fig:trace} gives an example.

We formalize the semantics of MLTL formulas in the function \isa{semantics\_mltl}, which takes a trace $\pi$ and an MLTL formula $\varphi$ and returns true if $\pi \models \varphi$ and false otherwise.
Our formal definition mirrors the mathematical definition, while making certain implicit assumptions explicit. 
For example, for atomic propositions, we explicitly ensure that if $\isasympi \models \isa{Prop\isactrlsub m p}$ holds, then $\isasympi$ is nonempty (for well-definedness).
Additionally, in the semantics of temporal operators, we explicitly ensure that the associated time bounds are well-defined (i.e, \isa{a \isasymle\ b}).
We now present the formal semantics for a representative subset of operators (\isa{Prop\isactrlsub m}, \isa{And\isactrlsub m}, \isa{G\isactrlsub m}, and  \isa{U\isactrlsub m}); the full semantics is in \rref{app:semantics}. 
\begin{isabelle}
\mltlsemanticspartial
\end{isabelle}

Here, for \isa{Prop\isactrlsub m}, we check that the atomic proposition \isa{q} is in the set of atomic propositions at the initial timestep (in Isabelle/HOL, this is {\isasympi}\ {\isacharbang}{\kern0pt}\ {\isadigit{0}}, as \isacharbang\ is the operator to access list elements).
For \isa{And\_mltl}, we check that trace \isasympi\ satisfies both subformulas \isasymphi, \isasympsi.

For \isa{G\isactrlsub m}, we check that the subformula \isasymphi\ is satisfied at every step in the trace.
More precisely, \isa{G\isactrlsub m[a,b] \isasymphi} holds automatically if the length of $\isasymphi$ is less than \isa{a}; otherwise, \isa{(drop\ i\ \isasympi)} $\models$ \isasymphi\ must hold for all \isa{i} between \isa{a} and \isa{b}.
Here, {\kern0pt}\isa{drop}\ i\ {\isasympi} encodes $\pi_i$, the suffix of the trace starting at timestep $i$, as \isa{drop} is Isabelle/HOL's operator to drop the first $i$ elements from a (zero-indexed) list.
We also ensure that \isa{a\isasymle b} for well-definedness.

For \isa{U\isactrlsub m}, we check \isa{a\isasymle b} for well-definedness, and then we check that there exists an \isa{i} between \isa{a} and \isa{b} with \isa{(drop\ i\ \isasympi)} $\models$ \isasympsi, and where \isasymphi\ is satisfied at all timesteps up to \isa{i}, i.e., \isa{(drop\ j\ \isasympi)} $\models$ \isasymphi {\kern0pt} for all \isa{j} between \isa{a} and \isa{i-1}.
In other words, the semantics for \isa{\isasymphi\ U\isactrlsub m[a,b] \isasympsi} checks that $\psi$ is satisfied at some point in the relevant interval [\isa{a},\isa{b}], and that $\varphi$ is satisfied at every timestep in [\isa{a},\isa{b}] before that point. 

Formalizing the semantics of MLTL helps to clarify the differences between MLTL and closely-related logics, and we envision formalizing bridges from MLTL to related logics and vice-versa (when possible) as an important topic for future work.
Perhaps the most closely-related is MTL-over-naturals \cite{OW08}.
Unlike MTL-over-naturals, MLTL traces are finite and MLTL intervals are finite, closed, and unitless (generic).
The semantics of Until is also different for MTL-over-naturals.\footnote{In MTL-over-naturals, $\pi \models \varphi \ \texttt{U}_{[a,b]}\psi$ iff $ |\pi| > a$ and $\exists i \in [a, \ b], i < |\pi|$ s.t. $\pi[i] \models \psi$ and \textbf{for all} $j < i$ (including $j < a$), $\pi[j] \models \varphi$.}
Note that other variants of MTL, such as Metric Interval Temporal Logic (MITL) \cite{DBLP:conf/podc/AlurFH91} treat the temporal operator intervals as continuous (rather than discrete) intervals that may be open-ended over real numbers (not integers), and even prohibit singular time intervals of the form $[a, a]$.
Bounded LTL (BLTL) \cite{DBLP:journals/apal/Kamide12} and $\text{LTL}_f$ \cite{DBLP:conf/ijcai/GiacomoV13} both reason over finite traces like MLTL, but share syntax with LTL; they have (unspecified) finite bounds on the time domain rather than intervals on each temporal operator and keep the $X$ operator. ($\text{LTL}_f$ adds a weak next operator as well.) 

\subsection{Properties}\label{sec:props}
To increase the usability of our formal semantics, we verify various essential properties thereof.
Our aim is to establish a reusable formalized library of MLTL properties.

In Isabelle/HOL, we formally define the \textit{semantic equivalence} of MLTL formulas in  $\mathtt{semantic\_equiv}$, which tests if \isasymphi\ and \isasympsi\ are equivalent by checking if their semantics match on all traces \isasympi:
\begin{isabelle}
  \semanticequiv
\end{isabelle}

We then consider properties involving relationships between operators.
While we define each core temporal operator  ($\texttt{F}$, $\texttt{G}$, $\texttt{U}$, $\texttt{R}$) in terms of its \textit{mathematical} definition, a common alternative approach (see, e.g.,  \cite{LVR19, Roz11, HJRW23}) is to formally define all MLTL operators via a functionally complete set, like $\{\neg, \land, \texttt{U}\}$ \cite{LVR19}, and then derive the standard mathematical definitions.
To establish that our encoding matches this alternative approach, we verify key equivalences between operators.
Many of these prove easily with Isabelle/HOL's built-in automation tactics.
For example, we establish that Future can be rewritten with Until in the lemma \isa{future\_as\_until} by showing that $\texttt{F}_{[a,b]} \varphi$ is equivalent to $\text{True} \ \texttt{U}_{[a,b]} \varphi$,
and we establish the duality of Future and Globally (i.e., that $\texttt{G}_{[a,b]} \varphi$ is equivalent to $\neg \texttt{F}_{[a,b]} \neg \varphi$) in \isa{globally\_future\_dual}.\footnote{Both of these properties are commonly used to define $\texttt{F}$ and $\texttt{G}$ in terms of $\texttt{U}$.}
In Isabelle/HOL, we have:
\begin{isabelle}
  \futureasuntil
\end{isabelle}
\begin{isabelle}
  \globallyfuturedual
\end{isabelle}

We also prove the duality of $\texttt{U}$ and $\texttt{R}$, i.e., that $\varphi \ \texttt{R}_{[a,b]} \psi$ is equivalent to $\neg (\neg \varphi \ \texttt{U}_{[a,b]} \neg \psi)$, in the following lemma:
\begin{isabelle}
  \releaseuntildual
\end{isabelle}
This proof is not immediately resolved by Isabelle's automation; we follow a proof sketch from the literature \cite{DBLP:conf/ifm/ElwingGSTWR23} which splits the biconditional into two implications.
In each direction, the proof considers three cases: \textcircled{1} when the length of the trace is less than $a$, \textcircled{2} when \isasympsi \ is satisfied for all timesteps in the interval $[a,b]$, and \textcircled{3} when some timestep in $[a,b]$ does not satisfy \isasympsi.
In the forward direction of \textcircled{3}, we found an error in the source material \cite{DBLP:conf/ifm/ElwingGSTWR23}, which incorrectly reduces the subgoal of showing that $\forall s \in [a, b]. (\pi_s \models \psi \lor (\exists t \in [a, s-1]. \ \pi_t \models \varphi))$ to showing that $\forall s \in [a, b]. (\exists t \in [a, s-1]. \ \pi_t \models \varphi)$. 
Fortunately, Sledgehammer \cite{DBLP:conf/lpar/PaulsonB10} directly proves the desired subgoal.

These duality lemmas, with DeMorgan's laws for $\land$ and $\lor$, are sufficient to establish that any well-defined MLTL formula can be rewritten using only the operators $\{\neg, \land, \texttt{U}\}$, which is defined in the literature \cite{LVR19} as the Backus Naur Form (BNF) for MLTL.
To make use of BNF, we verify a custom induction rule, \isa{bnf\_induct}:
\begin{isabelle}
  \bnfinduct
\end{isabelle}
To use this induction rule to prove a property \isasymP \ of an MLTL formula \isasymeta, we must prove not only the usual structural induction cases (in \isa{bnf\_induct}, these are captured in \isa{True}, \isa{False}, \isa{Prop}, \isa{Not}, \isa{And}, and \isa{Until}) but also two additional properties.
First, we must show that \isasymP\ is invariant on semantically equivalent formulas: i.e., for semantically equivalent \isasymphi 1 and \isasymphi 2, \isasymP \ holds on \isasymphi 1 if and only if \isasymP \ holds on \isasymphi 2\ (in \isa{bnf\_induct}, this is captured in the assumption \isa{PProp}).
Second, if \isasymeta \ contains any temporal operators, then their corresponding time intervals should be well-defined; for example, if \isasymeta \ is ``\isa{True\isactrlsub m\ U\isactrlsub m[a,b] False\isactrlsub m}'', then \isa{a} must be less than or equal to \isa{b}.
In \isa{bnf\_induct}, this is captured in the assumption \isa{IntervalsWellDef}, where the desired well-definedness property is formalized in \isa{intervals\_welldef}.\footnote{For the curious reader, we present this function in \rref{app:intervals}.}

In other words, given both \isa{IntervalsWellDef} and \isa{PProp}, the induction rule allows us to conclude \isa{\isasymP\ \isasymeta} if we can prove property \isasymP\ on MLTL formulas of the shapes \isa{\isasymeta\ = True\isactrlsub m}, \isa{\isasymeta\ = False\isactrlsub m}, and \isa{\isasymeta\ = Prop\isactrlsub m} (these are the base cases), and prove that \isasymP\ holds on formulas of the shapes \isa{\isasymeta\ = Not\isactrlsub m \isasymphi}, \isa{\isasymeta\ = \isasymphi\ And\isactrlsub m \isasympsi}, and \isa{\isasymeta\ = \isasymphi\ U\isactrlsub m[a,b] \isasympsi}, given the appropriate inductive hypotheses for each case.
When it applies (i.e., when \isa{IntervalsWellDef} and \isa{PProp} hold), \isa{bnf\_induct} is impactful as it considerably reduces the number of cases compared to the default structural induction rule for MLTL (which cases on each operator in the syntax).

We do not limit ourselves to establishing the standard functional completeness for MLTL, but prove a number of additional useful basic properties, like distributivity for $\texttt{U}$ and $\texttt{R}$ over $\land$ and $\lor$; Isabelle's automation makes most of these easy.

\subsection{Useful Functions and Custom Induction Rules}\label{sec:functions}
We augment our MLTL library by formalizing some useful functions on MLTL formulas.
Our first useful function, \isa{convert\_nnf}, makes use of the operator duality properties to convert an MLTL formula to its negation normal form (NNF), where negations are only in front of atomic propositions.
We prove that \isa{convert\_nnf} preserves the semantics of the input formula, and that it is idempotent:
\begin{isabelle}
  \convertnnfpreservessemantics
\end{isabelle}
\begin{isabelle}
  \convertnnfconvertnnf
\end{isabelle}

The proofs of these two lemmas work by induction on the \textit{depth} of the formula; accordingly, we formalize \isa{depth\_mltl}, a depth function for MLTL formulas.
We also define a function, \isa{nnf\_subformulas}, to compute the set of subformulas of an MLTL formula, and prove (by induction on formula depth) that any subformula of a formula in NNF is itself in NNF:
\begin{isabelle}
  \nnfsubformulas
\end{isabelle}

Using these functions and properties, we establish the following custom induction rule, \isa{nnf\_induct}, as another useful tool for proving properties about MLTL formulas.
In contrast to \isa{bnf\_induct}, which reduces the number of inductive cases to consider, \isa{nnf\_induct} allows us to consider \textit{simpler} cases---although \isa{nnf\_induct} still requires the desired property to hold on formulas of the each of the standard shapes, it limits the Not case to negations of atomic propositions (in case \isa{Not\_Prop}).
Happily, \isa{nnf\_induct} imposes no requirements on the desired property \isasymP\ (as opposed to \isa{bnf\_induct}, which requires \isasymP\ to be invariant on semantically equivalent formulas).
However, \isa{nnf\_induct} does require that the input formula \isasymeta\ is in NNF (in assumption \isa{nnf}).
Fortunately, as MLTL formulas can be efficiently converted to NNF \cite{DBLP:conf/ifm/ElwingGSTWR23}, this assumption is easily satisfied.

\begin{isabelle}
  \nnfinduct
\end{isabelle}

Finally, we formalize the \textit{computation length} of an MLTL formula. 
Mathematically, the computation length, $\texttt{complen}$, of an MLTL formula is recursively defined as follows, where $p \in \texttt{AP}$ is a atomic proposition, and $\varphi$ and $\psi$ are MLTL formulas \cite{DBLP:conf/ifm/ElwingGSTWR23}:
\begin{flalign*}
	&\texttt{complen}(p) = \texttt{complen}(\neg p) = 1,&\\
    &\texttt{complen}(\varphi \land \psi) = \texttt{complen}(\varphi \lor \psi) = \max(\texttt{complen}(\varphi), \texttt{complen}(\psi)),\\
    &\texttt{complen}(\texttt{G}_{[a,b]} \varphi) = \texttt{complen}(\texttt{F}_{[a,b]} \varphi) = b + \texttt{complen}(\varphi),&\\
    &\texttt{complen}(\varphi \ \texttt{U}_{[a,b]} \ \psi) = \texttt{complen}(\varphi \ \texttt{R}_{[a,b]} \ \psi) = b + \max(\texttt{complen}(\varphi)-1, \texttt{complen}(\psi))
\end{flalign*}

In Isabelle/HOL, we formalize this in \isa{complen\_mltl}, which maps an MLTL formula of arbitrary type to a natural number.
The formal definition mirrors the mathematical definition, modulo the requisite syntax changes, except that we define the computation length of \isa{True\_mltl} and \isa{False\_mltl} (to be 1); these definitions were missing in the literature \cite{DBLP:conf/ifm/ElwingGSTWR23}.
We present a few representative cases; the full formal definition is in \rref{app:A}.

\begin{isabelle}
\complenmltlpartial
\end{isabelle}

Computation length is a slightly optimized version of an earlier and closely-related\footnote{The definitions of computation length and wpd are mostly the same.
The main difference is that computation length includes a -1 term in the Until and Release operators; this is not present in wpd and it represents a small optimization.
Another difference is that the computation length of atomic propositions is defined to be 1, while the wpd of atomic propositions is 0; this difference is purely cosmetic, as the wpd captures the delay \textit{from the current timestep} (this is standard in runtime verification) while computation length captures the length of the total delay.
} notion of the \textit{worst-case propagation delay (wpd)} of a formula.
Worst-case propagation delay was introduced in the context of runtime verification \cite[Definition 5]{KZJZR20} and was designed to give the maximum number of timesteps an observer needs to \textit{wait from the current timestep} to decide the satisfaction of an MLTL formula so that the decision will not change with further information.
For example, if we try to evaluate the satisfaction of $\texttt{F}_{[0,2]} p$ on a trace of length 2, then it is possible that further information could change the verdict; e.g., $\texttt{F}_{[0,2]} p$ is false on the trace $[\{\}, \{\}]$ but true on the trace $[\{\}, \{\}, \{p\}]$.
However, after $\texttt{complen}(\texttt{F}_{[0,2]} p) = 3$ timesteps, the value of $p$ in subsequent timesteps will not change whether the trace satisfies $\texttt{F}_{[0,2]} p$.
In particular, the R2U2 tool \cite{KZJZR20}, which checks the satisfaction of MLTL formulas on traces, does not return an answer until it is certain that the answer will not change with further information. 
In this context, the wpd of the formula captures the duration of the worst possible delay of an answer from the current time.

Interestingly, computation length also plays a key role in the formal proofs of the formula progression algorithm.
Although the source material which we were following \cite{LR18} does not use or mention computation length, we found that this notion directly fixes some issues in the proofs.
We now turn to our formalization of formula progression; we further discuss computation length and some useful verified properties thereof in \rref{sec:Thm3}.

\section{Formula Progression}
\label{sec:fp}
\textit{Formula progression} is a common technique that steps through time to evaluate formulas; it dates back to the 1990's \cite{DBLP:conf/aaai/BacchusK96} and has been developed for various temporal logics including MTL \cite{DBLP:conf/aaai/LengH19}, $\text{LTL}_f$  \cite{DBLP:conf/seke/NiuXXXHL23}, and MLTL \cite{LR18}.
Formula progression represents a straightforward and generally easy-to-implement algorithm for evaluating whether a particular temporal trace satisfies a given formula, and is therefore utilized in many contexts, including satisfiability solving, validation, runtime verification, and benchmark generation.

\begin{figure*}
\centering
\includegraphics[scale=0.55]{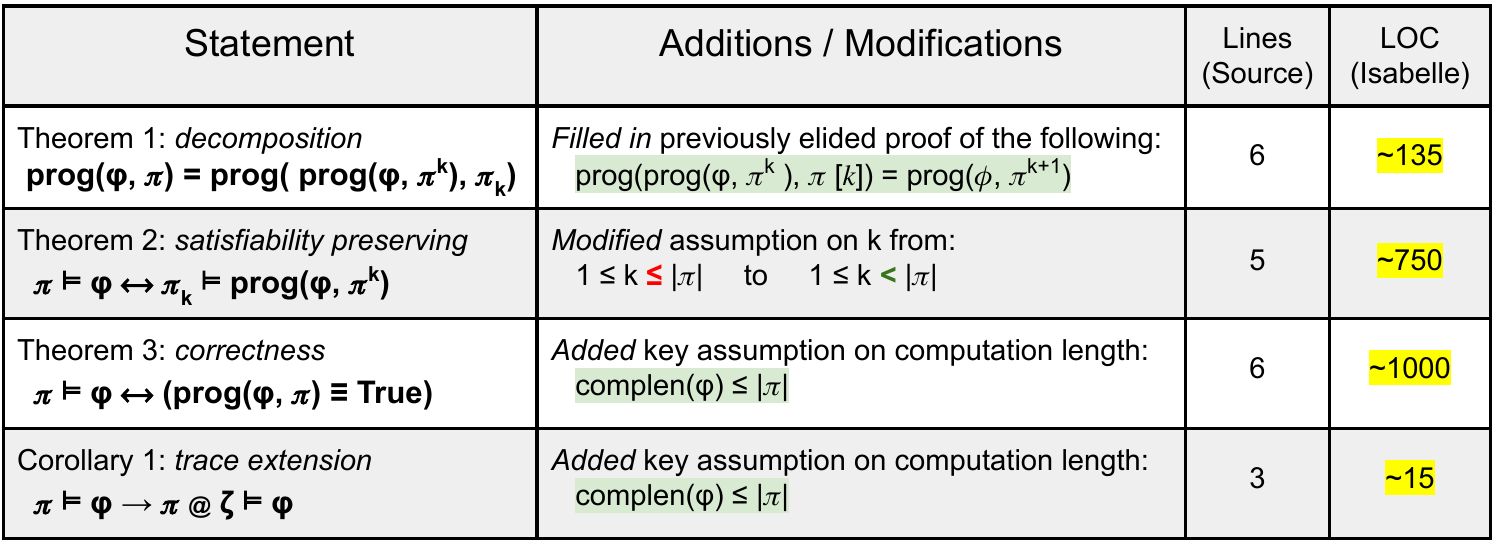}
\caption{We summarize the top-level results for formula progression and our modifications, comparing the number of lines in the existing proof sketches \cite{LR18} with the corresponding LOC in Isabelle/HOL. As our formal proof of Theorem 3 required identifying and formalizing many properties of computation length (not present in the source material), we count these in the LOC.}
\label{fig:FPOverview}
\end{figure*}

Armed with our encoding of MLTL, we formalize the MLTL formula progression algorithm and its associated correctness theorems \cite{LR18}.
While we mostly follow the original paper \cite{LR18} that developed the algorithm, we found and fixed several errors in the top-level correctness results.
In particular, the source material states three key theorems (and a corollary); \rref{fig:FPOverview} overviews our changes to these top-level results.
We now discuss the formula progression algorithm and each main result, with an emphasis on our modifications.

\subsection{Encoding the Formula Progression Algorithm}\label{sec:fpalg}
Formula progression takes an MLTL formula and steps through a trace.
At each timestep, formula progression partially evaluates the input formula, transforming it into a simplified formula which, under some generally unrestrictive conditions, is logically equivalent.
As MLTL is an inherently bounded logic, it is well-suited for formula progression; one can easily predetermine how many timesteps are needed to transform a formula into True or False.

The definitions for MLTL formula progression \cite[Definition 1]{LR18} reflect this compatibility and are straightforward to formalize. 
For a trace $\pi$ of length greater than 1, $\text{prog}(\isasymphi, \isasympi)$ is defined as $\text{prog}(\text{prog}(\isasymphi, \isasympi[0]), \isasymphi_1)$, which we formalize in \isa{formula\_progression}:
\begin{isabelle}
  \prog
\end{isabelle}
This function takes an MLTL formula with atoms of arbitrary type and a trace. 
To evaluate $\text{prog}(\isasymphi, \isasympi)$, it cases on the length of \isasympi.
If \isasympi\ is empty, it returns \isasymphi; if \isasympi\ has length 1, it calls the helper function \isa{formula\_progression\_len1}, which handles formula progression on traces of length 1.
When \isasympi\ has length longer than 1, it evaluates $\text{prog}(\isasymphi, \isasympi[0])$ as \isa{formula\_progression\_len1\ \isasymphi\ (\isasympi !0)}---here, $\isa{\isasympi !0}$ accesses the element at index 0 from \isasympi---and then passes the result to \isa{formula\_progression} on \isa{drop\ 1\ \isasympi} (which is $\isasympi_1$).

The helper function \isa{formula\_progression\_len1} cases on the structure of the input formula, closely following the associated mathematical definitions \cite[Definition 1]{LR18}.
As a representative example, we show one of the most complicated cases, for formulas of the shape $\varphi\ U_{[a, b]}\ \psi$; the full formalized function is in \rref{app:B}. Mathematically,
\begin{align*} \texttt{prog}(\varphi\ U_{[a, b]}\ \psi, \isasympi) = 
\left\{
	\begin{array}{ll}
		\varphi\ U_{[a-1, b-1]}\ \psi  & \mbox{if } 0 < a \leq b \\
		\texttt{prog}(\psi, \isasympi)\ \lor  (\texttt{prog}(\varphi, \isasympi)\ \land
		\varphi\ U_{[0, b-1]}\ \psi)
				& \mbox{if } 0 = a < b \\
        \texttt{prog}(\psi, \isasympi) & \mbox{if } 0 = a = b.
	\end{array}
\right.
\end{align*}
This is encoded in Isabelle/HOL as follows:
\begin{isabelle}
\proglenoneuntil
\end{isabelle}

Because our functions are executable, we can use the \isa{value} command (which invokes Isabelle's code generator \cite{codegen}) to evaluate formula progression on input formulas.
For example, running formula progression on the formula $\texttt{G}_{[0,2]} p$ and the trace $[\{p\}, \{p\}, \{\}]$ yields $\neg ((\neg \text{True}) \lor ((\neg \text{True}) \lor (\neg \text{False})))$, which is logically equivalent to $\text{False}$.
Note that the algorithm \textit{is not guaranteed} to produce an output in the simplest form.

\subsection{Verifying Theorem 1: Decomposition}

The first key theorem for formula progression informally states that performing formula progression along a trace can be split into first performing formula progression on the prefix, and then on the suffix, of the trace.
Mathematically, it states that  $\texttt{prog}(\varphi, \pi) = \texttt{prog}(\texttt{prog}(\varphi, \pi^k), \pi_k)$ for $1 \leq k \leq \texttt{length } \pi$.
In Isabelle/HOL, we prove the following theorem, where ``\isa{take\ k\ \isasympi}'' yields $\pi^k$ and ``\isa{drop\ k \isasympi}'' yields $\pi_k$ (cross-reference \rref{fig:PrefixSuffix}).
\begin{isabelle}
\theoremone
\end{isabelle}
Our formal proof of this theorem follows the proof sketch in the original source paper \cite{LR18} relatively closely.
We found it effective to split out the following identity used in the proof into a separate helper lemma: 
$\text{prog}(\text{prog}(\varphi, \pi^k), \pi[k]) =\text{prog}(\varphi, \pi^{k+1})$.
Although the source paper stated that this property holds by definition, our proof required induction on $k$.
This highlights the role of formalization, which necessitates making all handwaving rigorous.
Even so, Theorem 1 was the easiest of the three top-level formula progression theorems we prove; it only required $\sim$135 LOC (compared to 6 lines of proof in the source material).

\subsection{Verifying Theorem 2: Satisfiability Preservation}
 The second key theorem states that (in most cases) a trace $\pi$ satisfies a formula $\varphi$ iff the suffix of $\pi$ satisfies the formula progression of $\varphi$ on the prefix of $\pi$.
Mathematically, this is $\pi \models \varphi$ iff $\pi_k \models \text{prog}(\varphi, \pi^k)$; the source material \cite{LR18} stated this theorem for $1 \leq k \leq \text{length }\pi$, but our formal statement is for $1 \leq k < \text{length }\pi$.
In Isabelle/HOL, we have:

\begin{isabelle}
\theoremtwo
\end{isabelle}

In addition to altering the bound on \isa{k}, we add the requisite well-definedness assumption, \isa{intervals\_welldef\ \isasymphi}, which asserts that all temporal operators have well-defined timebounds (cross-reference \rref{sec:syntax}).
The well-definedness assumption is implicitly present in the source material.
The altered bound, however, is \textit{mathematically necessary} for the correctness of the result and was overlooked in the source material.
We uncovered this while formalizing the base case of this theorem.
The source material inducts on $k$ and states that the base case ($k = 1$) is by induction without providing details.
The issue we run into here is related to \textit{end-of-trace behavior}.
When both $k = 1$ and the length of $\pi$ is 1, the statement of Theorem 2 becomes $\pi \models \varphi \longleftrightarrow [] \models \texttt{prog}(\varphi, \pi)$, and this does not hold in general.

Specifically, consider a formula of the shape $\texttt{G}_{[0, b]} p$ for $b > 0$ with trace $\pi = [\{p\}]$ of length $1$.
If we allow $k=1$, then we would want to establish
$\pi \models \texttt{G}_{[0, b]} p \longleftrightarrow \pi_1 \models \texttt{prog}(\texttt{G}_{[0, b]} \varphi, \pi^1)$.
We have that $\pi_1 = []$ and $\pi^1 = \pi$, so this becomes $\pi \models \texttt{G}_{[0, b]} p \longleftrightarrow [] \models \texttt{prog}(\texttt{G}_{[0, b]} p, \pi)$.
Then using the semantics of MLTL on the left-hand side, $\pi \models \texttt{G}_{[0, b]} p$ reduces to $\forall i \in [0, b].\ \pi_i \models p$, which further reduces to $\forall i \in [0, b].(\pi_i \neq [] \land p \in \pi_i[0])$. 
However, this is \textit{false} because $\pi_1 = []$. 

Applying the definition of formula progression to the right-hand side, we have that:
\begin{align*}
\texttt{prog}(\texttt{G}_{[0, b]} p, \pi)
&= \neg \texttt{prog}(\texttt{F}_{[0, b]} \neg p, \pi)
= \neg (\neg \texttt{prog}(p , \pi) \lor \texttt{F}_{[0, b-1]} \neg p)\\
&= \neg (\neg true \lor \texttt{F}_{[0, b-1]} \neg p) = \neg \texttt{F}_{[0, b-1]} \neg p = \texttt{G}_{[0, b-1]} p
\end{align*}
Thus, the right-hand side becomes $[] \models \texttt{G}_{[0, b-1]} p$; using the semantics of MLTL this is $|[]| \leq 0 \lor (\forall i \in [0, b-1].\ []_i \models p)$, which is \textit{true} because the left disjunct is true.

Modifying the statement of Theorem 2 to insist $k < \text{length}\ \pi$ instead of $k \leq \text{length}\ \pi$ removes this issue (because, for $k = 1$, the traces $\pi$ must have length at least 2, so we do not encounter the above issues with the empty trace).
Further, and crucially, we are still able to use the modified version of Theorem 2 in the proofs of Theorem 3 and the top-level corollary; that is, our modification fixes Theorem 2 without overly weakening the result.

\subsection{Verifying Theorem 3 and Corollary 1: Correctness}\label{sec:Thm3}
The third key theorem for formula progression informally states that, for a sufficiently long input trace \isasympi, and for a well-defined input formula \isasymphi, \isasympi\ models formula \isasymphi\ if and only if the formula progression of \isasymphi\ over \isasympi\ is logically equivalent to True.
More precisely, we require (in the first assumption) that all intervals in \isasymphi\ are well-defined, and we require (in the second assumption) that the length of \isasympi\ is at least the \textit{computation length} of the formula (cross-reference \rref{sec:functions}, \cite{DBLP:conf/ifm/ElwingGSTWR23}).
In Isabelle, we have the following theorem, where \isa{semantic\_equiv} encodes logical equivalence of two MLTL formulas (cross-reference \rref{sec:functions}):

\begin{isabelle}
\theoremthree
\end{isabelle}

Here, the assumption that the length of \isasympi\ is greater than or equal to the computation length of the formula \isasymphi\ is \textit{mathematically crucial} for the correctness of the theorem.
Notably, this assumption was missing in the corresponding theorem in the original source material \cite[Theorem 3]{LR18}; identifying it is a contribution of our formalization.
Note that this assumption does not affect the practicality of formula progression; in practice, traces are typically long streams of data, so it is not overly restrictive to assume that a trace is sufficiently long.

We identified this assumption when attempting to prove the base case of Theorem 3.
The source material states that this is by structural induction and omits all mathematical details.
When initially attempting to formalize the base case \textit{without this critical assumption}, we quickly ran into issues in the \isa{Or\_mltl} case.
To see why, let us consider the proof on a mathematical level.
We are trying to prove that $(\pi \models \varphi \lor \psi) \longleftrightarrow (\text{prog} (\varphi \lor \psi, \pi) = \text{True})$ for a trace $\pi$ of length 1, given that $(\pi \models \varphi) \longleftrightarrow (\text{prog}(\varphi, \pi)= \text{True})$ and $(\pi \models \psi) \longleftrightarrow (\text{prog} (\psi, \pi) = \text{True})$.
Because $\text{prog} (\varphi \lor \psi, \pi)$  is defined as $\text{prog} (\varphi, \pi) \lor \text{prog} (\psi, \pi)$ when $\pi$ has length 1, and using that $\pi \models \varphi \lor \psi  \longleftrightarrow (\pi \models \varphi \lor \pi \models \psi)$, our goal becomes $(\pi \models \varphi \lor \pi \models \psi) \longleftrightarrow ((\text{prog} (\varphi, \pi) \lor \text{prog} (\varphi, \pi)) = \text{True})$.  Then, using our inductive hypotheses, we incur the goal: $(\text{prog} (\psi, \pi) = \text{True} \lor \text{prog} (\varphi, \pi) = \text{True}) \longleftrightarrow ((\text{prog} (\varphi, \pi) \lor \text{prog} (\psi, \pi)) = \text{True})$.
This does not hold; $\text{prog} (\varphi, \pi) \lor \text{prog} (\psi, \pi)$ can be logically equivalent to True without either $\text{prog} (\varphi, \pi)$ or $\text{prog} (\psi, \pi)$ being logically equivalent to True; all that is required is that $\text{prog} (\varphi, \pi)$ must hold in any states where $\text{prog} (\psi, \pi)$ is false and vice-versa.

By adding the assumption on the computation length, the base case of the formula progression theorem inherits the (structurally strong) assumption that $1$ is greater than or equal to the computation length of the input formula $\varphi$.  This assumption is strong enough to establish that the formula progression of $\varphi$ on any trace of length 1 is either globally true or globally false.
This solves the above issue in the \isa{Or\_mltl} case; since $\text{prog} (\varphi, \pi)$ and $\text{prog} (\psi, \pi)$ are now globally either True or False, we can indeed conclude $(\text{prog} (\psi, \pi) =  \text{True} \lor \text{prog} (\varphi, \pi) = \text{True}) \longleftrightarrow ((\text{prog} (\varphi, \pi) \lor \text{prog} (\psi, \pi)) =  \text{True})$.
We establish this useful property of the computation length in the following key lemma:
\begin{isabelle}
\complenboundedbyone
\end{isabelle}

This lemma is sufficient to prove our strengthened base case; however, we also need to ensure that the strengthened base case is \textit{usable} in the proof of Theorem 3.
This requires several additional properties of the computation length.
In particular, we establish the following two crucial properties:
\begin{isabelle}
\complenoneimpliesone
\end{isabelle}

\begin{isabelle}
\formulaprogressiondecreasescomplen
\end{isabelle}

This first lemma, \isa{complen\_one\_implies\_one}, proves that (for formulas \isasymphi\ with well-defined intervals), if the computation length of \isasymphi\ is 1, then the computation length of the formula progression of \isasymphi\ is also 1; i.e., formula progression does not increase the computation length for formulas that already have the minimum computation length.
The second key lemma, \isa{formula\_progression\_decreases\_complen}, establishes that performing formula progression on a formula \isasymphi\ over a trace \isasympi\ usually decreases the computation length of \isasymphi\ proportionally to the length of \isasympi.
More precisely, either the computation length of \isasymphi\ equals 1 (i.e., is already minimal), or the computation length of the formula progression of \isasymphi\ on trace \isasympi\ equals 1 (i.e., is minimal), or the computation length of the formula progression of \isasymphi\ on \isasympi\ is less than or equal to the computation length of \isasymphi\ minus the length of the trace \isasympi\ (i.e., the computation length of the progression is decreased by the length of \isasympi). 

We use \isa{complen\_one\_implies\_one} to prove \isa{formula} \isa{\_progression\_decreases\_complen}, and we use both lemmas in the proof of Theorem 3 to establish that 
$\text{complen}$ $\text{prog}(\varphi, (\pi^{\text{length } \pi - 1}))\\ \leq \text{length} (\pi_{\text{\ length } \pi - 1})$, which is equivalent to $\text{complen}$ $\text{prog}(\varphi, (\pi^{\text{length } \pi - 1})) \leq 1$.
This then allows us to use the base case of Theorem 3 with the formula $\text{prog}(\varphi, (\pi^{\text{length } \pi - 1}))$ and the trace $\pi_{\text{\ length } \pi - 1}$ to prove that $\text{prog} (\text{prog}(\varphi, (\pi^{\text{length } \pi - 1}), \pi_{\text{\ length } \pi - 1})) = \text{True}$ is equal to $\pi_{\ \text{length } \pi - 1} \models \text{prog}(\varphi, (\pi^{\text{length } \pi - 1}))$.
Then, following the source material \cite{LR18}, we can conclude that \textcircled{1} $\text{prog} (\text{prog}$ $(\varphi, (\pi^{\text{length } \pi - 1}), \pi_{\text{\ length } \pi - 1})) = \text{prog}(\varphi, \pi)$ using Theorem 1 and \textcircled{2} $\pi_{\text{\ length } \pi - 1} \models$ $\text{prog}(\varphi, (\pi^{\text{length } \pi - 1}))$ iff $\pi \models \varphi$ using Theorem 2.
Putting these pieces together closes the proof.

In total, \isa{formula\_progression\_correctness} took approximately 1000 lines of code to formalize (this includes about 800 lines for establishing the necessary properties of the computation length), as compared to 6 lines in the original source material.

Putting the three key theorems together, we achieve the following top-level corollary, which informally states that, for a well-defined MLTL formula \isasymphi\ and a sufficiently long trace \isasympi, if $\isasympi \models \isasymphi$, then any trace whose prefix is $\isasympi$ also models \isasymphi.
Formally, we prove the following result in Isabelle/HOL, where \isa{@}\ is Isabelle/HOL's syntax for appending lists.

\begin{isabelle}
\topcorollary
\end{isabelle}

The source material \cite[Corollary 1]{LR18} had stated this corollary without the key assumption on the length of the trace, but that is incorrect.
To see why, consider the following example.
We know that $[\{2\}]  \models G_{[1,3]} \text{False}$, as the length of the trace $[\{2\}]$ is $\leq 1$ (cross-reference the semantics of MLTL in \rref{sec:encoding}).
From this, without the assumption that the length of $[\{2\}]$ must be greater than or equal to the computation length of $ G_{[1,3]} \text{False}$, which is 4, we could obtain, for example, $[\{2\}, \{3\}, \{4\}, \{2, 3\}]  \models G_{[1,3]} \text{False}$, which contradicts MLTL semantics.
Fortunately, after adding in the necessary assumption, the (formal) proof of the corollary is quite straightforward and follows the proof in the source material \cite{LR18} without issue.

After formalizing this corollary, we realized that we could go one step further and formalize an alternate version of Theorem 3:
If a sufficiently long trace $\pi$ does not satisfy a formula $\varphi$, then the formula progression of $\varphi$ on $\pi$ is logically equivalent to False.
Although the proof of this exactly mirrors the proof of Theorem 3, it is not immediately implied by Theorem 3. Assuming that $\pi \not \models \varphi$, Theorem 3 only yields that the $\text{prog}(\varphi, \pi)$ is not semantically equivalent to True; this does not necessitate that the resulting formula is equivalent to False.
Combining the original and alternate versions of Theorem 3, we obtain the following important property of formula progression:
\begin{isabelle}
  \formulaprogressiontrueorfalse
\end{isabelle}
This allows us to conclude that for a sufficiently long trace $\pi$, the formula progression of $\varphi$ on $\pi$ is either logically equivalent to True or False.
In fact, then
\isa{complen\_bounded\_by\_1} lemma
is a special case of this more general property.

We also use our alternate version of Theorem 3 to prove the inverse of the top-level corollary, strengthening it to an if-and-only-if statement:
\begin{isabelle}
\complenproperty
\end{isabelle}
This says that for a sufficiently long trace $\pi$, extending $\pi$ with additional states (in $\zeta$) does not affect the satisfaction of formula $\varphi$. 
This formalizes the intuition for computation length---timesteps beyond the computation length of a formula do not affect the satisfaction of that formula.
Retrospectively, one bonus of our formalization is that it nicely integrates various related concepts (like formula progression and computation length) for MLTL in a single centralized and easily extensible library.

\section{Challenges and Insights}
\label{sec:challenges}
We view our formalization as emblematic of the potential benefits of formalizing source material that heavily relies on intuition.
The mathematical intuition for formula progression for MLTL is clear because of its boundedness.
Accordingly, the original source material heavily leaned into this intuition, and many important details were elided (possibly in light of space constraints), which introduced bugs.
In particular, the base cases of Theorem 2 and Theorem 3 claimed to hold by structural induction; mathematically, this seems quite intuitive.
However, formalization serves as a useful forcing function, where all intuition must be made precise, and we found that the base cases were incorrect and needed modifications; these then propagated into the top-level theorems and corollary.

One of the main creative challenges that we faced was identifying the necessary assumptions to add to the base cases.
The added assumptions needed to walk a fine line: They must be strong enough to prove the base cases but weak enough that they can still be used to prove the desired top-level result.
Additionally, they should maintain the practicality of formula progression.
We found that it was useful to keep the high-level context in mind throughout the course of our formalization; we frequently switched back and forth between working on a base case and using that base case to prove the overall desired result.

Similarly, it was only by analyzing the top-level corollary that we identified the main necessary assumption for Theorem 3 (that $\isa{length}\ \isasympi\ \isasymge$ $\isa{complen\_mltl}\ \isasymphi$).
We found it intuitively clear that this assumption is sufficient to establish the corollary and that it does not affect the practicality of formula progression (as traces in real-world systems tend to be long).
Interestingly, though, we were initially unsure whether it was strong enough to fix the proof of Theorem 3.
Establishing this required identifying and proving various key properties of the computation length, particularly those discussed in \rref{sec:fp}, which show that formula progression tends to decrease computation length.
Initially, we stated these in a series of (unproven) properties of the computation length that we \textit{wanted} to be able to use in our proofs; we were pleased when we later found that all of these properties held.

Another key challenge relates to end-of-trace behavior.
We noticed this especially in the proof of Theorem 2; the behavior of the empty trace caused issues in the theorem as it was originally stated.
In practice, MLTL traces are usually long, and we suspect that this makes it easy to overlook edge cases involving the empty trace.
Also, MLTL end-of-trace behavior is somewhat unintuitive; for example, $[] \models G_{[0, 1]} \text{False}$ according to MLTL semantics, because Globally is satisfied if the length of the trace is less than or equal to the lower bound on its associated interval.
Fortunately, formalization is excellent at making unintuitive edge cases precise and ensuring that they are correctly reasoned about.

Finally, we comment on our choice of the theorem prover Isabelle/HOL, which we benefited from throughout the formalization.
We particularly appreciate the automation afforded by Sledgehammer \cite{DBLP:conf/lpar/PaulsonB10} and the built-in support for unrolling definitions and functions; the latter was extremely helpful in a number of our proofs that rely on structural induction and reduce into detailed casework.\footnote{Most notably, in the base cases of Theorems 2 and 3 (where the trace has length one), formula progression for the temporal operators splits into different cases, making the proofs involved.}
We also appreciate Isabelle's well-developed libraries; though our work does not depend on any entries from the Archive of Formal Proofs, we loosely modeled our formalization of MLTL syntax and semantics on an existing formalization of LTL \cite{LTL-AFP}.

\section{Conclusion and Future Work}
\label{sec:conclusion}
In this work, we formalize MLTL syntax and semantics and develop a library of useful properties, including dualities, an NNF transformation, and custom induction rules.
We also verify MLTL formula progression \cite{LR18}, which necessitated fixing errors in the correctness proofs.
In the process, 
we develop the foundational notion of \textit{computation length} \cite{DBLP:conf/ifm/ElwingGSTWR23}.
The literature on computation length and its (earlier) counterpart, worst-case propagation delay \cite{KZJZR20}, is somewhat piecemeal.
We contribute a centralized body of verified lemmas about computation length.
Our work can both facilitate future formalizations involving MLTL and serve as a reference point for theoretical work.

A natural future direction is to continue to develop our library.
For example, verifying more properties of the computation length and related results from the literature involving worst-case propagation delay, e.g. \cite[Lemma 4]{ZADJR23} is an exciting topic. 
More broadly, we envision a suite of formally verified algorithms for MLTL that can be used to validate the existing (unverified) tools for analyzing MLTL formulas, or which could be developed into tools in their own right.
Indeed, we have already leveraged our formalization of MLTL to verify the MLTL to regular expressions algorithm \cite{WKR25} and validate the WEST tool \cite{WEST2024tool}, and (separately) to formalize some new theoretical results for MLTL \cite{NFM25}.
As our formula progression algorithm is executable, it could be used to validate existing (unverified) tool support, i.e., the FPROGG tool, which uses formula progression for benchmark generation \cite{RR25}. 
Finally, we believe that it would be of considerable value to bridge our formalization with existing formalizations of related logics; we are particularly interested in building a bridge from MLTL to VeriMon \cite{DBLP:conf/ictac/BasinDHHMKKMRST22} and Vydra \cite{DBLP:conf/atva/RaszykBT20}.

\section{Acknowledgements}
Thank you to Dmitriy Traytel for useful discussions on MLTL and related work.
Thanks also to Alec Rosentrater for useful comments and feedback on the paper and to Alexis Aurandt and Brian Kempa for useful discussions about worst-case propagation delay.
Thanks to NSF CAREER Award CNS-1552934 and NSF CCRI-2016592 for supporting this work. Zili Wang was supported by an NSF Graduate Research Fellowship under NSF:GRFP 2024364991 and by the Iowa State Bridging the Divide Grant.

\bibliography{MLTL}

\begin{thebibliography}{10}

\bibitem{RVBC2018}
{2018 Runtime Verification Benchmark Competition}.
\newblock \url{https://www.rv-competition.org/2018-2/}.

\bibitem{AH90}
R.~Alur and T.~A. Henzinger.
\newblock {R}eal-time {L}ogics: {C}omplexity and {E}xpressiveness.
\newblock In {\em LICS}, pages 390--401. IEEE, 1990.

\bibitem{DBLP:conf/podc/AlurFH91}
Rajeev Alur, Tom{\'{a}}s Feder, and Thomas~A. Henzinger.
\newblock The benefits of relaxing punctuality.
\newblock In Luigi Logrippo, editor, {\em PODC}, pages 139--152. {ACM}, 1991.
\newblock \href {https://doi.org/10.1145/112600.112613}
  {\path{doi:10.1145/112600.112613}}.

\bibitem{LTL3_Semantics-AFP}
Rayhana Amjad, Rob van Glabbeek, and Liam O'Connor.
\newblock Definitive set semantics for {LTL3}.
\newblock {\em Archive of Formal Proofs}, August 2024.
\newblock \url{https://isa-afp.org/entries/LTL3_Semantics.html}, Formal proof
  development.

\bibitem{Mav22}
{Anastasia Mavridou}.
\newblock Capturing and analyzing requirements with {FRET}.
\newblock Presentation, nfm, \url{https://github.com/NASA-SW-VnV/fret}, NASA,
  Pasadena, California, USA, May 2022.

\bibitem{AJR22}
Alexis Aurandt, Phillip Jones, and Kristin~Yvonne Rozier.
\newblock Runtime verification triggers real-time, autonomous fault recovery on
  the {CySat-I}.
\newblock In {\em NFM}, volume 13260 of {\em LNCS}, Caltech, California, USA,
  May 2022. Springer, Cham.
\newblock \href {https://doi.org/10.1007/978-3-031-06773-0\_45}
  {\path{doi:10.1007/978-3-031-06773-0\_45}}.

\bibitem{DBLP:conf/aaai/BacchusK96}
Fahiem Bacchus and Froduald Kabanza.
\newblock Planning for temporally extended goals.
\newblock In William~J. Clancey and Daniel~S. Weld, editors, {\em AAAI}, pages
  1215--1222. {AAAI} Press / The {MIT} Press, 1996.
\newblock URL: \url{http://www.aaai.org/Library/AAAI/1996/aaai96-180.php}.

\bibitem{DBLP:conf/ictac/BasinDHHMKKMRST22}
David~A. Basin, Thibault Dardinier, Nico Hauser, Lukas Heimes, Jonathan
  Juli{\'{a}}n~Huerta y~Munive, Nicolas Kaletsch, Srdan Krstic, Emanuele
  Marsicano, Martin Raszyk, Joshua Schneider, Dawit~Legesse Tirore, Dmitriy
  Traytel, and Sheila Zingg.
\newblock Veri{M}on: {A} formally verified monitoring tool.
\newblock In Helmut Seidl, Zhiming Liu, and Corina~S. Pasareanu, editors, {\em
  ICTAC}, volume 13572 of {\em LNCS}, pages 1--6. Springer, 2022.
\newblock \href {https://doi.org/10.1007/978-3-031-17715-6\_1}
  {\path{doi:10.1007/978-3-031-17715-6\_1}}.

\bibitem{CHHJR20}
Matthew Cauwels, Abigail Hammer, Benjamin Hertz, Phillip Jones, and
  Kristin~Yvonne Rozier.
\newblock Integrating runtime verification into an automated {UAS} traffic
  management system.
\newblock In {\em Proceedings of {DETECT: international workshop on moDeling,
  vErification and Testing of dEpendable CriTical systems}}, CCIS, L'Aquila,
  Italy, September 2020. Springer.
\newblock \href {https://doi.org/10.1007/978-3-030-59155-7\_26}
  {\path{doi:10.1007/978-3-030-59155-7\_26}}.

\bibitem{MTL_monitor_coq}
Agnishom Chattopadhyay and Konstantinos Mamouras.
\newblock A verified online monitor for metric temporal logic with quantitative
  semantics.
\newblock In Jyotirmoy Deshmukh and Dejan Ni{\v{c}}kovi{\'{c}}, editors, {\em
  RV}, pages 383--403, Cham, 2020. Springer International Publishing.

\bibitem{DBLP:conf/cpp/ConradTGPD22}
Esther Conrad, Laura Titolo, Dimitra Giannakopoulou, Thomas Pressburger, and
  Aaron Dutle.
\newblock A compositional proof framework for {FRET}ish requirements.
\newblock In Andrei Popescu and Steve Zdancewic, editors, {\em CPP}, pages
  68--81. {ACM}, 2022.
\newblock \href {https://doi.org/10.1145/3497775.3503685}
  {\path{doi:10.1145/3497775.3503685}}.

\bibitem{DBLP:journals/logcom/Coupet-Grimal03}
Solange Coupet{-}Grimal.
\newblock An axiomatization of linear temporal logic in the calculus of
  inductive constructions.
\newblock {\em J. Log. Comput.}, 13(6):801--813, 2003.
\newblock URL: \url{https://doi.org/10.1093/logcom/13.6.801}, \href
  {https://doi.org/10.1093/LOGCOM/13.6.801}
  {\path{doi:10.1093/LOGCOM/13.6.801}}.

\bibitem{DBR21}
James~B. Dabney, Julia~M. Badger, and Pavan Rajagopal.
\newblock Adding a verification view for an autonomous real-time system
  architecture.
\newblock In {\em Proceedings of SciTech Forum}, 2021-0566, page Online.
  {AIAA}, January 2021.
\newblock \href {https://doi.org/10.2514/6.2021-0566}
  {\path{doi:10.2514/6.2021-0566}}.

\bibitem{Dab21}
James~Bruster Dabney.
\newblock Using assume-guarantee contracts in autonomous spacecraft.
\newblock Flight Software Workshop (FSW) Online:
  \url{https://www.youtube.com/watch?v=zrtyiyNf674}, February 2021.

\bibitem{DRB22}
James~Bruster Dabney, Pavan Rajagopal, and Julia~M. Badger.
\newblock Using assume-guarantee contracts for developmental verification of
  autonomous spacecraft.
\newblock Flight Software Workshop (FSW) Online:
  \url{https://www.youtube.com/watch?v=HFnn6TzblPg}, February 2022.

\bibitem{DBLP:conf/aaai/LengH19}
Daniel de~Leng and Fredrik Heintz.
\newblock Approximate stream reasoning with metric temporal logic under
  uncertainty.
\newblock In {\em AAAI}, pages 2760--2767. {AAAI} Press, 2019.
\newblock URL: \url{https://doi.org/10.1609/aaai.v33i01.33012760}, \href
  {https://doi.org/10.1609/AAAI.V33I01.33012760}
  {\path{doi:10.1609/AAAI.V33I01.33012760}}.

\bibitem{DBLP:conf/ifm/ElwingGSTWR23}
Jenna Elwing, Laura Gamboa{-}Guzman, Jeremy Sorkin, Chiara Travesset, Zili
  Wang, and Kristin~Yvonne Rozier.
\newblock Mission-time {LTL} {(MLTL)} formula validation via regular
  expressions.
\newblock In Paula Herber and Anton Wijs, editors, {\em iFM}, volume 14300 of
  {\em LNCS}, pages 279--301. Springer, 2023.
\newblock \href {https://doi.org/10.1007/978-3-031-47705-8\_15}
  {\path{doi:10.1007/978-3-031-47705-8\_15}}.

\bibitem{CAVA_LTL_Modelchecker-AFP}
Javier Esparza, Peter Lammich, René Neumann, Tobias Nipkow, Alexander Schimpf,
  and Jan-Georg Smaus.
\newblock A fully verified executable {LTL} model checker.
\newblock {\em Archive of Formal Proofs}, May 2014.
\newblock \url{https://isa-afp.org/entries/CAVA_LTL_Modelchecker.html}, Formal
  proof development.

\bibitem{GRS14}
Johannes Geist, Kristin~Yvonne Rozier, and Johann Schumann.
\newblock {Runtime Observer Pairs and Bayesian Network Reasoners On-board
  FPGAs: Flight-Certifiable System Health Management for Embedded Systems}.
\newblock In {\em RV}, volume 8734, pages 215--230. Springer-Verlag, September
  2014.

\bibitem{DBLP:conf/ijcai/GiacomoV13}
Giuseppe~De Giacomo and Moshe~Y. Vardi.
\newblock Linear temporal logic and linear dynamic logic on finite traces.
\newblock In Francesca Rossi, editor, {\em IJCAI}, pages 854--860.
  {IJCAI/AAAI}, 2013.
\newblock URL:
  \url{http://www.aaai.org/ocs/index.php/IJCAI/IJCAI13/paper/view/6997}.

\bibitem{GMRPSS20}
Dimitra Giannakopoulou, Anastasia Mavridou, Julian Rhein, Thomas Pressburger,
  Johann Schumann, and Nija Shi.
\newblock Formal requirements elicitation with {FRET}.
\newblock In {\em International Working Conference on Requirements Engineering:
  Foundation for Software Quality (REFSQ-2020)}, number ARC-E-DAA-TN77785,
  2020.

\bibitem{codegen}
Florian Haftmann.
\newblock Code generation from {I}sabelle/{HOL} theories.
\newblock \url{https://isabelle.in.tum.de/doc/codegen.pdf}, 2024.
\newblock Tutorial, with contributions from Lukas Bulwahn and Tobias Nipkow.

\bibitem{HJRW23}
Gokul Hariharan, Phillip~H. Jones, Kristin~Yvonne Rozier, and Tichakorn
  Wongpiromsarn.
\newblock Maximum satisfiability of mission-time linear temporal logic.
\newblock In Laure Petrucci and Jeremy Sproston, editors, {\em FORMATS}, volume
  14138 of {\em LNCS}, pages 86--104. Springer, 2023.
\newblock \href {https://doi.org/10.1007/978-3-031-42626-1\_6}
  {\path{doi:10.1007/978-3-031-42626-1\_6}}.

\bibitem{HLR21}
Benjamin Hertz, Zachary Luppen, and Kristin~Yvonne Rozier.
\newblock Integrating runtime verification into a sounding rocket control
  system.
\newblock In {\em NFM}, May 2021.
\newblock Available online at \url{http://temporallogic.org/research/NFM21/}.

\bibitem{JJKRZ23}
Chris Johannsen, Phillip Jones, Brian Kempa, Kristin~Yvonne Rozier, and Pei
  Zhang.
\newblock {R2U2} version 3.0: Re-imagining a toolchain for specification,
  resource estimation, and optimized observer generation for runtime
  verification in hardware and software.
\newblock In Constantin Enea and Akash Lal, editors, {\em CAV}, pages 483--497,
  Cham, 2023. Springer Nature Switzerland.

\bibitem{JABDEGJKLMRVR21}
Christopher Johannsen, Marcella Anderson, William Burken, Ellie Diersen, John
  Edgren, Colton Glick, Stephanie Jou, Adhyaksh Kumar, John Levandowski, Evelyn
  Moyer, Taylor Roquet, Alexander VandeLoo, and Kristin~Yvonne Rozier.
\newblock {\em OpenUAS Version 1.0}.
\newblock IEEE, Athens, Greece (Virtual), June 2021.

\bibitem{DBLP:journals/apal/Kamide12}
Norihiro Kamide.
\newblock Bounded linear-time temporal logic: {A} proof-theoretic
  investigation.
\newblock {\em Ann. Pure Appl. Log.}, 163(4):439--466, 2012.
\newblock URL: \url{https://doi.org/10.1016/j.apal.2011.12.002}, \href
  {https://doi.org/10.1016/J.APAL.2011.12.002}
  {\path{doi:10.1016/J.APAL.2011.12.002}}.

\bibitem{KZJZR20}
Brian Kempa, Pei Zhang, Phillip~H. Jones, Joseph Zambreno, and Kristin~Yvonne
  Rozier.
\newblock {Embedding Online Runtime Verification for Fault Disambiguation on
  Robonaut2}.
\newblock In {\em FORMATS}, LNCS, pages 196--214, Vienna, Austria, September
  2020. Springer.
\newblock URL: \url{http://research.temporallogic.org/papers/KZJZR20.pdf}.

\bibitem{nuXmv-v1.1.0}
Fondazione~Bruno Kessler.
\newblock {nuXmv 1.1.0 (2016-05-10) Release Notes}.
\newblock \url{https://es-static.fbk.eu/tools/nuxmv/downloads/NEWS.txt}, 2016.

\bibitem{Mission_Time_LTL-AFP}
Katherine Kosaian, Zili Wang, and Elizabeth Sloan.
\newblock Mission-time linear temporal logic.
\newblock {\em Archive of Formal Proofs}, January 2025.
\newblock \url{https://isa-afp.org/entries/Mission_Time_LTL.html}, Formal proof
  development.

\bibitem{LR18}
Jianwen Li and Kristin~Y. Rozier.
\newblock {MLTL} benchmark generation via formula progression.
\newblock In Christian Colombo and Martin Leucker, editors, {\em RV}, volume
  11237 of {\em LNCS}, pages 426--433. Springer, 2018.
\newblock \href {https://doi.org/10.1007/978-3-030-03769-7\_25}
  {\path{doi:10.1007/978-3-030-03769-7\_25}}.

\bibitem{LVR19}
Jianwen Li, Moshe~Y. Vardi, and Kristin~Y. Rozier.
\newblock Satisfiability checking for mission-time ltl.
\newblock In {\em CAV}, {LNCS}, New York, NY, USA, July 2019. Springer.

\bibitem{LB15}
Michael Lowry and Anupa Bajwa.
\newblock {Autonomy Operating System (AOS) for UAVs}.
\newblock Proposal Presentation, NASA Ames Research Center, Moffett Field,
  California, June 2015.

\bibitem{LBQGRR17}
Michael Lowry, Anupa Bajwa, Patrick Quach, Gabor Karsai, Kristin~Yvonne Rozier,
  and Sanjai Rayadurgam.
\newblock {Autonomy Operating System for UAVs}.
\newblock Online:
  \url{https://nari.arc.nasa.gov/sites/default/files/attachments/15\%29\%20Mike\%20Lowry\%20SAEApril19-2017.Final_.pdf},
  April 2017.

\bibitem{LJBHCLR22}
Zachary Luppen, Michael Jacks, Nathan Baughman, Benjamin Hertz, James Cutler,
  Dae~Young Lee, and Kristin~Yvonne Rozier.
\newblock Elucidation and analysis of specification patterns in aerospace
  system telemetry.
\newblock In {\em NFM}, volume 13260 of {\em LNCS}, Caltech, California, USA,
  May 2022. Springer, Cham.
\newblock \href {https://doi.org/10.1007/978-3-031-06773-0\_28}
  {\path{doi:10.1007/978-3-031-06773-0\_28}}.

\bibitem{LLR21}
Zachary~A. Luppen, Dae~Young Lee, and Kristin~Yvonne Rozier.
\newblock A case study in formal specification and runtime verification of a
  {CubeSat} communications system.
\newblock In {\em {SciTech}}, Nashville, TN, USA, January 2021. {AIAA}.

\bibitem{MN04}
Oded Maler and Dejan Nickovic.
\newblock Monitoring temporal properties of continuous signals.
\newblock In {\em Formal Techniques, Modelling and Analysis of Timed and
  Fault-Tolerant Systems}, pages 152--166. Springer, 2004.

\bibitem{NASA-FRET}
{NASA Technology Transfer Program}.
\newblock {FRET : Formal Requirements Elicitation Tool (ARC-18066-1)}.
\newblock Online: \url{https://software.nasa.gov/software/ARC-18066-1}, 2024.

\bibitem{DBLP:books/sp/NipkowPW02}
Tobias Nipkow, Lawrence~C. Paulson, and Markus Wenzel.
\newblock {\em Isabelle/{HOL} - {A} Proof Assistant for Higher-Order Logic},
  volume 2283 of {\em LNCS}.
\newblock Springer, 2002.
\newblock \href {https://doi.org/10.1007/3-540-45949-9}
  {\path{doi:10.1007/3-540-45949-9}}.

\bibitem{DBLP:conf/seke/NiuXXXHL23}
Tong Niu, Yicong Xu, Shengping Xiao, Lili Xiao, Yanhong Huang, and Jianwen Li.
\newblock $\text{LTL}_f$ satisfiability checking via formula progression.
\newblock In Shi{-}Kuo Chang, editor, {\em SEKE}, pages 357--362. {KSI}
  Research Inc., 2023.
\newblock \href {https://doi.org/10.18293/SEKE2023-143}
  {\path{doi:10.18293/SEKE2023-143}}.

\bibitem{JAXA}
Naoko Okubo.
\newblock {Using R2U2 in JAXA program}.
\newblock Electronic correspondence, November--December 2020.
\newblock Correspondance.

\bibitem{OW08}
J.~Ouaknine and J.~Worrell.
\newblock Some recent results in metric temporal logic.
\newblock In Franck Cassez and Claude Jard, editors, {\em Formal Modeling and
  Analysis of Timed Systems}, pages 1--13, Berlin, Heidelberg, 2008. Springer
  Berlin Heidelberg.

\bibitem{DBLP:journals/jar/Paulson89}
Lawrence~C. Paulson.
\newblock The foundation of a generic theorem prover.
\newblock {\em J. Autom. Reason.}, 5(3):363--397, 1989.
\newblock \href {https://doi.org/10.1007/BF00248324}
  {\path{doi:10.1007/BF00248324}}.

\bibitem{DBLP:conf/lpar/PaulsonB10}
Lawrence~C. Paulson and Jasmin~Christian Blanchette.
\newblock Three years of experience with {S}ledgehammer, a practical link
  between automatic and interactive theorem provers.
\newblock In Geoff Sutcliffe, Stephan Schulz, and Eugenia Ternovska, editors,
  {\em IWIL}, volume~2 of {\em EPiC Series in Computing}, pages 1--11.
  EasyChair, 2010.
\newblock \href {https://doi.org/10.29007/36DT} {\path{doi:10.29007/36DT}}.

\bibitem{Per23}
Ivan Perez.
\newblock Runtime verification with {O}gma.
\newblock In {\em Invited Talk to University of California}, 2023.

\bibitem{ogma}
Ivan Perez and Alwyn Goodloe.
\newblock {OGMA}.
\newblock \url{https://github.com/nasa/ogma}, 2021.

\bibitem{PMPGG22}
Ivan Perez, Anastasia Mavridou, Tom Pressburger, Alwyn Goodloe, and Dimitra
  Giannakopoulou.
\newblock Automated translation of natural language requirements to runtime
  monitors.
\newblock In Dana Fisman and Grigore Rosu, editors, {\em TACAS}, pages
  387--395, Cham, 2022. Springer International Publishing.

\bibitem{Pnu77}
A.~Pnueli.
\newblock The temporal logic of programs.
\newblock In {\em FOCS, Proc. 18th IEEE Symp.}, pages 46--57, 1977.

\bibitem{DBLP:conf/birthday/PnueliA03}
Amir Pnueli and Tamarah Arons.
\newblock {TLPVS:} {A} {PVS}-based {LTL} verification system.
\newblock In Nachum Dershowitz, editor, {\em Verification: Theory and Practice,
  Essays Dedicated to Zohar Manna on the Occasion of His 64th Birthday}, volume
  2772 of {\em LNCS}, pages 598--625. Springer, 2003.
\newblock \href {https://doi.org/10.1007/978-3-540-39910-0\_26}
  {\path{doi:10.1007/978-3-540-39910-0\_26}}.

\bibitem{DBLP:conf/atva/RaszykBT20}
Martin Raszyk, David~A. Basin, and Dmitriy Traytel.
\newblock Multi-head monitoring of metric dynamic logic.
\newblock In Dang~Van Hung and Oleg Sokolsky, editors, {\em ATVA}, volume 12302
  of {\em LNCS}, pages 233--250. Springer, 2020.
\newblock \href {https://doi.org/10.1007/978-3-030-59152-6\_13}
  {\path{doi:10.1007/978-3-030-59152-6\_13}}.

\bibitem{RRS14}
Thomas Reinbacher, Kristin~Y. Rozier, and Johann Schumann.
\newblock Temporal-logic based runtime observer pairs for system health
  management of real-time systems.
\newblock In {\em TACAS}, volume 8413 of {\em LNCS}, pages 357--372.
  Springer-Verlag, April 2014.

\bibitem{DBLP:conf/vmcai/Roohi018}
Nima Roohi and Mahesh Viswanathan.
\newblock Revisiting {MITL} to fix decision procedures.
\newblock In Isil Dillig and Jens Palsberg, editors, {\em VMCAI}, volume 10747
  of {\em LNCS}, pages 474--494. Springer, 2018.
\newblock \href {https://doi.org/10.1007/978-3-319-73721-8\_22}
  {\path{doi:10.1007/978-3-319-73721-8\_22}}.

\bibitem{RR25}
Alec Rosentrater and Kristin~Y. Rozier.
\newblock {FPROGG}: {A} formula progression-based {MLTL} benchmark generator.
\newblock Under Submission, 2024.

\bibitem{NFM25}
Alec Rosentrater, Zili Wang, Katherine Kosaian, and Kristin Rozier.
\newblock Language partitioning for {M}ission-time {L}inear {T}emporal {L}ogic.
\newblock Accepted to NFM 2025, to appear.

\bibitem{Roz16b}
K.~Y. Rozier.
\newblock {R2U2} in space: System and software health management for small
  satellites.
\newblock In {\em {Spacecraft Flight Software Workshop (FSW)}}, December 2016.
\newblock \url{https://www.youtube.com/watch?v=OAgQFuEGSi8}.
\newblock URL: \url{https://www.youtube.com/watch?v=OAgQFuEGSi8}.

\bibitem{RSI15}
Kristin~Y. Rozier, Johann Schumann, and Corey Ippolito.
\newblock {Intelligent Hardware-Enabled Sensor and Software Safety and Health
  Management for Autonomous UAS}.
\newblock {Technical Memorandum} {NASA/TM-2015-218817}, NASA, NASA Ames
  Research Center, Moffett Field, CA 94035, USA, May 2015.

\bibitem{Roz11}
Kristin~Yvonne Rozier.
\newblock {L}inear temporal logic {S}ymbolic {M}odel {C}hecking.
\newblock {\em Computer Science Review Journal}, 5(2):163--203, May 2011.
\newblock URL: \url{http://dx.doi.org/10.1016/j.cosrev.2010.06.002}, \href
  {https://doi.org/doi:10.1016/j.cosrev.2010.06.002}
  {\path{doi:doi:10.1016/j.cosrev.2010.06.002}}.

\bibitem{Roz17}
Kristin~Yvonne Rozier.
\newblock On the evaluation and comparison of runtime verification tools for
  hardware and cyber-physical systems.
\newblock In {\em RV-CUBES}, volume~3, pages 123--137, Seattle, WA, USA,
  September 2017. Kalpa Publications.
\newblock URL: \url{https://easychair.org/publications/paper/877G}, \href
  {https://doi.org/10.29007/pld3} {\path{doi:10.29007/pld3}}.

\bibitem{RS17}
Kristin~Yvonne Rozier and Johann Schumann.
\newblock {R2U2: Tool Overview}.
\newblock In {\em RV-CUBES}, volume~3, pages 138--156, Seattle, WA, USA,
  September 2017. Kalpa Publications.
\newblock URL: \url{https://easychair.org/publications/paper/Vncw}.

\bibitem{LTL_to_GBA-AFP}
Alexander Schimpf and Peter Lammich.
\newblock Converting linear-time temporal logic to generalized {B}üchi
  automata.
\newblock {\em Archive of Formal Proofs}, May 2014.
\newblock \url{https://isa-afp.org/entries/LTL_to_GBA.html}, Formal proof
  development.

\bibitem{SMR16}
Johann Schumann, Patrick Moosbrugger, and Kristin~Y. Rozier.
\newblock Runtime analysis with {R2U2}: {A} tool exhibition report.
\newblock In {\em RV}, Madrid, Spain, September 2016. Springer-Verlag.

\bibitem{SRRMMI15}
Johann Schumann, Kristin~Y. Rozier, Thomas Reinbacher, Ole~J. Mengshoel, Timmy
  Mbaya, and Corey Ippolito.
\newblock Towards real-time, on-board, hardware-supported sensor and software
  health management for unmanned aerial systems.
\newblock {\em IJPHM}, 6(1):1--27, June 2015.

\bibitem{LTL_Master_Theorem-AFP}
Benedikt Seidl and Salomon Sickert.
\newblock A compositional and unified translation of {LTL} into
  $\omega$-automata.
\newblock {\em Archive of Formal Proofs}, April 2019.
\newblock \url{https://isa-afp.org/entries/LTL_Master_Theorem.html}, Formal
  proof development.

\bibitem{LTL_to_DRA-AFP}
Salomon Sickert.
\newblock Converting linear temporal logic to deterministic (generalized)
  {R}abin automata.
\newblock {\em Archive of Formal Proofs}, September 2015.
\newblock \url{https://isa-afp.org/entries/LTL_to_DRA.html}, Formal proof
  development.

\bibitem{LTL-AFP}
Salomon Sickert.
\newblock Linear temporal logic.
\newblock {\em Archive of Formal Proofs}, March 2016.
\newblock \url{https://isa-afp.org/entries/LTL.html}, Formal proof development.

\bibitem{LTL_Normal_Form-AFP}
Salomon Sickert.
\newblock An efficient normalisation procedure for linear temporal logic:
  {Isabelle/HOL} formalisation.
\newblock {\em Archive of Formal Proofs}, May 2020.
\newblock \url{https://isa-afp.org/entries/LTL_Normal_Form.html}, Formal proof
  development.

\bibitem{fret-proof-framework}
Laura Titolo, Esther Conrad, Dimitra Giannakopoulou, Thomas Pressburger, and
  Aaron Dutle.
\newblock {FRET Proof Framework}.
\newblock \url{https://lauratitolo.github.io/project/fret-proof-framework/},
  2022.

\bibitem{Var08}
Moshe~Y Vardi.
\newblock From {C}hurch and {P}rior to {PSL}.
\newblock In {\em 25 Years of Model Checking: History, Achievements,
  Perspectives}, pages 150--171. Springer, 2008.

\bibitem{WEST2024tool}
Zili Wang, Laura~P. Gamboa-Guzman, and Kristin~Yvonne Rozier.
\newblock {WEST: Interactive Validation of Mission-time Linear Temporal Logic
  (MLTL)}.
\newblock 2024.
\newblock URL: \url{https://temporallogic.org/research/WEST/}.

\bibitem{WKR25}
Zili Wang, Katherine Kosaian, and Kristin Rozier.
\newblock Formally verifying a transformation from {MLTL} formulas to regular
  expressions.
\newblock Accepted to TACAS 2025, preprint available at:
  https://arxiv.org/abs/2501.17444.

\bibitem{ZADJR23}
Pei Zhang, Alexis~A. Aurandt, Rohit Dureja, Phillip~H. Jones, and
  Kristin~Yvonne Rozier.
\newblock Model predictive runtime verification for cyber-physical systems with
  real-time deadlines.
\newblock In Laure Petrucci and Jeremy Sproston, editors, {\em FORMATS}, volume
  14138 of {\em LNCS}, pages 158--180. Springer, 2023.
\newblock \href {https://doi.org/10.1007/978-3-031-42626-1\_10}
  {\path{doi:10.1007/978-3-031-42626-1\_10}}.

\end{thebibliography}

\appendix

\section{Appendix A}\label{app:A}
In this appendix, we present some of the code snippets for functions that we did not include (or only partially included) in the main body of the paper.
These may also be found in our code, but we include them here for easy reference.

\subsection{MLTL Semantics}\label{app:semantics}
We present our full formal definition of the semantics of MLTL (cross-reference \rref{sec:encoding}).
\begin{isabelle}
  \mltlsemantics
\end{isabelle}

This semantics is designed to closely match the mathematical semantics of MLTL while also including the well-definedness assumption for temporal operators (that \isa{a} \isasymle \isa{b}).

\subsection{Well-Defined Intervals}\label{app:intervals}
We present our Isabelle/HOL function \isa{intervals\_welldef}, which checks if all intervals associated to temporal operators in an input formula are well-defined.
\begin{isabelle}
\intervalswelldef
\end{isabelle}
This function is recursively defined.
It returns True on formulas that clearly have no temporal operators (\isa{True\isactrlsub m}, \isa{False\isactrlsub m}, and \isa{Prop\isactrlsub m}).
For formulas of shapes \isa{And\isactrlsub m}, \isa{Not\isactrlsub m}, and \isa{Or\isactrlsub m}, \isa{intervals\_welldef} checks whether the relevant subformulas contain well-defined intervals.
For example, for \isa{And\isactrlsub m\ \isasymphi\ \isasympsi}, this function checks whether the intervals in \isasymphi\ and \isasympsi\ are well-defined.
For the temporal operators, \isa{intervals\_welldef} checks whether the top-level interval is well-defined and also whether the intervals of any relevant subformulas are well-defined. 
For example, for \isa{intervals\_welldef\ F\_m\ [a, b]\ \isasymphi} to hold, we must have both \isa{a \isasymle\ b} and also \isa{intervals\_welldef\ \isasymphi}.

\subsection{Computation Length}\label{app:complen}
We present our full encoding of the comptuation length of an MLTL formula \isa{complen\_mltl} (cross-reference \rref{sec:functions}).
\begin{isabelle}
  \complenmltl
\end{isabelle}

\section{Appendix B} \label{app:B}
We detail the full formula progression function, and we present the correspoding encoding of the corresponding function in Isabelle, \isa{formula\_progression\_len1} (cross-reference \rref{sec:fpalg}).

Let $\pi$ be a trace and $\varphi$ and $\psi$ be MLTL formulas. The formula progression of $\varphi$ on $\pi$, denoted $\texttt{prog}(\varphi, \pi)$, is defined recursively as follows \cite[Definition 1]{LR18}:

\begin{itemize}
  \item {If $|\pi| = 1$, then
  \begin{itemize}
    \item $\texttt{prog}(\text{True}, \pi) = \text{True}$ and $\texttt{prog}(\text{False}, \pi) = \text{False}$;
    \item if $\varphi = p$ is an atomic proposition, \\
    $\texttt{prog}(p, \pi) = \text{True}$ if and only if $p \in \pi[0]$;
    \item $\texttt{prog}(\neg \varphi, \pi) = \neg \texttt{prog}(\varphi, \pi)$;
    \item $\texttt{prog}(\varphi \lor \psi, \pi) = \texttt{prog}(\varphi, \pi) \lor \texttt{prog}(\psi, \pi)$;
    \item $\texttt{prog}(\varphi \land \psi, \pi) = \texttt{prog}(\varphi, \pi) \land \texttt{prog}(\psi, \pi)$;
    \item $\texttt{prog}(\varphi \texttt{U}_{[a,b]} \psi, \pi) = 
    \left\{
      \begin{array}{ll}
        \varphi\ U_{[a-1, b-1]}\ \psi  & \mbox{if } 0 < a \leq b; \\
        \texttt{prog}(\psi, \isasympi)\ \lor \\ (\texttt{prog}(\varphi, \isasympi)\ \land \\
        \varphi\ U_{[0, b-1]}\ \psi)
            & \mbox{if } 0 = a < b; \\
            \texttt{prog}(\psi, \isasympi) & \mbox{if } 0 = a = b;
      \end{array}
    \right.$
    \item $\texttt{prog}(\texttt{F}_{[a,b]} \varphi, \pi) = 
    \left\{
      \begin{array}{ll}
        \texttt{F}_{[a-1,b-1]}\varphi & \mbox{if } 0 < a \leq b; \\
        \texttt{prog}(\varphi, \pi) \lor \\
        \texttt{F}_{[0,b-1]}\varphi & \mbox{if } 0 = a < b; \\
        \texttt{prog}(\varphi, \pi) & \mbox{if } 0 = a = b;
      \end{array}
    \right.$
    \item $\texttt{prog}(\varphi \texttt{R}_{[a,b]} \psi, \pi) = \neg \texttt{prog}((\neg \varphi) \texttt{U}_{[a,b]} (\neg \psi), \pi)$;
    \item $\texttt{prog}(\texttt{G}_{[a,b]} \varphi, \pi) = \neg \texttt{prog}(\texttt{F}_{[a,b]} (\neg \varphi), \pi)$.
  \end{itemize} 
  }
  \item Else, $\texttt{prog}(\varphi, \pi) = \texttt{prog}(\texttt{prog}(\varphi, \pi[0]), \pi_1).$
\end{itemize}

The $|\pi| > 1$ case is described in detail in the Isabelle function \isa{formula\_progression} in \rref{sec:fpalg}.
We present the full $|\pi| = 1$ case as the function \isa{formula\_progression\_len1} in Isabelle as follows:

\begin{isabelle}
  \proglenone
\end{isabelle}

This function exactly captures the structural cases of the formula progression function on traces of length 1, so that there is a clear correspondence between the mathematical definition and the Isabelle encoding.
The Future case splits into cases very similar to the cases that Until splits into, and the Release and Globally cases are defined via duality in terms of the Until and Future cases, respectively.

\end{document}